\documentclass[twocolumn,dvipsnames]{aastex63}

\usepackage{makecell,CJK}
\usepackage{amsmath}
\usepackage{multirow}
\usepackage{makecell}

\newcommand\revision[1]{{{#1}}}

\providecommand\aap{Astron.\ Astrophys.} 
\providecommand\apjl{Astrophys.\ J.} 
\providecommand\apjs{Astrophys.\ J.\ Suppl.} 

\accepted{2023 Feb 18}
\submitjournal{PSJ}

\shorttitle{Observational Characterization of Main-Belt Comet Nuclei}
\shortauthors{Hsieh et al.}

\begin{document}
\begin{CJK*}{UTF8}{gbsn}

\title{Observational Characterization of Main-Belt Comet and Candidate Main-Belt Comet Nuclei}

\correspondingauthor{Henry H.\ Hsieh}
\email{hhsieh@psi.edu}

\author[0000-0001-7225-9271]{Henry H.\ Hsieh}
\affiliation{Planetary Science Institute, 1700 East Fort Lowell Rd., Suite 106, Tucson, AZ 85719, USA}
\affiliation{Institute of Astronomy and Astrophysics, Academia Sinica, P.O.\ Box 23-141, Taipei 10617, Taiwan}






\author[0000-0001-7895-8209]{Marco Micheli}
\affiliation{ESA PDO NEO Coordination Centre, Largo Galileo Galilei, 1, I-00044 Frascati (RM), Italy}

\author[0000-0002-6702-7676]{Michael S.\ P.\ Kelley}
\affiliation{Department of Astronomy, University of Maryland, 1113 Physical Sciences Complex, Building 415, College Park, MD 20742, USA}

\author[0000-0003-2781-6897]{Matthew M.\ Knight}
\affiliation{Physics Department, U.S. Naval Academy, 572C Holloway Rd., Annapolis, MD, 21402, USA}
\affiliation{Department of Astronomy, University of Maryland, 1113 Physical Sciences Complex, Building 415, College Park, MD 20742, USA}

\author[0000-0001-6765-6336]{Nicholas A.\ Moskovitz}
\affiliation{Lowell Observatory, 1400 W.\ Mars Hill Rd, Flagstaff, AZ 86001, USA}


\author[0000-0002-5736-1857]{Jana Pittichov\'a}
\affiliation{Jet Propulsion Laboratory, California Institute of Technology, 4800 Oak Grove Dr, Pasadena, CA 91109, USA}

\author[0000-0003-3145-8682]{Scott S.\ Sheppard}
\affiliation{Earth and Planets Laboratory, Carnegie Institution for Science, 5241 Broad Branch Road NW, Washington, DC 20015, USA}

\author[0000-0002-1506-4248]{Audrey Thirouin}
\affiliation{Lowell Observatory, 1400 W.\ Mars Hill Rd, Flagstaff, AZ 86001, USA}

\author[0000-0001-9859-0894]{Chadwick A.\ Trujillo}
\affiliation{Department of Astronomy and Planetary Science, Northern Arizona University, Flagstaff, AZ 86011, USA}

\author[0000-0002-1341-0952]{Richard J.\ Wainscoat}
\affiliation{Institute for Astronomy, University of Hawaii, 2680 Woodlawn Drive, Honolulu, HI 96822, USA}

\author[0000-0002-0439-9341]{Robert J.\ Weryk}
\affiliation{Institute for Astronomy, University of Hawaii, 2680 Woodlawn Drive, Honolulu, HI 96822, USA}
\affiliation{Physics and Astronomy, The University of Western Ontario, 1151 Richmond Street, London, ON N6A 3K7, Canada}

\author[0000-0002-4838-7676]{Quanzhi Ye (叶泉志)}
\affiliation{Department of Astronomy, University of Maryland, 1113 Physical Sciences Complex, Building 415, College Park, MD 20742, USA}
\affiliation{Center for Space Physics, Boston University, 725 Commonwealth Ave, Boston, MA 02215, USA}

\begin{abstract}
We report observations of \revision{nine} MBCs or candidate MBCs, most of which were obtained when the targets were apparently inactive. We find effective nucleus radii (assuming albedos of $p_V=0.05\pm0.02$) of $r_n=(0.24\pm0.05)$ km for 238P/Read, $r_n=(0.9\pm0.2)$ km for 313P/Gibbs, $r_n=(0.6\pm0.1)$ km for 324P/La Sagra, $r_n=(1.0\pm0.2)$ km for 426P/PANSTARRS, $r_n=(0.5\pm0.1)$ km for 427P/ATLAS, $r_n<(0.3\pm0.1)$ km for P/2016 J1-A (PANSTARRS), $r_n<(0.17\pm0.04)$ km for P/2016 J1-B (PANSTARRS), \revision{$r_n\leq(0.5\pm0.2)$ km for P/2017 S9 (PANSTARRS)}, and $r_n=(0.4\pm0.1)$ km for P/2019 A3 (PANSTARRS). We identify evidence of activity in observations of 238P in 2021, and find similar inferred activity onset times and net initial mass loss rates for 238P \revision{during perihelion approaches} in 2010, 2016, and 2021. P/2016 J1-A and P/2016 J1-B are also found to be active in 2021 and 2022, making them collectively the tenth MBC confirmed to be recurrently active \revision{near perihelion} and therefore likely to be exhibiting sublimation-driven activity. The nucleus of 313P is found to have colors of $g'-r'=0.52\pm0.05$ and $r'-i'=0.22\pm0.07$, consistent with 313P being a Lixiaohua family member. We also report non-detections of P/2015 X6 (PANSTARRS), where we conclude that its current nucleus size is likely below our detection limits ($r_n\lesssim0.3$ km). Lastly, we find that of \revision{17} MBCs or candidate MBCs for which nucleus sizes (or inferred parent body sizes) have been estimated, $>$80\% have $r_n\leq1.0$ km, pointing to an apparent physical preference toward small MBCs, where we suggest that YORP spin-up may play a significant role in triggering and/or facilitating MBC activity.
\end{abstract}



\section{Introduction\label{section:intro}}

\setlength{\tabcolsep}{5pt}
\setlength{\extrarowheight}{0em}
\begin{table*}[htb!]
\caption{Orbital Parameters$^a$}
\centering
\smallskip
\footnotesize
\begin{tabular}{lccrrcrr}
\hline\hline
\multicolumn{1}{c}{Object}
 & \multicolumn{1}{c}{$a$$^b$}
 & \multicolumn{1}{c}{$e$$^c$}
 & \multicolumn{1}{c}{$i$$^d$}
 & \multicolumn{1}{c}{$q$$^e$}
 & \multicolumn{1}{c}{$Q$$^f$}
 & \multicolumn{1}{c}{$T_J$$^g$}
 & \multicolumn{1}{c}{ID$^h$}
 \\
\hline
238P/Read (P/2005 U1)      & 3.166 & 0.252 &  1.264 & 2.369 & 3.963 & 3.153 & 37 \\ 
313P/Gibbs (P/2014 S4)     & 3.154 & 0.242 & 10.966 & 2.392 & 3.917 & 3.133 & 23 \\ 
324P/La Sagra (P/2010 R2)  & 3.094 & 0.154 & 21.420 & 2.618 & 3.570 & 3.100 & 31 \\ 
426P/PANSTARRS (P/2019 A7) & 3.188 & 0.161 & 17.774 & 2.675 & 3.700 & 3.104 & 10 \\ 
427P/ATLAS (P/2017 S5)     & 3.171 & 0.313 & 11.849 & 2.178 & 4.163 & 3.092 & 14 \\ 
P/2015 X6 (PANSTARRS)      & 2.755 & 0.170 &  4.558 & 2.287 & 3.222 & 3.319 & 8 \\ 
P/2016 J1-A (PANSTARRS)    & 3.172 & 0.228 & 14.330 & 2.448 & 3.896 & 3.113 & 10 \\ 
P/2016 J1-B (PANSTARRS)    & 3.172 & 0.228 & 14.331 & 2.448 & 3.896 & 3.113 & 8 \\ 
P/2017 S9 (PANSTARRS)      & \revision{3.155} & \revision{0.304} & \revision{14.138} & \revision{2.195} & \revision{4.115} & \revision{3.088} & \revision{3} \\ 
P/2019 A3 (PANSTARRS)      & 3.148 & 0.265 & 15.365 & 2.313 & 3.982 & 3.099 & 7 \\ 
\hline
\hline
\multicolumn{8}{l}{$^a$ Obtained from JPL's Small Body Database on 2023 January 10. Uncertainties of all} \\
\multicolumn{8}{l}{$~~~$ orbital elements are smaller than the listed precision.} \\
\multicolumn{8}{l}{$^b$ Semimajor axis, in au.} \\
\multicolumn{8}{l}{$^c$ Eccentricity.} \\
\multicolumn{8}{l}{$^d$ Inclination, in degrees.} \\
\multicolumn{8}{l}{$^e$ Perihelion distance, in au.} \\
\multicolumn{8}{l}{$^f$ Aphelion distance, in au.} \\
\multicolumn{8}{l}{$^g$ Tisserand parameter with respect to Jupiter.} \\
\multicolumn{8}{l}{$^h$ JPL orbit solution identification number.} \\
\end{tabular}
\phantomsection
\label{table:orbital_params}
\end{table*}

\subsection{Background\label{section:background}}

Active asteroids are small solar system bodies that are considered to have asteroid-like orbits based on their Tisserand parameter values, $T_J$, yet exhibit visible comet-like dust emission activity \citep{jewitt2015_actvasts_ast4}.  
\revision{Specifically, objects with $T_J>3$ are considered to have asteroid-like orbits while objects with $T_J<3$ are considered to have comet-like orbits \citep{vaghi1973_tisserand}, where $T_J$ parameterizes the relative velocity between an object and Jupiter at their closest approach and thus approximately characterizes an object's dynamical stability. For active asteroids, this threshold is sometimes set slightly higher \citep[e.g., $T_J>3.08$;][]{jewitt2015_actvasts_ast4} due to the identification of objects with $3.00<T_J<3.10$ that still exhibit comet-like dynamical behavior \citep[e.g.,][]{tancredi2014_asteroidcometclassification,hsieh2016_tisserand}.}
An important subset of this relatively recently recognized, and thus still poorly characterized, population are the main-belt comets (MBCs).  MBCs have orbits that specifically place them in the main asteroid belt (while some other active asteroids are on near-Earth object, or NEO, orbits) and exhibit activity that has been determined to be due, at least partially, to the sublimation of volatile material \citep{hsieh2006_mbcs}.

While it has not been possible to date to directly detect the presence of sublimation products in MBCs to confirm that activity is sublimation-driven \citep[e.g.,][]{snodgrass2017_mbcs}, \revision{imaging of} recurrent \revision{dust emission} activity near perihelion and inactivity elsewhere have been identified as being consistent with expectations for sublimation-driven active behavior and not easily explained by other proposed activity mechanisms such as impacts or rotational destabilization \citep[e.g.,][]{hsieh2004_133p,hsieh2012_scheila,jewitt2015_actvasts_ast4}.  
Such behavior is therefore considered strong evidence that a given active asteroid is a MBC.  \revision{In the future, JWST may be able to provide more direct evidence of sublimation-driven activity, via spectroscopic detection of the $\nu_3$ or $\nu_2$ fluorescence bands of water at 2.7~$\mu$m and 6.3~$\mu$m, respectively \citep{snodgrass2017_mbcs}, which will help to test the validity and reliability of this interpretation of imaging observations.}

Physical characterization of MBC nuclei is needed to understand the relationship of the population of active MBCs to the background population of inactive asteroids. Such studies of MBC nuclei indicate what characteristics dormant MBCs might have \citep[e.g.,][]{hsieh2014_324p}, provide improved inputs for thermal modeling studies \citep[e.g.,][]{schorghofer2008_mbaice,prialnik2009_mbaice,capria2012_mbcactivity}, and enable quantitative analyses of total dust production \citep[e.g.,][]{hsieh2018_238p288p,hsieh2021_259p} and photometric searches for low-level activity \citep[e.g.,][]{hsieh2011_176p,hsieh2015_324p,hsieh2018_358p}.

With the issues above in mind, we set out in this work to constrain the sizes of several likely MBCs and candidate MBCs \revision{(i.e., active asteroids whose likely activity mechanisms have not yet been definitively constrained but whose observed behavior so far is consistent with that of other MBCs)}.  In the following sections, we briefly summarize work reported to date related to the physical characterization of the targets considered in this work.
Table~\ref{table:orbital_params} lists orbital parameters for each target obtained from the Small Body Database\footnote{\url{https://ssd.jpl.nasa.gov/tools/sbdb_lookup.html}} maintained by the Jet Propulsion Laboratory (JPL).

\subsection{238P/Read\label{section:targets_238p}}

Comet 238P/Read (hereafter 238P), previously designated P/2005 U1, was discovered on 2005 October 24 by \citet{read2005_238p}
when the object was at a true anomaly of $\nu=26.4^{\circ}$ \revision{(where $\nu=0\degr$ corresponds to perihelion and $\nu=180\degr$ corresponds to aphelion)}, heliocentric distance of $r_h=2.416$~au, and geocentric distance of $\Delta=1.464$~au.  An observational analysis conducted shortly after its discovery indicated that its activity 
was consistent with sublimation-driven activity \citep{hsieh2009_238p}, while observations indicating that the comet had become active again in 2010 and 2016 corroborated this assessment \citep{hsieh2011_238p,hsieh2018_238p288p}.  Together, these observations made 238P the second active asteroid after 133P/Elst-Pizarro to be observed to exhibit recurrent activity, strongly suggesting that it is a MBC.  \citet{hsieh2011_238p} found a best-fit absolute magnitude of $H_R=19.05\pm0.05$ for 238P's nucleus, corresponding to an effective nucleus radius of $r_n\sim0.4$~km (assuming an $R$-band albedo of $p_R=0.05$).


\subsection{313P/Gibbs\label{section:targets_313p}}

Comet 313P/Gibbs (hereafter 313P), previously designated P/2014 S4, was discovered on 2014 September 25 
when the object was at $\nu=8.2^{\circ}$, $r_h=2.396$~au, and $\Delta=1.434$~au.  Archival data from the Sloan Digital Sky Survey showed that the comet was also active during a previous perihelion passage in 2003 \citep{hsieh2015_313p,hui2015_313p}, while a third active apparition was also subsequently observed in 2019 \citep{hsieh2019_313p}, making 313P's activity highly likely to be the result of the sublimation of volatile material, and therefore 313P to be a MBC.
\citet{hsieh2015_313p} reported that dust emission in both 2003 and 2014 persisted over at least three months, which was corroborated by a dust modeling analysis of 313P's 2014 activity conducted by \citet{pozuelos2015_313p} indicating that dust emission
lasted at least four months, consistent with expected behavior from sublimation-driven dust emission activity.

Archival observations from the Subaru Telescope in 2004 when the object was apparently inactive were used by \citet{hsieh2015_313p} to estimate a lower-limit $R$-band absolute magnitude for the nucleus of $H_R=17.1\pm0.3$, corresponding to an upper-limit effective nucleus radius of $r_n\sim(1.00\pm0.15)$~km, assuming an $R$-band albedo of $p_R=0.05$.  Meanwhile, \citet{jewitt2015_313p2} estimated a nucleus radius of $r_n=(0.7\pm0.1)$~km from high-resolution {\it Hubble Space Telescope} ({\it HST}) imaging observations obtained in 2015 when the object was nearly inactive.

\subsection{324P/La Sagra\label{section:target_324p}}

Comet 324P/La Sagra (hereafter 324P), previously designated P/2010 R2, was discovered on 2010 September 14 \citep{nomen2010_324p} when the object at $\nu=20.0^{\circ}$, $r_h=2.644$~au, and $\Delta=1.743$~au.  A dust modeling analysis of observations obtained shortly after its discovery in 2010 indicated that dust emission persisted over a period of at least $\sim7$ months following perihelion \citep{moreno2011_324p}, consistent with being driven by sublimation.  Observations in 2015 also confirmed that 324P had become active again while approaching perihelion, which was interpreted as strong evidence that it is indeed a MBC \citep{hsieh2015_324p}.  \citet{hsieh2014_324p} found best-fit $HG$ phase function parameters of $H_R=18.4\pm0.2$ and $G_R=0.17\pm0.10$, corresponding to an estimated nucleus radius of $r_N=(0.55\pm0.05)$~km (assuming $p_R=0.05$).


\subsection{426P/PANSTARRS\label{section:target_426p}}

Comet 426P/PANSTARRS (hereafter 426P), previously designated P/2019 A7, was discovered on 2019 January 8 by the Pan-STARRS1 (PS1) survey telescope \citep{ramanjooloo2019_p2019a7} when the object was at $\nu=81.1^{\circ}$, $r_h=3.030$~au, and $\Delta=2.089$~au.  As of \revision{2023 February 1}, no analyses of this object have appeared in published literature, although its activity near perihelion and semimajor axis placing it in the outer main asteroid belt are consistent with other MBCs.  We therefore consider 426P as a candidate MBC for the purposes of this work.

\subsection{427P/ATLAS\label{section:target_427p}}

Comet 427P/ATLAS (hereafter 427P), previously designated P/2017 S5, was discovered on 2017 September 27 by the Asteroid Terrestrial-impact Last Alert System (ATLAS) survey telescope \citep{heinze2017_p2017s5_cbet} when the object was at $\nu=21.3^{\circ}$, $r_h=2.214$~au, and $\Delta=1.267$~au.  \citet{jewitt2019_p2017s5} conducted an observational analysis shortly after its discovery,
finding that the comet's activity was due to a prolonged dust emission event occurring over $\sim$150 days, consistent with sublimation-driven emission.  Using follow-up observations obtained using HST the following year when the object was apparently inactive, \citet{jewitt2019_p2017s5} also estimated 427P's nucleus to have an effective radius of $450^{+100}_{-60}$~m, assuming an albedo of $p=0.06\pm0.02$. No evidence of rotational variation in the nucleus's brightness was found in the HST data when the object was apparently inactive, although the authors suggest that the nucleus could be a rapid rotator with a rotation period of $P_{\rm rot}\sim1.4$~h based on photometric variations observed during the object's active phase.  While 427P has not yet been confirmed to be recurrently active, based on dust modeling results indicating prolonged activity near perihelion in 2017 and its semimajor axis in the outer main asteroid belt, we consider it to be a candidate MBC for the purposes of this work.

\subsection{P/2015 X6 (PANSTARRS)\label{section:target_p2015x6}}

Comet P/2015 X6 (PANSTARRS) was discovered on 2015 December 7 by PS1 \citep{lilly2015_p2015x6_cbet} when the object was at $\nu=328.9^{\circ}$, $r_h=2.336$~au, and $\Delta=1.559$~au.  An observational and dust modeling analysis performed by \citet{moreno2016_p2015x6} indicated that its activity was produced by a sustained dust ejection event lasting for at least two months, consistent with the activity being driven by sublimation, or potentially rotational destabilization.  Activity was determined to have started between 18 and 26 days before discovery with an estimated average mass loss rate from that time until 2016 January 26 on the order of $\sim$1~kg~s$^{-1}$.
While P/2015 X6 has not yet been confirmed to be recurrently active, based on dust modeling results from \citet{moreno2016_p2015x6} indicating prolonged activity near perihelion in 2015-2016, we consider it to be a candidate MBC for the purposes of this work.


\subsection{P/2016 J1 (PANSTARRS)\label{section:target_p2016j1}}

Comet P/2016 J1 (PANSTARRS) was discovered on 2016 May 5 by PS1 consisting of two distinct fragments, P/2016 J1-A and P/2016 J1-B \citep{weryk2016_p2016j1_cbet,wainscoat2016_p2016j1_mpec}, when the object was at $\nu=345.8^{\circ}$, $r_h=2.462$~au, and $\Delta=1.479$~au.  Using observations obtained on 2016 August 4 when activity was still present, \citet{hui2017_p2016j1} estimated component radii of $r_n \lesssim 900~$m for P/2016 J1-A and $r_n \lesssim 400$~m for P/2016 J1-B.  They furthermore reported that a syndyne-synchrone analysis indicated that both components had been active for 3-6 months prior to their observations, suggesting the action of volatile sublimation, with ejection speeds of $\sim$0.5~m~s$^{-1}$ and mass loss rates of $\sim$1~kg~s$^{-1}$ and $\sim$0.1~kg~s$^{-1}$ for fragments A and B, respectively.  Meanwhile, \citet{moreno2017_p2016j1} determined that activity likely started $\sim$8 months prior to the current perihelion passage with similar maximum mass loss rates of $\sim$0.7~kg~s$^{-1}$ and $\sim$0.5~kg~s$^{-1}$ for fragments A and B, respectively.  
While prior to this work, neither P/2016 J1-A nor P/2016 J1-B had been confirmed to be recurrently active, based on dust modeling results from \citet{moreno2017_p2016j1} indicating prolonged activity near perihelion in 2016, we considered it to be a candidate MBC for the purposes of this work.



\subsection{P/2017 S9 (PANSTARRS)\label{section:target_p2017s9}}

Comet P/2019 S9 (PANSTARRS) was discovered on 2017 September 30 by PS1 \citep{weryk2017_p2017s9} when the object was at $\nu=23.6^{\circ}$, $r_h=2.239$~au, and $\Delta=1.639$~au.  As of \revision{2023 February 1}, no analyses of this object have appeared in refereed published literature.  We note, however, that its activity near perihelion and semimajor axis placing it in the outer main asteroid belt are consistent with other MBCs.  We therefore consider P/2017 S9 as a candidate MBC for the purposes of this work.

\subsection{P/2019 A3 (PANSTARRS)\label{section:target_p2019a3}}

Comet P/2019 A3 (PANSTARRS) was discovered on 2019 January 3 by PS1  \citep{weryk2019_p2019a3} when the object was at $\nu=46.4^{\circ}$, $r_h=2.474$~au, and $\Delta=2.147$~au.  As of \revision{2023 February 1}, no analyses of this object have appeared in refereed published literature. We note, however, that its activity near perihelion and semimajor axis placing it in the outer main asteroid belt are consistent with other MBCs.  We therefore consider P/2019 A3 as a candidate MBC for the purposes of this work.




\section{Observations}\label{section:observations}

Observations were obtained with the 8.1~m Gemini North (Gemini-N) telescope (programs GN-2013A-Q-102, GN-2016A-Q-88, GN-2016B-LP-11, GN-2017A-LP-11, GN-2020B-LP-104, GN-2021A-LP-104, GN-2022A-LP-104, and GN-2022B-Q-307), using the Gemini Multi-Object Spectrograph - North \citep[GMOS-N;][]{hook2004_gmos} in imaging mode, and the 3.54~m Canada-France-Hawaii Telescope (CFHT; program 15BT12), using MegaCam \citep{boulade2003_megacam}, on Maunakea in Hawaii, the 8.1~m Gemini South (Gemini-S) telescope (programs GS-2020A-LP-104, GS-2021A-LP-104, GS-2021B-LP-104, GS-2022A-LP-104, GS-2022B-LP-104, and GS-2022B-Q-111), using the Gemini Multi-Object Spectrograph - South \citep[GMOS-S;][]{gimeno2016_gmoss} in imaging mode, at Cerro Pachon in Chile, the 6.5~m Baade Magellan telescope, using the Inamori Magellan Areal Camera and Spectrograph \citep[IMACS;][]{dressler2011_imacs}, at Las Campanas in Chile, the 5.1~m Hale Telescope, using the Wafer-Scale camera for Prime \citep[WaSP;][]{nikzad2017_detectors} wide field prime focus camera, at Palomar Observatory in California, and Lowell Observatory's 4.3~m Lowell Discovery Telescope (LDT), using the Large Monolithic Imager \citep[LMI;][]{bida2014_dct} at Happy Jack, Arizona.  \revision{Details of all instrumentation are shown in Table~\ref{table:instrumentation}.}

\setlength{\tabcolsep}{4.5pt}
\setlength{\extrarowheight}{0em}
\begin{table*}[htb!]
\caption{\revision{Observing Instrumentation Characteristics}}
\centering
\smallskip
\footnotesize
\begin{tabular}{lccccc}
\hline\hline
\multicolumn{1}{c}{Telescope}
 & \multicolumn{1}{c}{Instrument}
 & \multicolumn{1}{c}{FOV$^a$}
 & \multicolumn{1}{c}{Pixel Scale}
 & \multicolumn{1}{c}{Binning}
 & \multicolumn{1}{c}{Filters}
 \\[2pt]
\hline
 Gemini-N & GMOS-N  & $5\farcm5\times5\farcm5$ & $0\farcs1614$ & 2$\times$2 & $r'$ \\
 Gemini-S & GMOS-S  & $5\farcm5\times5\farcm5$ & $0\farcs1614$ & 2$\times$2 & $r'$ \\
 Baade    & IMACS   & $15\farcm4\times15\farcm4$ & $0\farcs20$ & 1$\times$1 & $r'$,$R$ \\
 Hale     & WaSP    & $18\farcm4\times18\farcm5$ & $0\farcs18$ & 1$\times$1 & $r'$ \\
 LDT      & LMI     & $12\farcm3\times12\farcm3$ & $0\farcs24$ & 2$\times$2 & $g'$,$r'$,$i'$ \\
 CFHT     & MegaCam & $0\fdg96\times0\fdg94$   & $0\farcs187$  & 1$\times$1 & $r'$ \\
\hline
\hline
\multicolumn{6}{l}{$^a$ Field of view} \\
\end{tabular}
\label{table:instrumentation}
\end{table*}

All observations were conducted using non-sidereal tracking and at airmasses of $<2.0$, \revision{with typical seeing conditions of $0\farcs7<\theta_s<1\farcs5$}, where dither offsets of up to $10''$ east or west, and north or south were applied to each individual exposure.  \revision{A minimum of 3 exposures per target was obtained during each visit in order to verify the identity of each object (and any associated activity) from its non-sidereal motion.  In some cases, however, certain detections in a sequence were discarded due to being too close to background sources for photometry to be reliable (see below), leading to fewer than 3 exposures per target on a night being reported here.}

Standard bias subtraction, flatfield correction, and cosmic ray removal were performed for all images using Python 3 code utilizing the {\tt ccdproc} package\footnote{\url{https://ccdproc.readthedocs.io/}} \citep{craig2017_ccdproc} in Astropy\footnote{\url{http://www.astropy.org}} \citep{astropy2018_astropy} and the {\tt L.A.Cosmic} python module\footnote{Written for python by Maltes Tewes; \\{\url{https://github.com/RyleighFitz/LACosmics}}} \citep{vandokkum2001_lacosmic,vandokkum2012_lacosmic}.
Photometry measurements of the target object and at least one background reference star were performed using IRAF \citep{tody1986_iraf,tody1993_iraf} \revision{and pyraf software\footnote{\url{https://pypi.org/project/pyraf/}} \citep{stsci2012_pyraf}},
\revision{where photometry of reference stars was obtained by measuring net fluxes within circular apertures with sizes chosen using curve-of-growth analyses of representative stars, with background sampled from surrounding circular annuli.
Meanwhile, photometry of target objects was performed using circular apertures with sizes chosen using curve-of-growth analyses of each target object detection, where background statistics were measured in nearby but non-adjacent regions of blank sky to avoid potential dust contamination from the object or nearby field stars.}

Absolute photometric calibration was performed using field star magnitudes from the {\tt refcat} all-sky stellar reference catalog \citep{tonry2018_refcat}.  We aimed to use 5-30 \revision{well-isolated} reference stars \revision{(i.e., field stars with no other neighboring sources within the photometry aperture used for those data, and ideally, within the annuli used to measure sky background as well)} for photometric calibration where possible.
In some cases, however, only a few \revision{suitable} reference stars, or even just one, were available due to the small margin between the limiting magnitude of the {\tt refcat} catalog, especially at large southern declinations, and the saturation limit of many of our observations. \revision{Dense background star fields also limited the availability of suitable reference stars in some cases}.  
\revision{In all cases, when detections of our target objects themselves were deemed to be too close to background sources for photometry to be reliable (as determined from curve-of-growth analyses), photometric measurements of those detections were rejected.}

Conversion of $r'$-band Gemini and PS1 photometry to $R$-band was accomplished using transformations derived by \citet{tonry2012_ps1} and by R.\ Lupton\footnote{{\url{http://www.sdss.org/}}}.  

To maximize signal-to-noise ratios for the purposes of searching for possible faint activity (which we would like to avoid for phase function determination purposes), we construct composite images of the object for each night of data by shifting and aligning individual images on the object's photocenter using linear interpolation and then adding the images together.  

\begin{figure*}
\figurenum{1}
\plotone{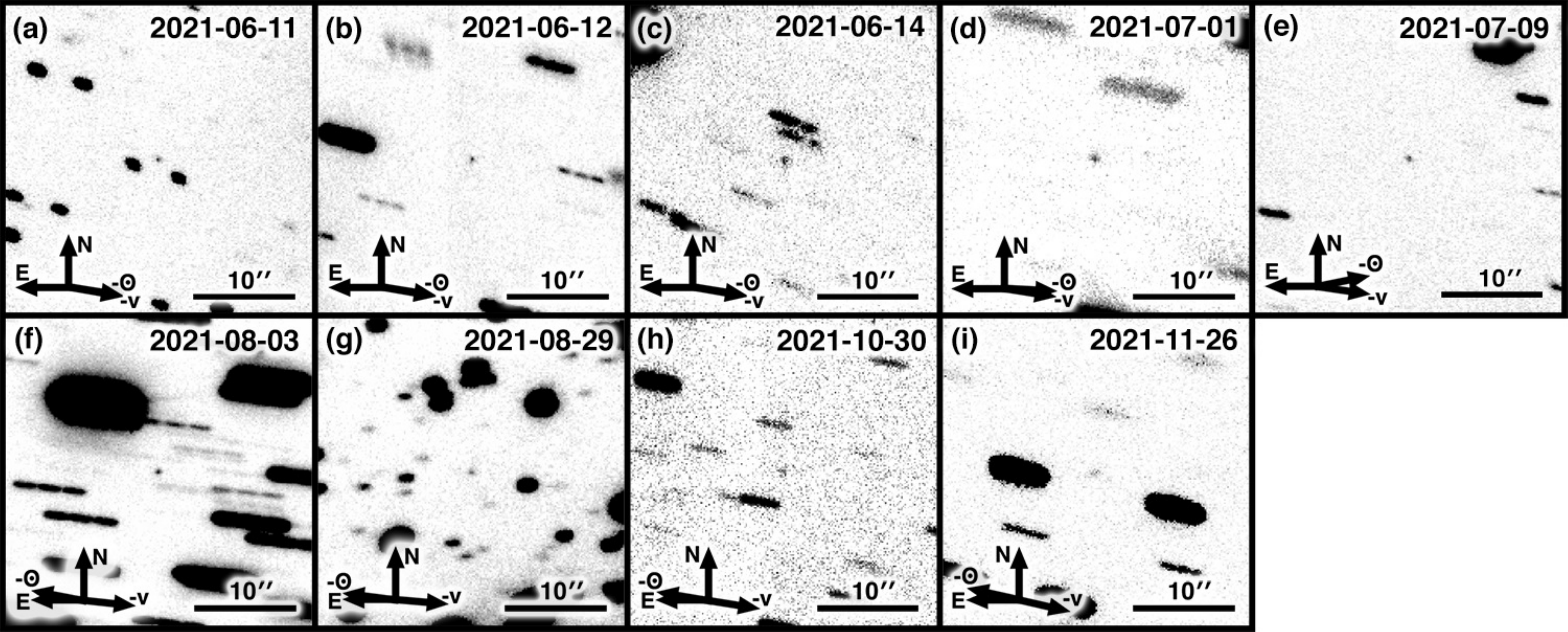}
\caption{\revision{Composite images of 238P constructed from data detailed in Table~\ref{table:observations}.  Scale bars indicate the size of each panel.  North (N), East (E), the antisolar direction ($-\odot$), and the negative heliocentric velocity direction ($-v$) are indicated in each panel.  The object is located at the center of each panel. The complete figure set (9 images) is available in the online journal.}}
\label{figure:observations_238p}
\end{figure*}

Successful detections were obtained for 238P, 313P, 426P, 427P, P/2016 J1-A, P/2016 J1-B, P/2017 S9, and P/2019 A3 \revision{(see Table~\ref{table:observations} for all observation details)}.
\revision{An example set of nightly composite images for 238P is shown in Figure~\ref{figure:observations_238p}, while sets of nightly composite images for other targets are available in Figure Set 1.}

\setlength{\tabcolsep}{3.5pt}
\setlength{\extrarowheight}{0em}
\begin{table*}[htb!]
\caption{\revision{Observations}}
\centering
\smallskip
\footnotesize
\tablecomments{\revision{Table~\ref{table:observations} is published online in its entirety in machine-readable format.  A portion is shown here for guidance regarding its form and content.}}
\begin{tabular}{llccrcrccrcccccc}
\hline\hline
\multicolumn{1}{c}{Target}
 & \multicolumn{1}{c}{UT Date$^a$}
 & \multicolumn{1}{c}{Tel.$^b$}
 & \multicolumn{1}{c}{$N$$^c$}
 & \multicolumn{1}{c}{$t$$^d$}
 & \multicolumn{1}{c}{Filt.}
 & \multicolumn{1}{c}{$\nu$$^e$}
 & \multicolumn{1}{c}{$r_h$$^f$}
 & \multicolumn{1}{c}{$\Delta$$^g$}
 & \multicolumn{1}{c}{$\alpha$$^h$}
 & \multicolumn{1}{c}{$m_{\rm app}$$^i$}
 & \multicolumn{1}{c}{$m_{{\rm app},r}$$^j$}
 & \multicolumn{1}{c}{$m_{r}(1,1,\alpha)$$^k$}
 & \multicolumn{1}{c}{$m_{r}(1,1,0)$$^l$}
 & \multicolumn{1}{c}{PhFn?$^m$}
 & \multicolumn{1}{c}{Ref.$^n$}
 \\
\hline
238P & 2021-06-11 & GS &  2 &  600 & $r'$ & 268.3 & 2.987 & 2.099 & 11.3 & 24.85$\pm$0.10 & n/a & 20.86$\pm$0.10 & 20.41$\pm$0.19 & y & n/a \\
238P & 2021-06-12 & GS &  3 &  900 & $r'$ & 268.5 & 2.985 & 2.089 & 11.0 & 24.89$\pm$0.08 & n/a & 20.92$\pm$0.08 & 20.47$\pm$0.18 & y & n/a \\
238P & 2021-06-14 & GS &  2 &  600 & $r'$ & 268.9 & 2.980 & 2.069 & 10.4 & 24.62$\pm$0.12 & n/a & 20.67$\pm$0.12 & 20.24$\pm$0.20 & y & n/a \\
238P & 2021-07-01 & GS &  2 &  600 & $r'$ & 272.2 & 2.937 & 1.938 &  4.6 & 24.50$\pm$0.15 & n/a & 20.72$\pm$0.15 & 20.47$\pm$0.18 & y & n/a \\
238P & 2021-07-09 & GS &  4 &  600 & $r'$ & 273.8 & 2.917 & 1.902 &  1.6 & 24.32$\pm$0.07 & n/a & 20.60$\pm$0.07 & 20.47$\pm$0.09 & y & n/a \\
\hline
\hline
\multicolumn{16}{l}{$^a$ UT data in YYYY-MM-DD format.} \\
\multicolumn{16}{l}{$^b$ Telescope (GS: 8.1 m Gemini South telescope).} \\
\multicolumn{16}{l}{$^c$ Number of exposures.} \\
\multicolumn{16}{l}{$^d$ Total integration time, in seconds.} \\
\multicolumn{16}{l}{$^e$ True anomaly, in degrees.} \\
\multicolumn{16}{l}{$^f$ Heliocentric distance, in au.} \\
\multicolumn{16}{l}{$^g$ Geocentric distance, in au.} \\
\multicolumn{16}{l}{$^h$ Solar phase angle (Sun-object-Earth), in degrees.} \\
\multicolumn{16}{l}{$^i$ Mean apparent magnitude in specified filter.} \\
\multicolumn{16}{l}{$^j$ Equivalent apparent $r'$-band magnitude for non-$r'$-band observations, assuming solar colors \citep[using][]{holmberg2006_solarcolors}.} \\
\multicolumn{16}{l}{$^k$ Reduced magnitude (normalized to $r_h=\Delta=1$~au) computed from measured apparent magnitude.} \\
\multicolumn{16}{l}{$^l$ Absolute magnitude (normalized to $r_h=\Delta=1$~au) computed using best-fit phase function parameters (Table~\ref{table:nucleus_phase_function_params}).} \\
\multicolumn{16}{l}{$^m$ Data used in this work for phase function fitting? (y: yes; n: no).} \\
\multicolumn{16}{l}{$^n$ Reference for previously reported data.} \\
\end{tabular}
\label{table:observations}
\end{table*}

\section{Results and Analysis}\label{section:results}

\subsection{Phase Function Fitting\label{section:phase_functions}}

We use a Monte Carlo-style approach to obtain phase function parameter fits for targets for which we were able to successfully observe and measure photometry.  In this approach, we generate a large number of test data sets based on the original photometric measurements obtained for each object, perform individual phase function fits to each test data set, and then identify the median best-fit values and 1-$\sigma$ intervals
for the phase function from the distribution of the best-fit values derived from the set of fitting runs.  Individual test data sets were generated by starting with the original mean magnitudes measured during each visit to each object (Table~\ref{table:observations}), and applying Gaussian-distributed offsets to each photometric point characterized by 1-$\sigma$ values equal to the measured uncertainties of each photometric point.
Phase function fits were then performed using the Levenberg-Marquardt algorithm for performing least-squares fitting of data to a specified model as implemented by the {\tt LevMarLSQFitter} function\footnote{\url{https://docs.astropy.org/en/stable/api/astropy.modeling.fitting.LevMarLSQFitter.html}} in {\tt astropy}.
We then adopt median values from the resulting distributions of best-fit values determined from individual fitting runs as our nominal solutions, and also use those distributions of best-fit values to determine upper and lower 1-$\sigma$ uncertainty intervals (i.e., intervals enclosing 34.135\% of the total sample of test run results above and below the computed median value).

\begin{figure*}[htb!]
\centerline{\includegraphics[width=7.0in]{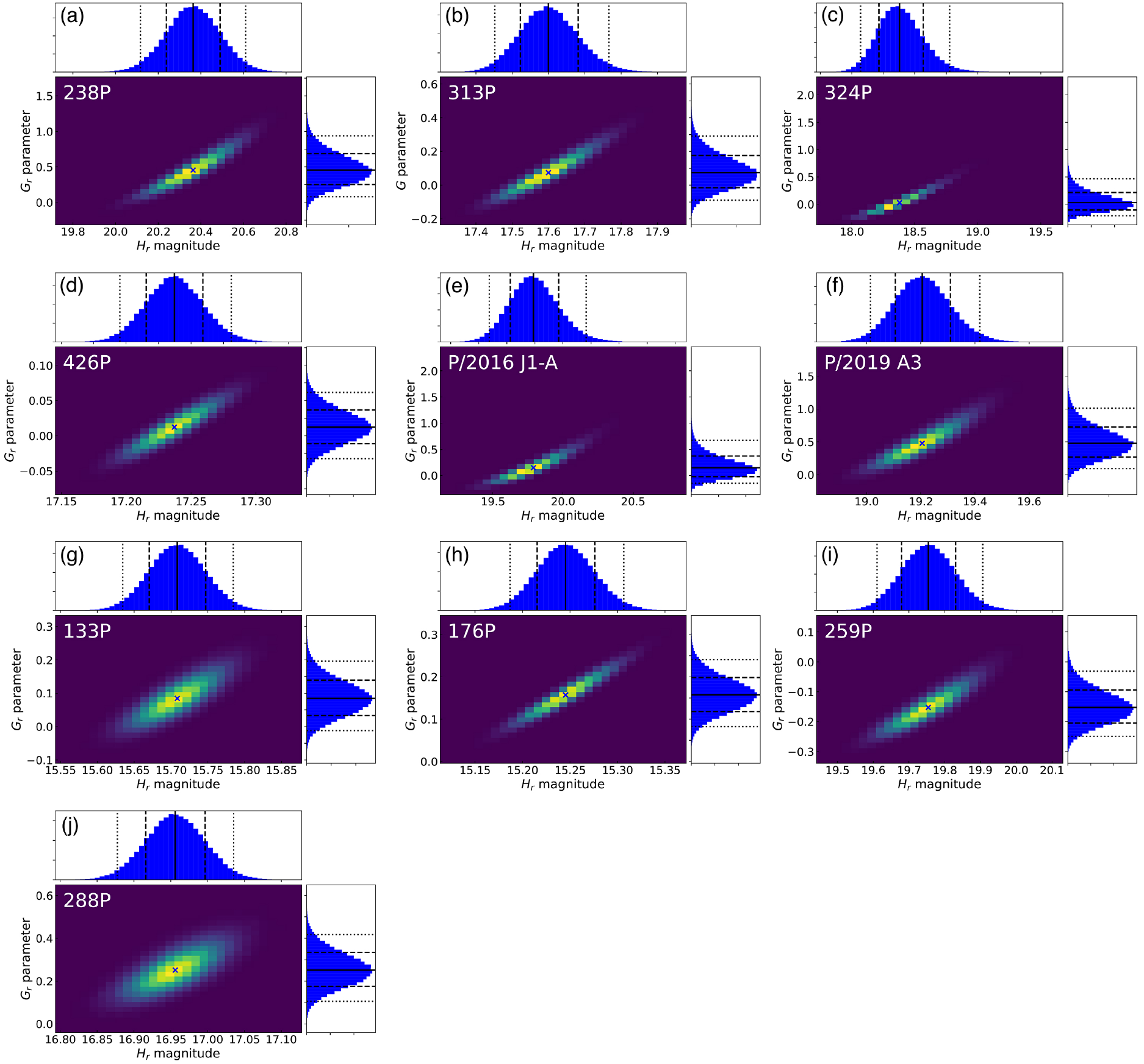}}
\caption{\small Two-dimensional heat map (main panel at lower left) showing the distribution of best-fit $H_{r}$ and $G_{r}$ parameters found for (a) 238P, (b) 313P, (c) 324P, (d) 426P, (e) P/2016 J1-A, (f) P/2019 A3, \revision{(g) 133P, (h) 176P, (i) 259P, and (j) 288P} from 100\,000 individual fitting runs as described in the text, where brighter colors (e.g., yellow) show regions of higher occurrences of values and darker colors (e.g., dark purple) show regions of lower occurrences of values.  One-dimensional histograms showing the individual distributions of $H_r$ and $G_r$ parameter values appear above and to the right of the main panel, respectively, where black lines show the center of each distribution (i.e., the most likely best-fit value), dashed lines enclose the 1-$\sigma$ interval for each parameter value, and dotted lines enclose the 95\% confidence interval for each parameter value. \revision{Panels (a)-(f) are based on new photometric data reported in this work, while panels (g)-(j) are based on previously reported photometric data.
}
}
\label{figure:hg_distributions}
\end{figure*}

\begin{figure*}[htb!]
\centerline{\includegraphics[width=6.5in]{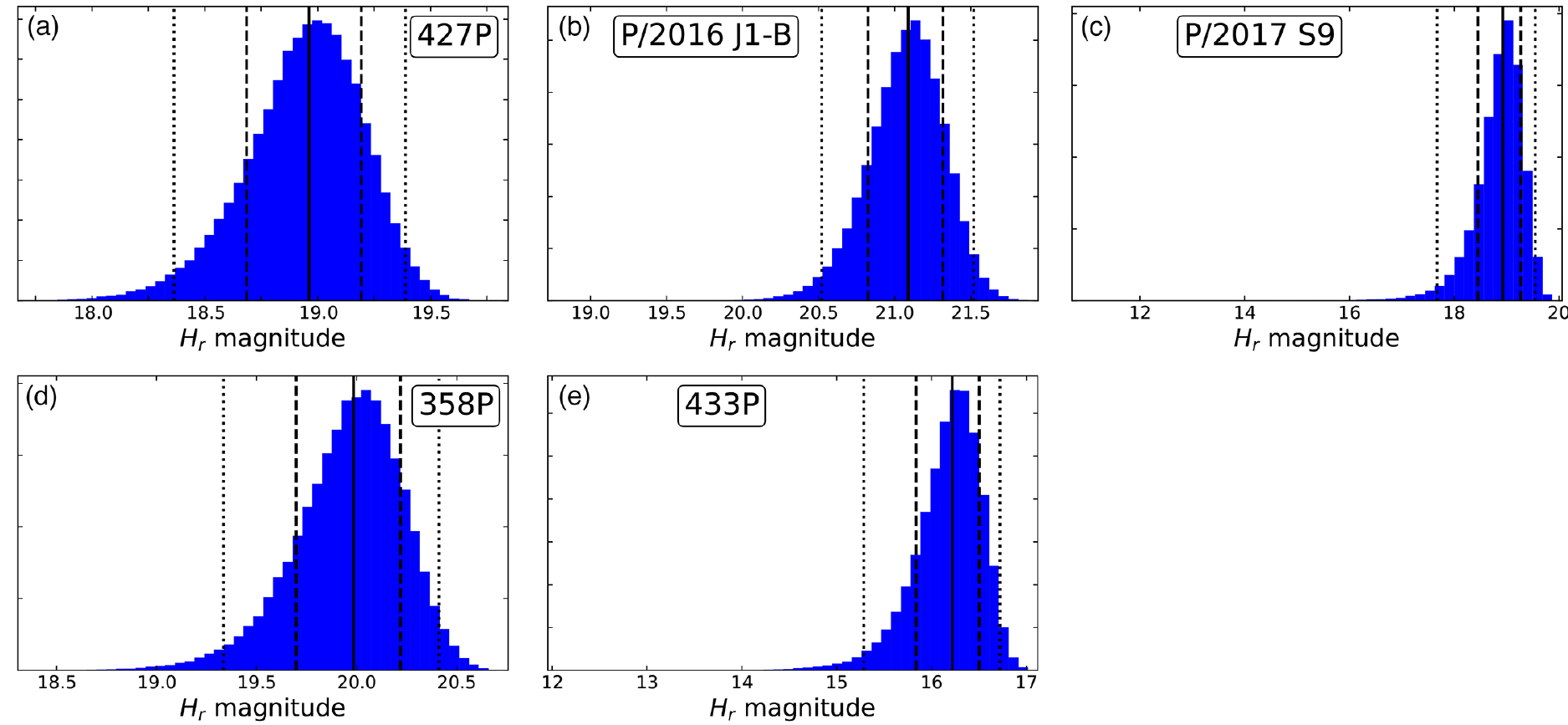}}
\caption{\small Histograms showing the distribution of best-fit lower-limit $H_{r}$ magnitudes (assuming $G_{r}=0.15$) found for (a) 427P, (b) P/2016 J1-B, \revision{(c) P/2017 S9, (d) 358P, and (e) 433P} from 100\,000 individual fitting runs as described in the text.  Vertical black lines in each panel show the center of each distribution (i.e., the most likely best-fit value), dashed lines enclose the 1-$\sigma$ interval for each parameter value, and dotted lines enclose the 95\% confidence interval for each parameter value.
\revision{Panels (a)-(c) are based on new photometric data reported in this work, while panels (d) and (e) are based on previously
reported photometric data.}
}
\label{figure:hgf_distributions}
\end{figure*}



For the purposes of this analysis, mean magnitudes for each visit of each object were used instead of photometric measurements for individual exposures because the short duration of most of our same-night observing sequences meant that they likely did not sample significant portions of each object's rotational lightcurve.  As such, it was judged to be more beneficial to use mean magnitudes measured for each visit in order to improve the signal-to-noise ratio of the input data employed for phase function fitting, rather than attempting to preserve the greater temporal sampling of the individual photometric points.

\begin{figure*}[htb!]
\centerline{\includegraphics[width=6.7in]{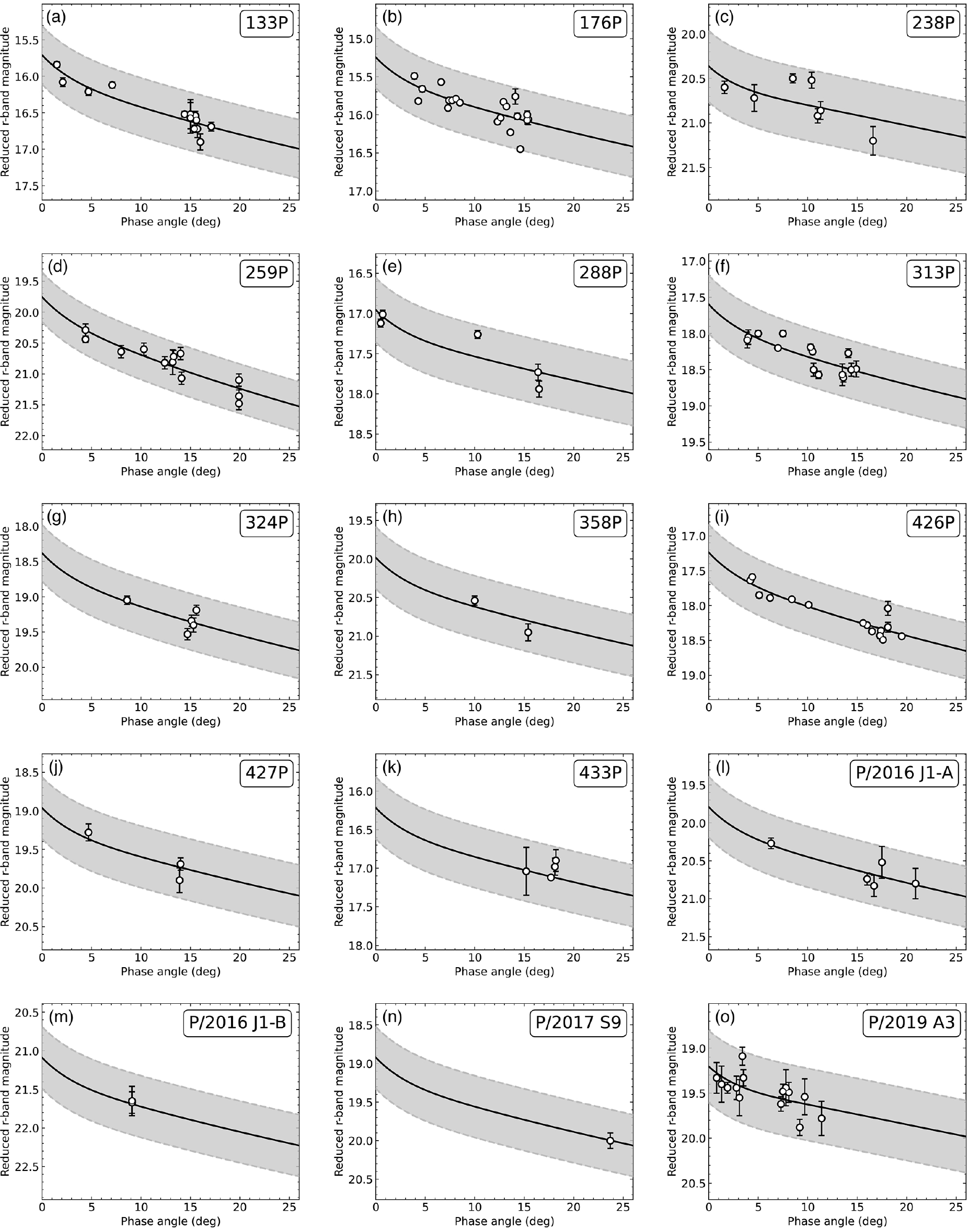}}
\caption{\small \revision{Best-fit $HG$ phase functions (solid lines) for (a) 133P, (b) 176P, (c) 238P, (d) 259P, (e) 288P, (f) 313P, (g) 324P, (h) 358P, (i) 426P, (j) 427P, (k) 433P, (l) P/2016 J1-A, (m) P/2016 J1-B, (n) P/2017 S9, and (o) P/2019 A3, based on photometric data shown in Table \ref{table:observations}, where the gray region bounded by dashed lines shows the range of possible photometric variations due to rotation, assuming a maximum lightcurve amplitude of $\Delta m=0.4$~mag (i.e., the maximum lightcurve amplitude inferred from photometry of 176P in this work).}
}
\label{figure:HG_phase_functions}
\end{figure*}

Except for P/2016 J1-A and P/2016 J1-B (discussed further \revision{below, as well as} in Section~\ref{section:activity_detection}), all of our targets appear inactive in all of the observations reported here (see Figure Set 1).  However, in an effort to further minimize the impact of any photometric contamination from faint dust emission activity, we focus our primary fitting efforts on photometric data obtained when our objects were between true anomalies of $\nu=140^{\circ}$ and $\nu=290^{\circ}$, i.e., along orbit position ranges over which activity has not previously been seen for any known MBCs \citep[see][and references within]{hsieh2018_238p288p}, \revision{except in the cases of P/2016 J1-B and P/2017 S9.  In both of these cases, only data outside our preferred true anomaly range were available, and so, we used the data that were available.  As the possibility that these objects could be active at these points in their orbits was considered significant, however, we report absolute magnitudes and nucleus sizes derived from these data as lower and upper limits, respectively.  In the case of P/2016 J1-B, we further limited our phase function fitting to post-perihelion data only due to the visible presence of activity in our pre-perihelion data for the object.
As the presence of early activity in P/2016 J1-B suggests that we may not be able to rule out the presence of similarly early activity in P/2016 J1-A during our reported observations, we report computed absolute magnitudes and nucleus radii for this target as lower and upper limits, respectively, as well.}

For reference and comparison, we also reanalyze previously reported photometric data for inactive MBC nuclei using our Monte Carlo approach to better characterize uncertainties in best-fit parameter values.  Specifically, we reanalyze data for 133P/Elst-Pizarro, 176P/LINEAR, 259P/Garradd, 288P/(300163) 2006 VW$_{139}$, 358P/PANSTARRS, and 433P/(238470) 2005 QN$_{173}$, detailed in Table~\ref{table:observations}.

Most of those data were already used to compute previously reported phase function parameters for these objects.  However, those previous analyses did not all adhere to the requirement we adopted for this work that all photometric data used for phase function parameter fitting must have been obtained between true anomalies of $\nu=140^{\circ}$ and $\nu=290^{\circ}$.
As such, in addition to reanalyzing these data using a different fitting approach, in many cases, we also restrict our updated fitting analyses to subsets (marked in Table~\ref{table:observations}) of the original data sets used in previous fitting efforts, based on the same true anomaly restrictions we apply to our analyses of newly reported photometric data presented in this work.  All previously reported data were converted to equivalent $r'$-band magnitudes \citep[assuming solar colors of $(g'-r')_\odot=0.45\pm0.02$, $(r'-i')_\odot=0.12\pm0.01$, $(i'-z')_\odot=0.04\pm0.02$;][]{holmberg2006_solarcolors} prior to phase function parameter fitting.  

For targets for which photometry from $\geq5$ individual visits was available (133P, 176P, 238P, 259P, 288P, 313P, 324P, 426P, P/2016 J1-A, and P/2019 A3), we find best-fit values for both the $r$-band absolute magnitude, $H_r$, and $r$-band slope parameter, $G_r$, using the $HG$ phase function model available from the {\tt photometry} module of {\tt sbpy}\footnote{\url{https://sbpy.org/}} \citep{mommert2019_sbpy}. For targets for which photometry from $<5$ visits are available (427P, P/2016 J1-B, \revision{and P/2017 S9}), we instead adopt the average $G_V$ parameter value and standard deviation measured for approximately 8000 C-type asteroids by \citet{veres2015_ps1phasefunctions} of $G_V=0.18\pm0.28$ as an assumed value for $G_r$,
and only solve for $H_r$ in the $HG$ phase function model.  

This use of $G$ parameter values measured for C-type asteroids is based on past work finding that most MBCs are associated with asteroid families dominated by members with primitive (e.g., C-type) taxonomic classifications, and thus are likely to have similar classifications themselves \citep{hsieh2018_activeastfamilies}.
For these $H_r$-only fits, we also apply our Monte Carlo approach to a selection of $G_r$ parameter values for individual test runs solving for $H_r$, applying Gaussian-distributed offsets to the nominal assumed value for $G_r$ based on the reported standard deviation.  In this way, we seek to find more realistic uncertainties on our best-fit results for $H_r$ than would be derived using a single fixed value such as $G=0.15$ with no associated uncertainties, as is commonly assumed for objects for which $G$ has not been explicitly determined \citep[e.g., see][]{veres2015_ps1phasefunctions}.

Two-dimensional histograms of phase function parameter solutions for targets for which we solved for both $H_r$ and $G_r$ are shown in Figure~\ref{figure:hg_distributions}, while one-dimensional histograms of phase function parameter solutions for targets for which we solved only for $H_r$ are shown in Figure~\ref{figure:hgf_distributions}.  In test runs, we found that repeated sets of 100\,000 individual fitting runs per object produced consistent best-fit results to three decimal places, and as such, used 100\,000 individual fitting runs for each object studied in this work.
Best-fit $HG$ phase function solutions are plotted in Figures~\ref{figure:HG_phase_functions}. Best-fit phase function parameters for all targets are listed in Table~\ref{table:nucleus_phase_function_params}.



In the cases of 324P and P/2016 J1-A, even though $\geq5$ photometric points are available for each object, the phase angle coverage of each data set is not ideal (i.e., with a majority of points confined to a small phase angle range), and so for reference, we perform additional fits for these objects where we only solve for $H_r$ and assume $G_r=0.18\pm0.28$.  In these cases, we show results of both full $HG$ fits and $H$-only fits in Table~\ref{table:nucleus_phase_function_params}, although only show histograms and best-fit phase function plots for the $HG$ solutions.  Ultimately, however, we find that $H_V$ and $r_n$ results for both our $HG$ and $H$-only fits are consistent with each other within uncertainties for both objects.

\setlength{\tabcolsep}{4pt}
\setlength{\extrarowheight}{0.4em}
\begin{table*}[htb!]
\caption{Phase Function Fit Results}
\centering
\smallskip
\footnotesize
\begin{tabular}{lrcccrcc>{\raggedright\arraybackslash}p{1.8cm}}
\hline\hline
\multicolumn{1}{c}{Object}
 & \multicolumn{1}{c}{$n$$^a$}
 & \multicolumn{1}{c}{$\alpha$ range$^b$}
 & \multicolumn{1}{c}{$\nu$ range$^c$}
 & \multicolumn{1}{c}{$r_h$ range$^d$}
 & \multicolumn{1}{c}{$H_{r}$$^e$}
 & \multicolumn{1}{c}{$G_{r}$$^f$}
 & \multicolumn{1}{c}{$H_V$$^g$}
 & \multicolumn{1}{c}{$r_n$$^h$}
 \\[2pt]
\hline
238P$^{\dagger}$         &  7 & $1.6^{\circ}-16.6^{\circ}$  & $268.3^{\circ}-284.4^{\circ}$ & $2.79-2.99$ & $20.36^{+0.13}_{-0.13}$  & $0.46^{+0.23}_{-0.20}$       & $~~~20.53^{+0.13}_{-0.13}$ & $~~~0.24^{+0.07}_{-0.04}$ \\
313P$^{\dagger}$         & 14 & $3.9^{\circ}-14.9^{\circ}$  & $161.5^{\circ}-218.6^{\circ}$ & $3.67-3.86$ & $17.60^{+0.19}_{-0.17}$  & $0.07^{+0.24}_{-0.18}$       &  ~~~$17.77^{+0.19}_{-0.17}$  & ~~~$0.85^{+0.25}_{-0.15}$ \\ 
324P$^{\dagger\ddagger}$ &  5 & $8.6^{\circ}-15.6^{\circ}$  & $178.8^{\circ}-184.8^{\circ}$ & $3.57$ & 18.38$^{+0.19}_{-0.17}$  & 0.03$^{+0.19}_{-0.14}$     & ~~~18.55$^{+0.19}_{-0.17}$  &  ~~~0.59$^{+0.18}_{-0.10}$ \\ 
 & --- & --- & --- & --- & $18.55^{+0.24}_{-0.31}$ & \multicolumn{1}{c}{(0.18$\pm$0.28)$^i$} & ~~~18.72$^{+0.24}_{-0.31}$ & ~~~0.56$^{+0.19}_{-0.11}$ \\
426P$^{\dagger}$         & 15 & $4.4^{\circ}-19.5^{\circ}$  & $204.1^{\circ}-286.4^{\circ}$ & $2.97-3.64$ & $17.24^{+0.02}_{-0.02}$  & $0.01^{+0.02}_{-0.02}$      &  ~~~$17.41^{+0.02}_{-0.02}$ & ~~~$1.00^{+0.28}_{-0.15}$ \\
427P$^{\dagger}$         &  3 & $4.7^{\circ}-14.0^{\circ}$  & $171.3^{\circ}-219.6^{\circ}$ & $3.77-4.14$ & $18.96^{+0.23}_{-0.28}$  & \multicolumn{1}{c}{(0.18$\pm$0.28)$^i$} & ~~~19.13$^{+0.23}_{-0.28}$ & ~~~0.46$^{+0.15}_{-0.09}$ \\
P/2016 J1-A$^{\dagger}$  &  5 & $6.3^{\circ}-17.5^{\circ}$  & $260.0^{\circ}-281.5^{\circ}$ & $2.81-3.13$ & $>19.77^{+0.18}_{-0.16}$  & $0.13^{+0.23}_{-0.17}$ & $>19.95^{+0.18}_{-0.16}$ & $<0.31^{+0.09}_{-0.05}$ \\ 
 & --- & --- & --- & --- & $>19.87^{+0.24}_{-0.28}$ & \multicolumn{1}{c}{(0.18$\pm$0.28)$^i$} & $>20.04$$^{+0.24}_{-0.28}$ & $<0.30$$^{+0.10}_{-0.06}$ \\
P/2016 J1-B$^{\dagger}$  &  2 & $9.1^{\circ}$               & $37.9^{\circ}$                & 2.55 & $>21.09^{+0.23}_{-0.26}$ & \multicolumn{1}{c}{(0.18$\pm$0.28)$^i$} & $>21.26$$^{+0.23}_{-0.26}$ & $<0.17$$^{+0.06}_{-0.03}$ \\ 
\revision{P/2017 S9$^{\dagger}$}    &  \revision{1} & \revision{$23.7^{\circ}$} & \revision{$338.2^{\circ}$} & \revision{2.23} & \revision{$\geq18.92^{+0.35}_{-0.48}$} & \multicolumn{1}{c}{\revision{(0.18$\pm$0.28)$^i$}} & \revision{$\geq19.10^{+0.35}_{-0.48}$} & \revision{$\leq0.48^{+0.20}_{-0.11}$} \\
P/2019 A3$^{\dagger}$    & 14 & $0.8^{\circ}-11.4^{\circ}$  & $172.5^{\circ}-222.3^{\circ}$ & $3.64-3.97$ & $19.21^{+0.11}_{-0.10}$  & $0.48^{+0.25}_{-0.21}$ & ~~~$19.37^{+0.11}_{-0.10}$ & ~~~$0.41^{+0.12}_{-0.07}$ \\[2pt] 
\hline
133P$^{\ddagger}$        & 13 & $1.5^{\circ}-17.1^{\circ}$  & $144.2^{\circ}-270.9^{\circ}$ & $3.07-3.68$ & $15.71^{+0.04}_{-0.04}$  & $0.09^{+0.05}_{-0.05}$ & ~~~$15.88^{+0.04}_{-0.04}$ & ~~~$2.03^{+0.57}_{-0.32}$ \\
176P$^{\ddagger}$        & 20 & $3.9^{\circ}-15.4^{\circ}$  & $173.2^{\circ}-288.1^{\circ}$ & $2.90-3.80$ & $15.25^{+0.03}_{-0.03}$  & $0.16^{+0.04}_{-0.04}$  & ~~~15.42$^{+0.03}_{-0.03}$ & ~~~2.29$^{+0.51}_{-0.31}$$^j$ \\
259P$^{\ddagger}$        & 12 & $4.4^{\circ}-19.9^{\circ}$  & $194.5^{\circ}-265.1^{\circ}$ & $2.48-3.60$ & $19.75^{+0.08}_{-0.08}$  & $-0.15^{+0.06}_{-0.05}$ & ~~~19.92$^{+0.08}_{-0.08}$ & ~~~0.32$^{+0.09}_{-0.05}$ \\
288P$^{\ddagger k}$        &  5 & $0.5^{\circ}-16.5^{\circ}$  & $140.0^{\circ}-240.1^{\circ}$ & $3.25-3.46$ & $16.96^{+0.04}_{-0.04}$  & $0.25^{+0.08}_{-0.08}$  & ~~~17.13$^{+0.04}_{-0.04}$ & ~~~{0.86$\pm$0.17$^l$  $~~~~$0.61$\pm$0.12} \\
358P$^{\ddagger}$        &  2 & $10.0^{\circ}-15.4^{\circ}$ & $284.9^{\circ}-289.3^{\circ}$ & $2.76-2.80$ & $19.98^{+0.24}_{-0.29}$  & \multicolumn{1}{c}{(0.18$\pm$0.28)$^i$} & ~~~20.15$^{+0.24}_{-0.29}$ & ~~~0.29$^{+0.10}_{-0.06}$ \\
433P$^{\ddagger}$        &  4 & $15.2^{\circ}-18.2^{\circ}$ & $191.7^{\circ}-287.5^{\circ}$ & $2.74-3.73$ & $16.22^{+0.29}_{-0.38}$  & \multicolumn{1}{c}{(0.18$\pm$0.28)$^i$} & ~~~16.39$^{+0.29}_{-0.38}$ & ~~~1.61$^{+0.60}_{-0.34}$ \\[2pt]
\hline
\hline
\multicolumn{9}{l}{$^a$ Number of photometric points used in phase function solution.} \\
\multicolumn{9}{l}{$^b$ Solar phase angle range of observations included in phase function solution.} \\
\multicolumn{9}{l}{$^c$ True anomaly range of observations included in phase function solution.} \\
\multicolumn{9}{l}{$^d$ Heliocentric distance range, in au, of observations included in phase function solution.} \\
\multicolumn{9}{l}{$^e$ Best-fit $r'$-band absolute magnitude, using a $HG$ phase function model.} \\
\multicolumn{9}{l}{$^f$ Best-fit $r'$-band $G$ parameter value, using a $HG$ phase function model.} \\
\multicolumn{9}{l}{$^g$ Equivalent absolute $V$-band magnitude, assuming solar colors, corresponding to derived $H_{r}$ value.} \\
\multicolumn{9}{l}{$^h$ Estimated nucleus radius, in km, based on derived $H_V$ value and an assumed $V$-band albedo of $p_V=0.05\pm0.02$,} \\
\multicolumn{9}{l}{$~~~$ except where otherwise specified.} \\
\multicolumn{9}{l}{$^i$ Assumed value based on average $G$ value and standard deviation for C-type asteroids found by \citet{veres2015_ps1phasefunctions}.} \\
\multicolumn{9}{l}{$^j$ Computed assuming $p_V=p_R=0.06\pm0.02$ \citep{hsieh2009_albedos}.} \\
\multicolumn{9}{l}{$^k$ Identified as a binary system by \citet{agarwal2017_288p}.} \\
\multicolumn{9}{l}{$^l$ Computed for a binary system with a component scattering cross-sectional area ratio of $\sim$2:1, based on the results} \\
\multicolumn{9}{l}{$~~~$  of \citet{agarwal2020_288p}, with the same total scattering cross-section as a single body with the computed $H_r$ magnitude.} \\
\multicolumn{9}{l}{$^\dagger$ Phase function solution based on data reported in this work.} \\
\multicolumn{9}{l}{$^\ddagger$ Phase function solution based on data reported in previous works: 133P \citep{hsieh2009_albedos}; 176P \citep{hsieh2009_albedos};} \\
\multicolumn{9}{l}{$~~~$ 259P \citep{maclennan2012_259p}; 288P \citep{hsieh2018_238p288p}; 324P \citep{hsieh2014_324p}; 358P \citep{hsieh2018_358p}; } \\
\multicolumn{9}{l}{$~~~$ 433P \citep{hsieh2021_433P}} \\
\end{tabular}
\label{table:nucleus_phase_function_params}
\end{table*}

\subsection{Nucleus Sizes\label{section:nucleus_sizes}}

In order to estimate physical radii for our target objects, we first compute equivalent $V$-band absolute magnitudes, $H_V$, from our computed $r'$-band absolute magnitudes using
\begin{equation}
    H_V = H_{r} + 0.733(V-R) - 0.088
\end{equation}
from \citet{jordi2006_filtertransformations}, assuming approximately solar colors for all objects, i.e., $(V-R)_{\odot}=0.354$ \citep{holmberg2006_solarcolors}. 
We then estimate effective nucleus radii for our targets using
\begin{equation}
r_n = \left( {2.24\times10^{22}\over p_V} \times 10^{0.4(m_{\odot,V} - H_V)} \right)^{1/2}
\label{equation:nucleus_size}
\end{equation}
where we use $m_{\odot,V}=-26.71\pm0.03$ for the apparent $V$-band magnitude of the Sun \citep{hardorp1980_sun3}.  
\revision{Assuming similar $R$-band and $V$-band albedo values
\citep[as supported by Hapke photometric modeling of the surface of the primitive-type asteroid (101955) Bennu showing minimal geometric albedo variation over the wavelength range covered by the $V$- and $R$-band filters;][]{golish2021_bennuphotometrymodeling}, we assume $p_V=p_R=0.05\pm0.02$ for most of our targets, using the $R$-band albedo ($p_R$) computed from {\it Spitzer Space Telescope} observations of 133P \citep{hsieh2009_albedos}. The one exception  is the case of 176P, for which we use $p_R=0.06\pm0.02$, which was separately measured in the same work.}

These calculations are performed as part of the Monte Carlo-style phase function fitting procedures described in Section~\ref{section:phase_functions}, where we apply Gaussian-distributed offsets to the nominal assumed albedo based on the specified uncertainty for each test run.
Resulting $H_V$ values and nucleus size estimates for our target objects are shown in Table~\ref{table:nucleus_phase_function_params}.

\begin{figure*}[htb!]
\centerline{\includegraphics[width=5.0in]{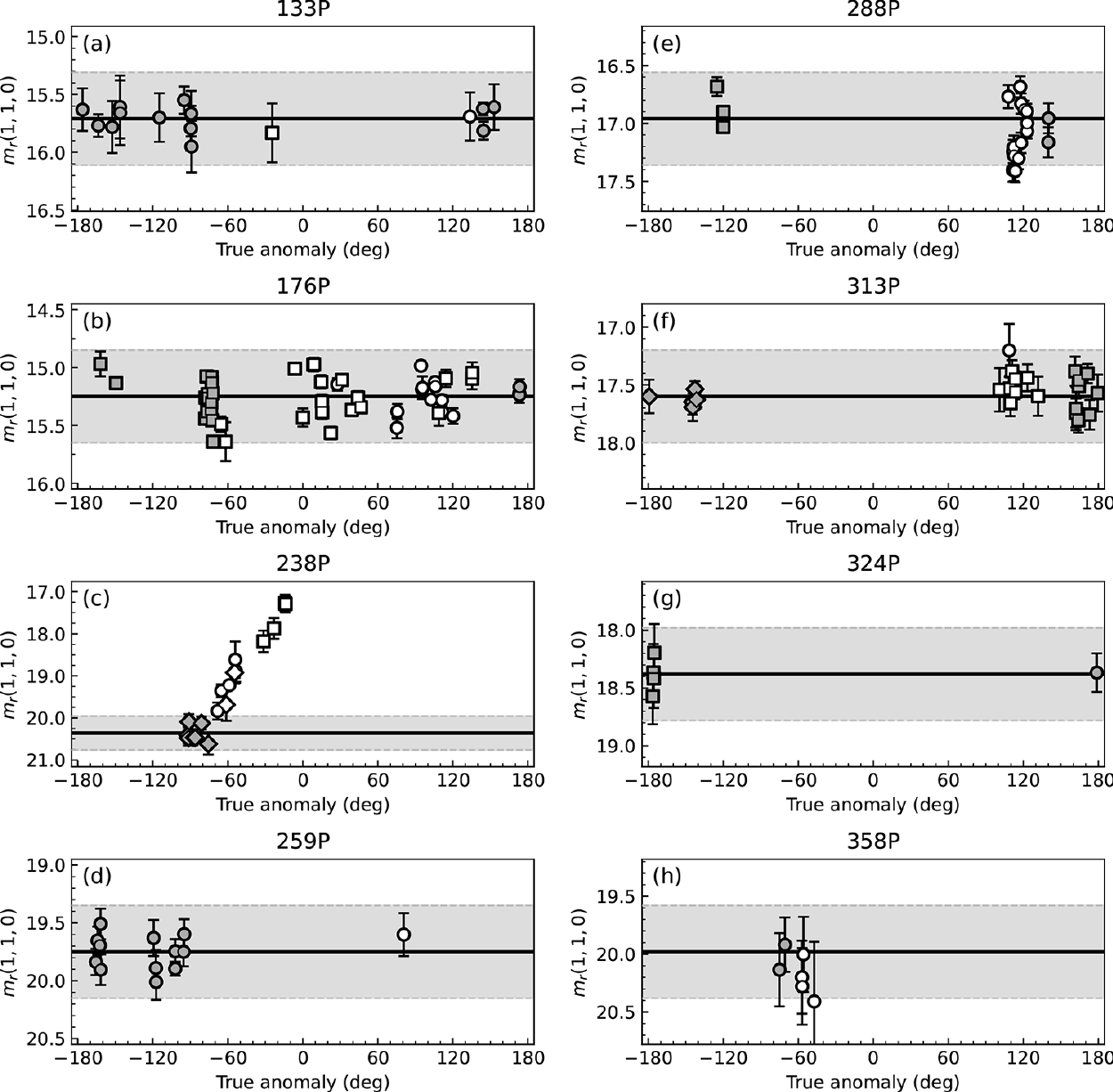}}
\caption{\small Plots of equivalent $r'$-band absolute magnitudes corresponding to photometry reported for (a) 133P/Elst-Pizarro, (b) 176P/LINEAR, (c) 238P/Read, (d) 259P/Garradd, (e) 288P/(300163) 2006 VW$_{139}$, (f) 313P/Gibbs, (g) 324P/La Sagra, and (h) 358P/PANSTARRS at times when each object appeared to be inactive.  Filled symbols indicate photometry data that were used to compute phase function parameters in this work (i.e., obtained within the true anomaly range $140^{\circ}<\nu<290^{\circ}$), while open symbols indicate photometry data that were excluded from the phase function parameter fitting analysis in this work but were still obtained at times when each object appeared to be inactive.  Shaded regions indicate the range of brightness variations that could be expected from a rotational lightcurve with an amplitude of $\Delta m=0.4$~mag.
In panel (a), circular and square symbols mark data obtained between aphelia in February 1999 and September 2004, and between aphelia in September 2004 and April 2010, respectively.
In panel (b), circular and square symbols mark data obtained between aphelia in December 2002 and August 2008, and between aphelia in August 2008 and May 2014, respectively.
In panel (c), circular, square, and diamond symbols mark data obtained between aphelia in May 2008 and December 2013, between aphelia in December 2013 and August 2019, and between aphelia in August 2019 and March 2025, respectively.
In panel (d), circular symbols mark data obtained between aphelia in October 2010 and April 2015.
In panel (e), circular and square symbols mark data obtained between aphelia in November 2008 and March 2014, and between aphelia in March 2014 and July 2019, respectively.
In panel (f), circular, square, and diamond symbols mark data obtained between aphelia in September 2000 and April 2006, between aphelia in November 2011 and June 2017, and between aphelia in June 2017 and February 2023.
In panel (g), circular and square symbols mark data obtained between aphelia in October 2007 and March 2013, and between aphelia in March 2013 and August 2018, respectively.
In panel (h), circular symbols mark data obtained between aphelia in June 2015 and January 2021.
}
\label{figure:activity_search_1}
\end{figure*}

\subsection{Activity Detection\label{section:activity_detection}}

\subsubsection{Overview\label{section:activity_detection_overview}}

One of the major motivations for the precise characterization of MBC nucleus sizes is the ability to carry out photometric searches for low-level activity \citep[e.g.,][]{hsieh2015_324p,hsieh2014_176p,hsieh2018_358p,hsieh2021_259p}.  While activity detection and characterization are not the primary focus of the work presented here, we are nonetheless interested in both determining whether any of our targets were active during any of the observations we report here, and also whether any of those targets were active during previously reported observations when they were presumed to be inactive (usually due to the lack of any visible activity such as a coma or tail).

\begin{figure*}[htb!]
\centerline{\includegraphics[width=5.0in]{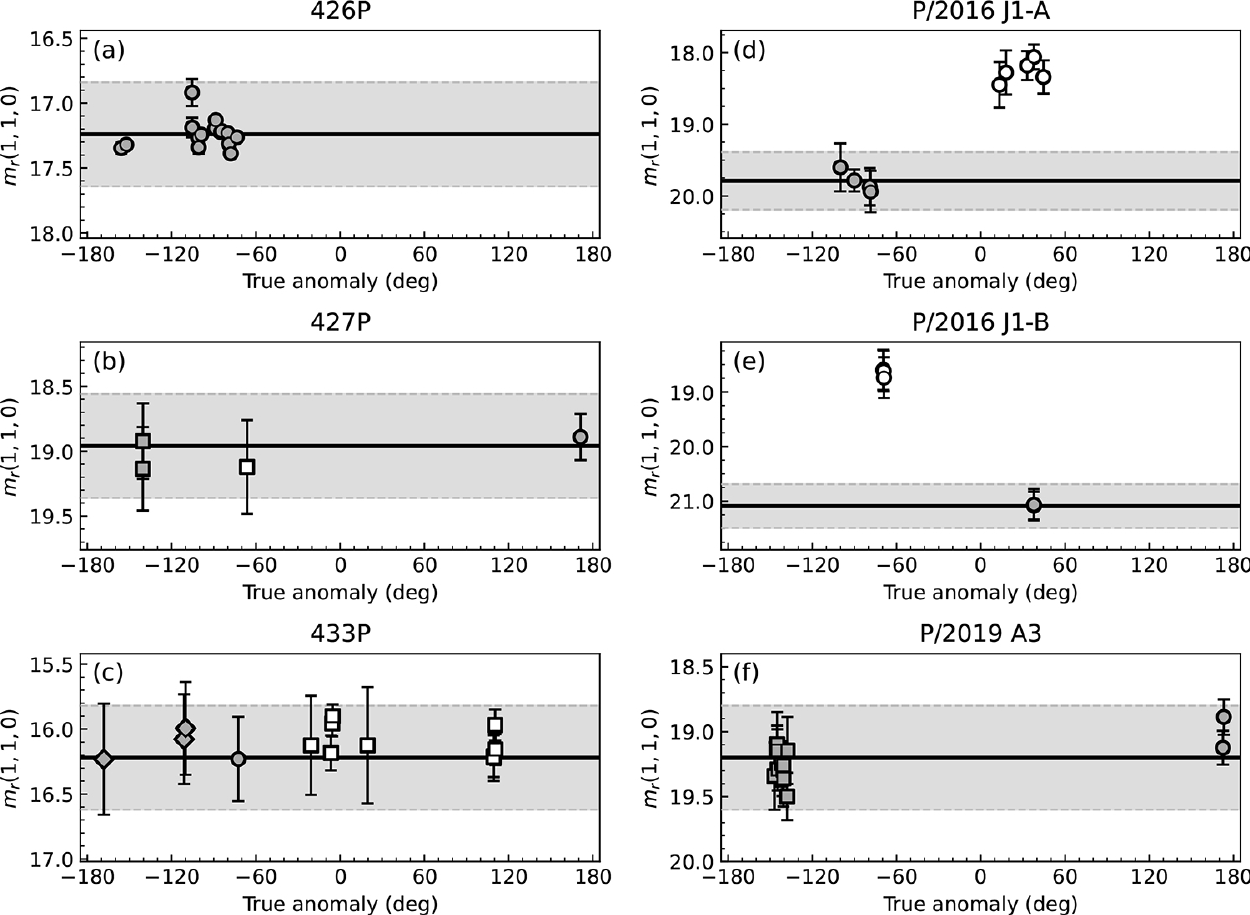}}
\caption{\small Plots of equivalent $r'$-band absolute magnitudes corresponding to photometry reported for (a) 426P/PANSTARRS, (b) 427P/ATLAS, (c) 433P/(248370) 2005 QN$_{173}$, (d) P/2016 J1-A (PANSTARRS), and (e) P/2016 J1-B (PANSTARRS) at times when each object appeared to be inactive.  Filled symbols indicate photometry data that were used to compute phase function parameters in this work (i.e., obtained within the true anomaly range $140^{\circ}<\nu<290^{\circ}$), while open symbols indicate photometry data that were excluded from the phase function parameter fitting analysis in this work but were still obtained at times when each object appeared to be inactive.
Shaded regions indicate the range of brightness variations that could be expected from a rotational lightcurve with an amplitude of $\Delta m=0.4$~mag.
In panel (a) circular symbols mark data obtained between aphelia in November 2020 and July 2026.
In panel (b), circular and square symbols mark data obtained between aphelia in October 2014 and May 2020, and between aphelia in May 2020 and January 2026, respectively.
In panel (c), circular, square, and diamond symbols mark data obtained between aphelia in July 2002 and December 2007, between aphelia in December 2007 and April 2013, and between aphelia in September 2018 and January 2024, respectively.
In panels (d) and (e), circular symbols mark data obtained between aphelia in 2019 April and December 2024.
In panel (f), circular and square symbols mark data obtained between aphelia in October 2015 and May 2021, and between aphelia in May 2021 and December 2026, respectively.
}
\label{figure:activity_search_2}
\end{figure*}

The task of detecting activity using photometry can take two forms: direct detection of activity in individual observations and statistical detection of activity over multiple observations over extended periods of time.  In the first case, we search for individual photometric points that are significantly brighter than expected from the best-fit phase function parameters for that object's inactive nucleus, where in practice, ``significantly brighter'' is generally interpreted as being brighter by a margin larger than both the 3-$\sigma$ uncertainty of the nucleus brightness prediction itself as well as potential brightness variations due to the object's rotational lightcurve.

Meanwhile, in the second case, while photometric points may not individually exceed the target's expected brightness at the time to conclusively indicate the presence of activity, repeated photometry measurements that are consistently brighter than the expected nucleus brightness in excess of photometric and brightness prediction uncertainties can also indicate the presence of activity. This method is inherently less conclusive than the first method we describe, since while consistently unknowingly sampling the brighter portion of an object's rotational lightcurve by chance when conducting short sequences of observations is unlikely, it is possible.

The possibility of inadvertently consistently sampling the brighter portion of an object's rotational lightcurve by chance can be mitigated by the use of large numbers of data points (which should decrease the likelihood that all observations will occur when the object is brighter than average), or preferably, by observing an object over one or more full rotation periods to fully eliminate uncertainties arising from only sampling a limited portion of the rotational lightcurve.  The latter approach of course presents other challenges, however, such as the significantly larger amounts of observing time required to sample full rotational lightcurves and the fact that the faintness of most MBC nuclei (particularly when inactive) have made it extremely difficult to measure rotation periods, meaning that only a very small number of MBC rotation periods are known to date \citep[e.g.,][]{hsieh2004_133p,hsieh2010_133p,hsieh2011_176p}.  There is also the caveat that absolute magnitudes derived from photometric data obtained at different apparitions can sometimes vary due to changes in aspect angle, depending on an object's shape \citep[e.g.,][]{mahlke2021_atlasphasefunctions}

To carry out our activity search, using the phase function parameters determined earlier in this work, we compute the equivalent $r'$-band absolute magnitudes (see Table~\ref{table:observations}) corresponding to all photometry reported here for our targets and plot them as functions of true anomaly (Figures~\ref{figure:activity_search_1} and \ref{figure:activity_search_2}).  We additionally plot equivalent $r'$-band absolute magnitudes (see Table~\ref{table:observations}) for photometry reported in previous works where the target was presumed to be inactive at the time to assess the validity of those presumptions.  In these plots, we use filled and open symbols to indicate observations that are included and excluded, respectively, from our phase function fitting analysis based on their true anomalies, as we are particularly interested in noting if there are any signs of weak activity within the excluded true anomaly range.  That said, signs of activity at non-excluded true anomalies would also be extremely interesting, as they could indicate the action of non-sublimation-driven activity processes like intermittent mass loss from rotational destabilization, such as the case of (6478) Gault \citep{chandler2019_gault}.

We find clear evidence of activity in three of the targets observed in this work --- 238P, P/2016 J1-A, and P/2016 J1-B --- which we briefly analyze and discuss below.


\subsubsection{238P/Read\label{section:activity_detection_238p}}

We find that 238P likely became active between observations obtained on UT 2021 August 29 and UT 2021 October 30\revision{, as it approached its UT 2022 June 5 perihelion passage}.  As can be seen in Figure~\ref{figure:activity_search_1}c, photometric points obtained on UT 2021 October 30 and UT 2021 November 26 (open square symbols at $\nu=-61.3^{\circ}$ and $\nu=-54.6^{\circ}$, or $\nu=298.7^{\circ}$ and $\nu=305.4^{\circ}$, respectively) are significantly brighter than expected based on the phase function derived using the other 2021 data points for this object.  We \revision{estimate} the total mass, $M_d$, of visible ejected dust on those two dates inferred from the excess flux observed on those dates, following the procedure detailed in \citet{hsieh2014_324p}, where $M_d$ is given by
\begin{equation}
M_d = {4\over 3}\pi r^2_n {\bar a}\rho_d \left({1-10^{0.4(H_{R,t}-H_R)} \over 10^{0.4(H_{R,t}-H_R)} }\right)
\end{equation}
where we assume dust grain densities of $\rho_d$$\,=\,$2500~kg~m$^{-3}$, consistent with CI and CM carbonaceous chondrites, which are associated with C-type asteroids like the MBCs \citep{britt2002_astdensities_ast3}, and effective mean dust grain radii of ${\bar a}$$\,\sim\,$1~mm.

\revision{Following \citet{jewitt2014_133p} and \citet{hsieh2014_324p}, this effective mean dust grain radius (by mass), ${\bar a}$, is weighted by the size distribution, scattering cross-section, and residence time, and is approximated using
\begin{equation}
{\bar a} \sim {a_{\rm max}\over\ln(a_{\rm max}/a_{\rm min})}
\label{equation:mean_dust_radius}
\end{equation}
where $a_{\rm min}$ and $a_{\rm max}$ are the lower and upper bounds of the particle size distribution, which is assumed to be a power law with an index of $q=3.5$, and where $a_{\rm max}\gg a_{\rm min}$ is assumed.  Dust modeling of 238P's 2005 active apparition has been previously performed by \citet{hsieh2009_238p}, who found $a_{\rm min}=10$~$\mu$m and $a_{\rm max}=(1-10)$~mm, corresponding to ${\bar a}\sim (0.2-1.4)$~mm using Equation~\ref{equation:mean_dust_radius}.  As indicated above, we adopt ${\bar a}\sim1$~mm for simplicity, but given the directly proportional relationship of computed dust masses ($M_d$) to ${\bar a}$, the $M_d$ values reported here should only be considered to be order-of-magnitude estimates.}

\setlength{\tabcolsep}{4.5pt}
\setlength{\extrarowheight}{0.4em}
\begin{table*}[htb!]
\caption{Best-Fit Activity Results for 238P}
\centering
\footnotesize
\begin{tabular}{lccccclccc}
\hline\hline
\multicolumn{3}{c}{Observational Data}
 && \multicolumn{4}{c}{Best-Fit Activity Onset Date/Position} \\
\multicolumn{1}{c}{UT Dates$^a$}
 & \multicolumn{1}{c}{$n_{\rm obs}$$^b$}
 & \multicolumn{1}{c}{$\nu$$^c$}
 && \multicolumn{1}{c}{Days to Peri.$^d$}
 & \multicolumn{1}{c}{UT Date$^e$}
 & \multicolumn{1}{c}{$\nu$$^f$}
 & \multicolumn{1}{c}{$r_h$$^g$}
 && \multicolumn{1}{c}{$\dot M$$^h$}
 \\
\hline
2010 Jul 07 -- 2010 Aug 15 & 6 & $291.8^{\circ}-306.4^{\circ}$ & ~ & $263^{+8}_{-12}$ & 2010 Jun 20$^{+8}_{-12}$ & $288^{+2}_{-3}$ & $2.75^{-0.02}_{+0.03}$ & ~ & $0.3^{+0.1}_{-0.1}$ \\
2016 Jul 08 -- 2016 Sep 06 & 4 & $328.5^{\circ}-346.1^{\circ}$ & ~ & $183^{+29}_{-73}$ & 2016 Apr 22$^{+29}_{-73}$ & $308^{+8}_{-18}$ & $2.57^{-0.06}_{+0.16}$ & ~ & $0.5^{+0.3}_{-0.2}$ \\
2021 Aug 29 -- 2021 Oct 30 & 2 & $298.7^{\circ}-305.4^{\circ}$ & ~ & $231^{+8}_{-20}$ & 2021 Oct 18$^{+8}_{-20}$ & $296^{+2}_{-5}$ & $2.67^{-0.02}_{+0.05}$ & ~ &  $0.4^{+0.2}_{-0.2}$ \\[2pt]
\hline
\hline
\multicolumn{10}{l}{$^a$ UT date range of observations used for best-fit analysis.} \\
\multicolumn{10}{l}{$^b$ Number of photometric points used for best-fit analysis.} \\
\multicolumn{10}{l}{$^c$ True anomaly range of observations used for best-fit analysis.} \\
\multicolumn{10}{l}{$^d$ Days prior to perihelion at the time of the best-fit activity onset point.} \\
\multicolumn{10}{l}{$^e$ UT date at the time of the best-fit activity onset point.} \\
\multicolumn{10}{l}{$^f$ True anomaly, in degrees, of the best-fit activity onset point.} \\
\multicolumn{10}{l}{$^g$ Heliocentric distance, in au, of the best-fit activity onset point.} \\
\multicolumn{10}{l}{$^h$ Best-fit median mass loss rate, in kg~s$^{-1}$, over the specified time period.} \\
\end{tabular}
\label{table:obs_238p_activity}
\end{table*}

\revision{We also note that the following analysis assumes that the particle size distribution (and therefore, ${\bar a}$) for 238P remains constant within each active apparition as well as between active apparitions, although variations in that size distribution within a single active apparition or between active apparitions are entirely possible and plausible.  It is not possible to ascertain the time evolution of the particle size distribution of 238P's activity from available observational constraints, however, so we simply highlight this point as a caveat that should be taken into account when interpreting the following results.
}

Using a Monte Carlo approach similar to the one we used for phase function fitting, we performed repeated fits of a linear function to our computed dust masses and solved for the time at which that function's value was zero, i.e., the start of activity, taking into account uncertainties on the best-fit $HG$ phase function parameters and $r_n$ determined above for 238P, as well as on the dust masses computed here.
We find a median start time of activity of $231^{+8}_{-20}$~days prior to perihelion, corresponding to UT 2021 October 18$^{+8}_{-20}$, $\nu=(296^{+2}_{-5})^{\circ}$, and $r_h=2.67^{-0.02}_{+0.05}$~au.  By noting the slope of the best-fit linear function to our two post-reactivation observations, we also estimate net dust production rates over this time period of ${\dot M}_d=0.4^{+0.2}_{-0.2}$~kg~s$^{-1}$.  These results are also summarized in Table~\ref{table:obs_238p_activity}.

Given our new best-fit solution for the absolute magnitude of 238P's nucleus, we also briefly revisit the analysis of previous active apparitions of the object reported by \citet{hsieh2018_238p288p}.  That work reported a decline in initial net dust production rate between 2010 and 2016 from ${\dot M}_d=(1.4\pm0.3)$~kg~s$^{-1}$ to ${\dot M}_d=(0.7\pm0.3)$~kg~s$^{-1}$, and activity onset times $\sim(205\pm50)$ days prior to perihelion in 2010 and $\sim(225\pm85)$ days prior to perihelion in 2016.  Data used in that analysis covered a different true anomaly range than the 2021 data reported here, having been obtained over true anomaly ranges of $305.8^{\circ}<\nu<332.4^{\circ}$ in 2010 and $328.5^{\circ}<\nu<356^{\circ}$ in 2016, where we note that the true anomaly ranges covered by the 2010 and 2016 data were actually also mostly non-overlapping.

From our work here, we now know that 238P was actually active when the photometric data used by \citet{hsieh2011_238p} to derive best-fit phase function parameters for 238P's nucleus were obtained (see open circular symbols in Figure~\ref{figure:activity_search_1}c).  These data were obtained between 2010 July 7 and August 15, overlapping the true anomaly range covered by our 2021 observations when the object was seen to be active (see Table~\ref{table:obs_238p_activity}).
Performing the same analysis on these data (listed in Table~\ref{table:observations}) as described above, we find a new median start time of activity of $263^{+8}_{-12}$~days prior to perihelion and a new net dust production rate over this time period of ${\dot M}_d=0.3^{+0.1}_{-0.1}$~kg~s$^{-1}$ (see Table~\ref{table:obs_238p_activity} for additional details).
Thus, we now find that, within 3-$\sigma$ uncertainties, both the initial net dust production rates and onset times of 238P's activity in 2010 and 2021 were consistent with each other over comparable orbit arcs.

There are unfortunately no available observational data for 238P from 2016 over the same orbit arc as our 2021 observations.  As such, we cannot compare 238P's early activity from that apparition as directly to our 2021 observations as we were able to for its 2010 apparition.  For reference, however, we still consider the earliest data available from that perihelion passage, obtained on UT 2016 July 8 ($\nu=328.5^{\circ}$) and UT 2016 August 6 ($\nu=346.1^{\circ}$), when the object was unambiguously visibly active but the coma and tail were still relatively compact.  Applying the same procedures described above for the analysis of 238P's 2010 and 2021 early activity, we find a median start time of 238P's activity of $183^{+29}_{-73}$~days prior to perihelion, and a net dust production rate over this time period of ${\dot M}_d=0.5^{+0.3}_{-0.2}$~kg~s$^{-1}$ (see Table~\ref{table:obs_238p_activity} for additional details). 

In summary, we find effectively no change within 3-$\sigma$ uncertainties in 238P's activity onset time or initial net mass loss rate between 2010, 2016, and 2021.  This conclusion represents a revision of the results of \citet{hsieh2018_238p288p}, who found a decline in the initial net mass loss rate between 2010 and 2016, which may be the result of our use here of more direct comparable data obtained over similar orbit arcs.  While extrapolation of data sets when comparing active behavior between different active apparitions may be an unavoidable necessity when only data from non-overlapping orbit arcs are available, the results presented here clearly reaffirm that it is preferable to compare data obtained over overlapping orbit arcs when possible.  



As part of our ongoing MBC observing campaign, we have continued to monitor 238P's activity since the latest observations reported here from 2021.  A detailed analysis of that activity and comparison to previous active apparitions is beyond the scope of this work, however, and so will appear in a future publication.

\subsubsection{P/2016 J1-A/B (PANSTARRS)\label{section:activity_detection_p2016j1ab}}

Next, we find that both P/2016 J1-A and P/2016 J1-B were also active during the observations we report here. 
P/2016 J1-A appeared inactive in observations from UT 2020 December 24 (when the object was at $\nu=260.0^{\circ}$) to UT 2021 May 17 ($\nu=287.9^{\circ}$), before then exhibiting photometric evidence of activity (with measured photometry $>$1.3~mag brighter than expected for an inactive nucleus, which is a much larger photometric difference than could be reasonably expected from rotational lightcurve variations) from UT 2022 April 8 ($\nu=13.3^{\circ}$) to UT 2022 August 2 ($\nu=44.6^{\circ}$).  This photometric detection of activity for P/2016 J1-A in 2022 is confirmed by visual evidence of activity (e.g., Figures~1.6h and 1.6i).  Meanwhile, P/2016 J1-B appeared visibly active in images obtained on UT 2021 May 29, 30, and 31 ($290.5^{\circ}<\nu<291.0^{\circ}$; Figures~1.7a - 1.7c), but less clearly active in images obtained on UT 2022 July 7 ($\nu=37.9^{\circ}$; Figure~1.7d and 1.7e).  While the lack of clearly visible evidence of activity in P/2016 J1-B in 2022 could be a result of the object appearing fainter in 2022 as compared to 2021, photometric analysis shows that the object's activity strength in fact apparently declined significantly between the two observing periods (Figure~\ref{figure:activity_search_2}e).

\setlength{\tabcolsep}{4.5pt}
\setlength{\extrarowheight}{0em}
\begin{table*}[htb!]
\caption{Multi-filter 313P Observations}
\centering
\smallskip
\footnotesize
\begin{tabular}{lccr}
\hline\hline
\multicolumn{1}{c}{UT Date}
 & \multicolumn{1}{c}{JD$^a$}
 & \multicolumn{1}{c}{$f$$^b$}
 & \multicolumn{1}{c}{$m_f$$^c$}
 \\
\hline
2018 May 08 & 2458246.7572 & $g'$ & 23.47$\pm$0.05 \\
2018 May 08 & 2458246.7609 & $r'$ & 23.00$\pm$0.05 \\
2018 May 08 & 2458246.7647 & $i'$ & 22.82$\pm$0.08 \\
2018 May 08 & 2458246.7685 & $g'$ & 23.68$\pm$0.06 \\
2018 May 08 & 2458246.8714 & $g'$ & 23.47$\pm$0.06 \\
2018 May 08 & 2458246.8752 & $r'$ & 23.01$\pm$0.06 \\
2018 May 08 & 2458246.8789 & $i'$ & 22.87$\pm$0.09 \\
2018 May 08 & 2458246.8827 & $g'$ & 23.64$\pm$0.08 \\
\hline
\hline
\multicolumn{4}{l}{$^a$ Julian day.} \\
\multicolumn{4}{l}{$^b$ Filter.} \\
\multicolumn{4}{l}{$^c$ Apparent magnitude in specified filter.} \\
\end{tabular}
\label{table:obs_313p_colors}
\end{table*}

These observations confirm the recurrent nature of P/2016 J1-A's and P/2016 J1-B's activity for the first time, 
making P/2016 J1 the tenth MBC confirmed to be recurrently active, after 133P \citep{hsieh2004_133p}, 238P \citep{hsieh2011_238p}, 259P \citep{hsieh2021_259p}, 288P \citep{hsieh2018_238p288p}, 313P \citep{hsieh2015_313p}, 324P \citep{hsieh2015_324p}, 358P \citep{hsieh2018_358p}, 432P \citep{weryk2021_432p}, and 433P \citep{chandler2021_2005qn173}.
This confirmation makes the activity of the two P/2016 J1 fragments highly likely to be sublimation-driven \citep[e.g.,][]{jewitt2015_actvasts_ast4}.  Interestingly, however, the active behavior of P/2016 J1-B --- peaking prior to perihelion and declining post-perihelion --- is unlike that of its companion component, P/2016 J1-A, as well as all other MBCs with well-characterized active behaviors around perihelion \citep[e.g.,][]{hsieh2012_324p,hsieh2018_238p288p}.  This unusual behavior could indicate that rather than P/2016 J1-B's activity being primarily modulated by heliocentric distance, as appears to be the case for most MBCs, seasonal effects may also play a significant contributing role in activity modulation, e.g., as was previously hypothesized for 133P \citep{hsieh2004_133p} and is observed on other comets \citep[e.g.,][]{marschall2020_cometcomasurfacelinks}.  
Detailed observational monitoring during P/2016 J1-B's next perihelion approach in early 2027 as well as characterization of the object's rotational lightcurve, shape, and pole orientation could help to clarify the role of seasonal modulation on its activity, although given the component's extremely small size (and thus extreme faintness when inactive), the latter tasks will likely be quite challenging.





\subsubsection{Other Targets\label{section:activity_detection_other}}

Of the other MBCs observed as part of this work, none exhibited clear evidence of activity in any of the observations reported here.  We see possible evidence of low-level activity across multiple nights of observations of 313P in Figure~\ref{figure:activity_search_1}f in what appears to be persistently brighter-than-expected magnitudes measured between UT 2015 October 18 and UT 2016 May 7 ($101.0^{\circ}\leq\nu\leq131.8^{\circ}$).  We calculate the average of the equivalent absolute magnitudes (computed using the best-fit $G_r$ parameter value found for 313P; Table~\ref{table:nucleus_phase_function_params}) corresponding to these data to be $m_r(1,1,0)=17.52\pm0.03$, or about 0.1~mag brighter than our computed best-fit $r'$-band absolute magnitude for the object of $H_{r}=17.60\pm0.08$.  Similarly, we find an average equivalent $r'$-band absolute magnitude of 433P between UT 2010 June 14 and UT 2010 October 30 ($-20.7^{\circ}\leq\nu\leq19.6^{\circ}$ of $m_r(1,1,0)=16.03\pm0.02$, and an average equivalent absolute magnitude of $m_r(1,1,0)=16.12\pm0.05$ for data obtained between UT 2011 November 24 and UT 2011 December 1 ($109.4^{\circ}\leq\nu\leq110.6^{\circ}$), where both are slightly brighter than our computed best-fit $r'$-band absolute magnitude of $H_r=16.18\pm0.03$.  As the average equivalent absolute magnitudes of 313P and 433P during periods of suspected activity are all consistent with our computed best-fit absolute magnitudes for the respective nuclei of these targets within 3-$\sigma$ uncertainties, however, we conclude that there is no definitive evidence of activity for these targets over those time periods.


Meanwhile, we do not find convincing evidence of activity in other targets, including 133P, 176P, 259P, 288P, 324P, 358P, and 427P.  In the cases of 133P, 259P, 358P, and 427P, too few observations are available in the true anomaly range of interest to make any conclusive inferences about the presence or absence of activity.  In the case of 176P, we find average $r'$-band absolute magnitudes of $m_r(1,1,0)=15.26\pm0.01$ for data obtained between UT 2006 February 3 ($\nu=27.7^{\circ}$) and UT 2007 May 19 ($\nu=120.2^{\circ}$), and $m_r(1,1,0)=15.28\pm0.01$ for data obtained between UT 2010 October 5 ($\nu=294.5^{\circ}$) and UT 2013 May 13 ($\nu=135.5^{\circ}$), both of which are consistent with the best-fit absolute magnitude value of $H_r=15.25^{+0.03}_{-0.03}$ computed here from data obtained between $\nu=140^{\circ}$ and $\nu=290^{\circ}$, validating the results of \citet{hsieh2011_176p} who concluded that 176P's 2005 activity had ceased by UT 2006 February 3, as well as the results of \citet{hsieh2014_176p} who concluded that no evidence of activity was present during 176P's 2011 perihelion passage.



\subsection{313P Nucleus Colors\label{section:313p_colors}}

Multi-filter observations were obtained on UT 2018 May 08 (Table~\ref{table:obs_313p_colors}) in two $g'r'i'g'$ sequences so that the $g'$-band observations that bracket each sequence can be used to compute interpolated $g'$-band magnitudes at the time of the bracketed $r'$- and $i'$-band observations in order to control for any brightness variations due to rotation that may have occurred during the observation sequence.  We use this approach to compute $g'-r'$ and $g'-i'$ broadband colors, where $r'-i'$ colors can then be computed from $(g'-i')- (g'-r')$.

Following \citet{demeo2013_asteroidtaxonomy}, we transform these computed $g'-r'$ and $g'-i'$ colors to reflectance values (normalized to the central wavelength of the $g'$-band filter, or 468.6~nm, and with solar colors subtracted) using
\begin{equation}
    R_f = 10^{-0.4\left[(g'-f)_{\odot}-(g'-f)\right]}
\label{equation:reflectance}
\end{equation}
where $g'-f$ and $(g'-f)_{\odot}$ are the colors of the object and the Sun, respectively, in terms of the difference between magnitudes measured in $g'$-band and in a certain filter, $f$, and $(g'-r')_{\odot}=0.45\pm0.02$ and $(g'-i')_{\odot}=0.55\pm0.03$ are used for the colors of the Sun \citep{holmberg2006_solarcolors}.  Averaging $r'$-band and $i'$-band reflectance values for the two multi-filter observation sequences that we obtained and fitting a linear function to those average reflectance values, we find a mean spectral slope over the $g'r'i'$ wavelength region of $S'_{gri}=(5.91\pm0.01)$~\%/100~nm.  
Inserting our computed mean reflectance values into an inverted form of Equation~\ref{equation:reflectance}, we also find equivalent mean broadband colors of of $g'-r'=0.52\pm0.05$ and $r'-i'=0.22\pm0.07$.

The spectral slope we find for 313P is within 1-$\sigma$ of the mean visible slope of $(3.58\pm3.21)$\%/1000\AA\ found for members of the Lixiaohua \revision{asteroid} family by \citet{depra2020_lixiaohua}, \revision{and in particular, is close to the mean visible slope of $(5.99\pm1.00)$\%/1000\AA\ found for T-type asteroids in their sample,} and so is consistent with 313P being a member of that family.
\revision{This slope is also bluer than the lower end of the range of $S'_{gri}$ values ($6.0<S'_{gri}<25.0$) associated with D-type asteroids \citep{demeo2013_asteroidtaxonomy}, which are commonly regarded as spectroscopically similar to classical comet nuclei \citep[see ][and references therein]{kelley2017_midIRcomets}, suggesting that 313P is compositionally distinct from classical comets.  Without $z'$-band data, we cannot conclusively assign 313P a taxonomic classification, but we do note that, under the criteria outlined by \citet{demeo2013_asteroidtaxonomy}, 313P's spectral slope is within the range of $S'_{gri}$ values consistent with B-type asteroids, a taxonomic classification that has also been suggested for fellow MBCs 133P/Elst-Pizarro and 176P/LINEAR \citep{licandro2011_133p176p}.
}

\subsection{Non-Detections\label{section:nondetections}}

\setlength{\tabcolsep}{4.5pt}
\setlength{\extrarowheight}{0em}
\begin{table*}[htb!]
\caption{P/2015 X6 Non-Detections}
\centering
\smallskip
\footnotesize
\begin{tabular}{lccrcrrrrcrrcc}
\hline\hline
\multicolumn{1}{c}{UT Date}
 & \multicolumn{1}{c}{Telescope$^a$}
 & \multicolumn{1}{c}{$N$$^b$}
 & \multicolumn{1}{c}{$t$$^c$}
 & \multicolumn{1}{c}{Filter}
 & \multicolumn{1}{c}{$\nu$$^d$}
 & \multicolumn{1}{c}{$r_h$$^e$}
 & \multicolumn{1}{c}{$\Delta$$^f$}
 & \multicolumn{1}{c}{$\alpha$$^g$}
 & \multicolumn{1}{c}{$T$-mag$^h$}
 & \multicolumn{1}{c}{$\sigma_{\rm RA}$$^i$}
 & \multicolumn{1}{c}{$\sigma_{\rm Dec}$$^j$}
 & \multicolumn{1}{c}{$m_{r,{\rm lim}}$$^k$}
 & \multicolumn{1}{c}{\revision{$r_{n,{\rm lim}}$$^l$}}
 \\
\hline
2019 Aug 04 & Gemini-S &  4 & 1200$~~\,$ & $r'$ & 246.1 & 2.872 & 1.890 & 6.3 & 22.1 & 243.6 & 76.0 & 25.9$~~$ & \revision{0.13$\pm$0.03} \\
2019 Aug 05 & Gemini-S &  3 &  900$~~\,$ & $r'$ & 246.3 & 2.870 & 1.884 & 5.9 & 22.1 & 244.3 & 75.9 & 25.7$~~$ & \revision{0.14$\pm$0.03} \\
2019 Aug 06 & Gemini-S &  6 & 1800$~~\,$ & $r'$ & 246.5 & 2.868 & 1.879 & 5.6 & 22.0 & 245.1 & 75.9 & 26.1$~~$ & \revision{0.11$\pm$0.02} \\
2019 Sep 29 & Gemini-S &  6 & 1800$~~\,$ & $r'$ & 257.3 & 2.776 & 1.991 & 15.2 & 22.0 & 220.3 & 52.8 & 26.1$~~$ & \revision{0.14$\pm$0.03} \\
2019 Nov 21 & Gemini-S &  3 & 1800$~~\,$ & $r'$ & 268.8 & 2.680 & 2.565 & 21.6 & 22.4 & 171.5 & 46.8 & 26.1$~~$ & \revision{0.19$\pm$0.05} \\
2019 Nov 23 & Gemini-S &  3 & 1800$~~\,$ & $r'$ & 269.3 & 2.677 & 2.589 & 21.5 & 22.4 & 170.5 & 47.1 & 26.1$~~$ & \revision{0.19$\pm$0.05} \\
2019 Nov 24 & Gemini-S &  3 & 1800$~~\,$ & $r'$ & 269.5 & 2.675 & 2.601 & 21.5 & 22.4 & 169.9 & 47.2 & 26.1$~~$ & \revision{0.19$\pm$0.05} \\
2020 Aug 17 & Gemini-N &  3 &  900$~~\,$ & $r'$ & 341.3 & 2.296 & 2.563 & 23.2 & 21.8 & 251.7 & 36.0 & 25.7$~~$ & \revision{0.20$\pm$0.05} \\
2020 Aug 20 & LDT      &  2 &  600$~~\,$ & $VR$ & 342.1 & 2.294 & 2.531 & 23.6 & 21.7 & 254.4 & 34.5 & 25.3$~~$ & \revision{0.23$\pm$0.06} \\ 
2020 Sep 14 & CFHT     &  4 &  360$~~\,$ & $gri$ & 349.9 & 2.283 & 2.242 & 25.7 & 21.4 & 282.3 & 20.6 & 24.8$~~$ & \revision{0.27$\pm$0.07} \\
2020 Sep 16 & CFHT     &  4 &  360$~~\,$ & $gri$ & 350.5 & 2.283 & 2.219 & 25.8 & 21.4 & 284.9 & 19.5 & 24.8$~~$ & \revision{0.26$\pm$0.07} \\
2020 Oct 27 & Gemini-S &  2 &  600$~~\,$ & $r'$ & 3.1 & 2.279 & 1.742 & 24.2 &  20.9 & 360.5 &  2.1 & 25.5$~~$ & \revision{0.15$\pm$0.04} \\
2020 Dec 12 & Gemini-S &  3 &  900$~~\,$ & $r'$ & 17.3 & 2.293 & 1.358 & 10.0 & 20.4 & 489.5 &  9.8 & 25.7$~~$ & \revision{0.08$\pm$0.02} \\
2020 Dec 17 & Gemini-S &  9 & 3$\times$900$^m$ & $r'$ & 18.8 & 2.296 & 1.339 &  7.5 & 20.3 & 499.3 &  7.8 & 25.7$^n$ & \revision{0.08$\pm$0.02} \\
2021 Jan 09 & Gemini-S &  3 &  450$~~\,$ & $r'$ & 25.8 & 2.312 & 1.339 &  4.5 & 20.4 & 502.8 &  8.4 & 25.3$~~$ & \revision{0.09$\pm$0.02} \\
2021 Jan 10 & Gemini-S &  4 &  600$~~\,$ & $r'$ & 26.1 & 2.313 & 1.342 &  5.1 & 20.4 & 501.3 &  9.2 & 25.5$~~$ & \revision{0.08$\pm$0.02} \\
2021 Feb 04 & Gemini-S &  4 & 1200$~~\,$ & $r'$ & 33.5 & 2.335 & 1.499 & 16.0 & 20.7 & 434.3 & 20.9 & 25.9$~~$ & \revision{0.10$\pm$0.02} \\
\hline
\hline
\multicolumn{14}{l}{$^a$ Telescope used (CFHT: Canada-France-Hawaii Telescope; Gemini-N: Gemini North telescope;} \\
\multicolumn{12}{l}{~~~ Gemini-S: Gemini South telescope; LDT: Lowell Discovery Telescope).} \\
\multicolumn{14}{l}{$^b$ Number of exposures.} \\
\multicolumn{14}{l}{$^c$ Total integration time, in seconds.} \\
\multicolumn{14}{l}{$^d$ True anomaly, in degrees.} \\
\multicolumn{14}{l}{$^e$ Heliocentric distance, in au.} \\
\multicolumn{14}{l}{$^f$ Geocentric distance, in au.} \\
\multicolumn{14}{l}{$^g$ Solar phase angle (Sun-object-Earth), in degrees.} \\
\multicolumn{14}{l}{$^h$ Predicted apparent visual ($V$-band) total magnitude from JPL Horizons.} \\
\multicolumn{14}{l}{$^i$ 1-$\sigma$ ephemeris uncertainty, in arcsec, in right ascension from JPL Horizons, at the time of observations.} \\
\multicolumn{14}{l}{$^j$ 1-$\sigma$ ephemeris uncertainty, in arcsec, in declination from JPL Horizons, at the time of observations.} \\
\multicolumn{14}{l}{$^k$ Estimated 3-$\sigma$ limiting apparent magnitude scaled from limiting magnitude measured from data obtained on } \\
\multicolumn{14}{l}{$~~~$ UT 2020 December 17.} \\
\multicolumn{14}{l}{\revision{$^l$ Upper limit effective nucleus radius corresponding to estimated limiting apparent magnitude,}} \\
\multicolumn{14}{l}{\revision{$~~~$ assuming $G_r=0.18\pm0.28$, solar colors, and $p_V=0.05\pm0.02$.}} \\
\multicolumn{14}{l}{$^m$ \revision{Total integration times per field for observations of three adjacent fields.}} \\
\multicolumn{14}{l}{$^n$ 3-$\sigma$ limiting magnitude directly determined from data.} \\
\end{tabular}
\label{table:obs_p2015x6}
\end{table*}


\subsubsection{\revision{\it{Overview}}\label{section:nondetections_overview}}

Despite making multiple attempts to observe P/2015 X6 \revision{from 2019 to 2021}, we were not able to secure definitive recovery detections of the object. 
\revision{We also made several unsuccessful attempts to recover P/2017 S9 between June 2021 and June 2022, before the object was finally found in PS1 and Pan-STARRS2 (PS2) survey data in December 2022 \citep{weryk2022_p2017s9}.}

Attempted observations of P/2016 X6 were obtained using Gemini-South, LDT, and CFHT (Table~\ref{table:obs_p2015x6}), while attempted observations for P/2019 S9 were obtained using Gemini-North, Gemini-South, and the Palomar Hale telescope (Table~\ref{table:obs_p2017s9_nondetections}).  Observations with the 8.1~m Gemini telescopes of course provided the most sensitive images, but unfortunately, also had the smallest FOVs at $5\farcm5\times5\farcm5$ (Table~\ref{table:instrumentation}), while most observations with the 3.54~m CFHT, 4.3~m LDT, and 5.1~m Palomar Hale telescope were not as sensitive as those from Gemini but had \revision{significantly larger FOVs (see Table~\ref{table:instrumentation})}.  We also were able to use non-standard ``wide-band'' filters spanning multiple standard broadband filter bandpasses for most of our \revision{CFHT ($gri$), LDT ($VR$), and Palomar ($RI$) observations}.

To maximize the sensitivity of our data, we created composite images for each night of observations by shifting and aligning individual images based on each object's expected non-sidereal rates of apparent motion in right ascension and declination (obtained from JPL Horizons) using linear interpolation and then adding the images together. \revision{These composite images could then be searched for point-source-like detections corresponding to objects just below the detection limit of individual images that are moving at the non-sidereal rates matching the target of each set of observations.  We however were unable to detect either object in any of the observations listed in Tables~\ref{table:obs_p2015x6} and \ref{table:obs_p2017s9_nondetections} using this method.}


We note that given the magnitudes predicted by JPL Horizons (see Tables~\ref{table:obs_p2015x6} and \ref{table:obs_p2017s9_nondetections}) as well as our own estimates from previous observations (generally obtained when the objects were active, and as such, are highly uncertain), our targets should have been bright enough to detect with the telescopes and exposure times used for these observations.  For reference, we provide approximate 3-$\sigma$ $r'$-band limiting magnitudes for our observations 
\revision{of P/2015 X6 (see Section~\ref{section:nondetections_p2015x6}) and P/2017 S9 (see Section~\ref{section:nondetections_p2017s9}),
as well as corresponding upper limit effective nucleus radii \revision{(using Equation~\ref{equation:nucleus_size})} corresponding to these limiting magnitudes in Tables~\ref{table:obs_p2015x6} and \ref{table:obs_p2017s9_nondetections}.}

We note that these limiting magnitude estimates are based on the best-case-scenario assumption that the targets were not obscured by background sources at any point during their observing sequences.  To account for the possibility of our target objects being hidden by field stars or galaxies during part of their observing sequences, we also inspected individual images and constructed composite images from subsets of data from single observing nights, also finding no likely detections.


\setlength{\tabcolsep}{5.7pt}
\setlength{\extrarowheight}{0em}
\begin{table*}[htb!]
\caption{P/2017 S9 Non-Detections$^a$}
\centering
\smallskip
\footnotesize
\begin{tabular}{lcrrcrrrrcrrcc}
\hline\hline
\multicolumn{1}{c}{UT Date}
 & \multicolumn{1}{c}{Telescope$^b$}
 & \multicolumn{1}{c}{$N$}
 & \multicolumn{1}{c}{$t$}
 & \multicolumn{1}{c}{Filter}
 & \multicolumn{1}{c}{$\nu$}
 & \multicolumn{1}{c}{$r_h$}
 & \multicolumn{1}{c}{$\Delta$}
 & \multicolumn{1}{c}{$\alpha$}
 & \multicolumn{1}{c}{$T$-mag}
 & \multicolumn{1}{c}{$\sigma_{\rm RA}$}
 & \multicolumn{1}{c}{$\sigma_{\rm Dec}$}
 & \multicolumn{1}{c}{$m_{r,{\rm lim}}$}
 & \multicolumn{1}{c}{\revision{$r_{n,{\rm lim}}$}}
 \\
\hline
2021 Jun 03$^c$ & Gemini-S &  3 & 1119 & $r'$ & 220.9 & 3.716 & 2.754 &  5.7 & 24.7 & 1915.8 & 339.9 & \revision{---} & \revision{---} \\
2021 Jul 12 & Palomar  & 17 & 5100 & $RI$ & 225.8 & 3.633 & 2.951 & 13.2 & 24.8 & 1879.4 & 336.1 & 25.4$^d$ & \revision{0.35$\pm$0.08} \\ 
2021 Aug 05 & Palomar  & 16 & 4800 & $r'$ & 228.9 & 3.578 & 3.201 & 16.0 & 24.9 & 1523.6 & 334.8 & 24.8$^e$ & \revision{0.52$\pm$0.13} \\
2022 May 26$^c$ & Gemini-N &  3 &  900 & $r'$ & 278.4 & 2.738 & 2.365 & 21.3 & 23.0 & 3662.7 & 258.8 & \revision{---} & \revision{---} \\
2022 Jun 01$^c$ & Gemini-N &  3 &  900 & $r'$ & 279.7 & 2.720 & 2.271 & 21.1 & 22.9 & 3833.2 & 286.9 & \revision{---} & \revision{---} \\
2022 Jun 06$^c$ & Gemini-N &  3 &  900 & $r'$ & 280.9 & 2.705 & 2.195 & 20.8 & 22.8 & 3985.2 & 311.5 & \revision{---} & \revision{---} \\
2022 Jun 07$^c$ & Gemini-N &  3 &  900 & $r'$ & 280.9 & 2.705 & 2.195 & 20.8 & 22.8 & 4017.6 & 316.8 & \revision{---} & \revision{---} \\
2022 Jun 09$^c$ & Gemini-N &  3 &  900 & $r'$ & 281.5 & 2.696 & 2.150 & 20.5 & 22.8 & 4080.8 & 326.9 & \revision{---} & \revision{---} \\
\hline
\hline
\multicolumn{14}{l}{$^a$ See Table~\ref{table:obs_p2015x6} for column heading explanations} \\
\multicolumn{14}{l}{$^b$ Telescope used (Gemini-N: Gemini North telescope; Gemini-S: Gemini South telescope; Palomar: Palomar Hale telescope).} \\
\multicolumn{14}{l}{$^c$ \revision{Object's expected position using updated orbital elements not within field of view of observations.}} \\
\multicolumn{14}{l}{$^d$ Estimated 3-$\sigma$ limiting magnitude \revision{at expected position of object using updated orbital elements,} scaled from } \\
\multicolumn{14}{l}{$~~~$ limiting magnitude measured from data obtained on UT 2021 August 5.} \\
\multicolumn{14}{l}{$^e$ 3-$\sigma$ limiting magnitude \revision{at expected position of object using updated orbital elements} directly determined } \\
\multicolumn{14}{l}{$~~~$ from data} \\
\end{tabular}
\label{table:obs_p2017s9_nondetections}
\end{table*}

\subsubsection{\revision{\it{P/2015 X6 (PANSTARRS)}}\label{section:nondetections_p2015x6}}

In the case of P/2015 X6, most of our observations were obtained using Gemini-North or Gemini-South and thus focused on maximizing image depth in the immediate vicinity of the predicted position of the object and did not cover the full 1-$\sigma$ ephemeris uncertainty region as specified by JPL Horizons.  We however also obtained one set of Gemini-South observations that covered most of P/2016 X6's 1-$\sigma$ ephemeris uncertainty region at the time, one set of LDT observations that fully covered the object's 1-$\sigma$ ephemeris uncertainty region, and two sets of CFHT observations that fully covered the object's 3-$\sigma$ ephemeris uncertainty region.

The Gemini-South observations \revision{that covered most of P/2016 X6's 1-$\sigma$ ephemeris uncertainty region} were conducted on UT 2020 December 17 when P/2015 X6's nucleus should have been at its brightest point during our recovery campaign, and in fact, was even expected to be active again if a dust modeling analysis indicating that its 2015 activity was sublimation-driven was correct \citep{moreno2016_p2015x6}.  We utilized a mosaic search approach for these observations \revision{to cover the object's 1-$\sigma$ ephemeris uncertainty region ($\pm8\farcm32$ in the right ascension direction and $\pm0\farcm1$ in the declination direction; see Table~\ref{table:obs_p2015x6})}, where we observed one field centered on the object's expected position and then also observed one field offset by $5\farcm5$ to the East and another offset by $5\farcm5$ to the West, where each field was observed to the same depth as our other sets of Gemini observations\revision{, giving a total coverage of $\pm8\farcm25$ in the right ascension direction and $\pm2\farcm75$ in the declination direction relative to the object's predicted position}.
We measure 3-$\sigma$ detection limits of $m_r=24.9$ for single exposures, and $m_r=25.7$ for composite image stacks of three exposures each.  This detection limit corresponds to a lower-limit absolute magnitude of $H_r=22.7\pm0.2$ (assuming $G_r=0.18\pm0.28$; see Section~\ref{section:phase_functions}), or $H_V=22.9\pm0.2$, corresponding to a nucleus radius of $r_n<0.08\pm0.02$~km (assuming $p_V=0.05\pm0.02$; see Section~\ref{section:nucleus_sizes}).

\revision{Limiting magnitudes for other Gemini observations listed in Table~\ref{table:obs_p2015x6} are estimated by scaling the directly measured limiting magnitude found for our UT 2020 December 17 Gemini observations by relative total integration times, while the limiting magnitude of our LDT observations is estimated by scaling from those UT 2020 December 17 Gemini observations based on both relative total integration time and relative primary mirror aperture size.}
Determining a precise limiting magnitude for our CFHT observations is not straightforward given the use of a non-standard filter for these observations, but scaling \revision{by relative total integration time and relative primary mirror aperture size} from our Gemini observations, and finding from the MegaCam Exposure Time Calculator\footnote{https://etc.cfht.hawaii.edu/mp/} that $gri$-band observations are approximately 0.5~mag deeper than $r'$-band observations, we estimate that MegaCam observations on UT 2020 September 14 and 16 had a total 3-$\sigma$ limiting magnitude of $m_r\sim24.8$~mag on each night.  This detection limit corresponds to a lower-limit absolute magnitude of $H_r=20.1\pm0.4$ (assuming $G_r=0.18\pm0.28$), or $H_V=20.3\pm0.4$, corresponding to a nucleus radius of $r_n<0.27\pm0.07$~km (assuming $p_V=0.05$), which sets a less stringent nucleus size limit than our Gemini observations described above, but as discussed above, applies to P/2015 X6's entire 3-$\sigma$ ephemeris uncertainty region.

We conclude that P/2015 X6's current nucleus size is below our detection limits (i.e., $r\lesssim0.1$~km and $r\lesssim0.3$~km for its 1-$\sigma$ and 3-$\sigma$ ephemeris uncertainty regions, respectively).  We further suggest that the object may have even disintegrated following its 2015 apparition given our unsuccessful recovery attempts in 2020 \revision{when the nucleus should have been at its brightest, and was even expected to potentially become active again}.  That said, we cannot exclude the possibility that a combination of uncertainty in the astrometric measurements used to derive the object's orbit and possible non-gravitational perturbations from asymmetric mass loss \citep[e.g., see][]{hui2017_activeastsnongrav} since 2015 means that the orbit solution for the nucleus is simply no longer accurate enough to recover the object.  However, the lack of any serendipitous recoveries by all-sky surveys as a new solar system object (\revision{e.g., like in the case of P/2017 S9}) during the object's 2020 perihelion passage during which it should have reactivated (assuming that its 2015 activity was sublimation-driven) appears to favor the distintegration scenario.  

\subsubsection{\revision{\it{P/2017 S9 (PANSTARRS)}}\label{section:nondetections_p2017s9}}

\revision{As discussed in Section~\ref{section:nondetections_overview}, we made several unsuccessful attempts to recover P/2017 S9 in 2021 and early 2022 when the object's ephemeris uncertainties were extremely large (see Table~\ref{table:obs_p2017s9_nondetections}).  However, the successful recovery of the object in December 2022 \citep{weryk2022_p2017s9} allowed us to revisit those observations to determine if they should have been able to detect the object.  Using updated ephemeris predictions (which now have sub-arcsec uncertainties), we find that P/2017 S9's expected position was outside the FOVs of all of our Gemini-North and Gemini-South observations, but within the FOVs of our Palomar images.  However, we still did not detect the object even in composite images constructed from those Palomar data.}

\revision{We directly measure a 3-$\sigma$ $r'$-band detection limit of $m_r=24.8$ for the composite image constructed from our UT 2021 August 05 Palomar data.  Scaling this limiting magnitude by the slightly larger total exposure time of our UT 2021 July 12 Palomar observations and estimating an additional 0.5~mag of sensitivity due to the broadband $RI$ filter used to obtain those data, we estimate a 3-$\sigma$ $r'$-band detection limit of $m_r=25.4$ for those observations.}

\revision{Using the lower-limit absolute magnitude we compute from our detection with Gemini-South on UT 2022 December 24 after the object had been recovered (see Table~\ref{table:nucleus_phase_function_params}), we calculate that the object should have had lower-limit apparent $r'$-band magnitudes of $m_r=24.8\pm0.6$ and $m_r=25.1\pm0.6$ during our UT 2021 July 12 and 2021 August 5 Palomar observations, respectively, close to the limiting magnitudes determined or estimated for those data ($m_{r,{\rm lim}}=25.4$ and $m_{r,{\rm lim}}=24.8$, respectively).  If unresolved ejected dust was present during our 2022 Gemini-South observations (obtained when the object was at $\nu=338.2^{\circ}$) and not present during our 2021 Palomar observations (obtained at $225^{\circ}<\nu<230^{\circ}$), that could explain these non-detections in the Palomar data.  These non-detections could also be explained by rotational lightcurve variability, i.e., if our 2022 Gemini-South observations occurred near a lightcurve maximum for the object while our 2021 Palomar observations occurred near lightcurve minima.  We therefore conclude that there is insufficient evidence at the present time to determine if P/2017 S9 was active at the time of our 2022 Gemini-South observations.
}



\subsubsection{\revision{\it{Lessons Learned}}\label{section:nondetections_lessons}}

The results presented above highlight the need to secure recovery observations of newly discovered MBCs during as much of their discovery apparitions as possible while they are still bright enough to detect, as well as during the crucial second apparition following their discoveries, when possible (which, to be fair, may be extremely challenging for objects whose brightnesses decrease significantly once they become inactive).  These steps will help to maximize the initial observation arcs of newly discovered objects before ephemeris uncertainties grow so large that the objects become unrecoverable.  \revision{Such efforts will preserve our ability to observe these rare objects in the future (e.g., for carrying out long-term activity evolution studies) or otherwise at least set meaningful upper limits on an object's brightness and size based on non-detections.  Even after an object's orbit is relatively secure, ongoing periodic recoveries will also aid in the identification of deviations in an object's long-term orbit evolution due to non-gravitational perturbations from asymmetric mass loss \citep[e.g., see][]{hui2017_activeastsnongrav}, which will be useful both for scientific purposes and further ensuring that the object remains observable in the future. }




We note that efforts to recover P/2015 X6 and \revision{further characterize} P/2017 S9's physical and dynamical properties could be supplemented by searches for detections in archival telescope data using search tools such as the Canadian Astronomy Data Centre's Solar System Object Image Search tool\footnote{\url{http://www.cadc-ccda.hia-iha.nrc-cnrc.gc.ca/en/ssois/}} \citep{gwyn2012_ssois}, the NASA Planetary Data System (PDS) Small Bodies Node's Comet Asteroid Telescopic Catalog Hub (CATCH) tool\footnote{\url{https://catch.astro.umd.edu/}}, and the Zwicky Transient Facility data search page\footnote{\url{https://irsa.ipac.caltech.edu/applications/ztf/}}.  Analyses of the observations identified by such tools as potentially containing detections of our target objects are beyond the scope of the work presented here, however, and so have not been performed at this time.

\setlength{\tabcolsep}{4.5pt}
\setlength{\extrarowheight}{0.4em}
\begin{table*}[htb!]
\caption{Previously Reported and New MBC Nucleus Size Estimates}
\centering
\smallskip
\footnotesize
\begin{tabular}{lc>{\centering\arraybackslash}p{2.0cm}>{\centering\arraybackslash}p{2.0cm}cc>{\centering\arraybackslash}p{1.8cm}cl}
\hline\hline
 && \multicolumn{2}{c}{Previous Work}
 && \multicolumn{2}{c}{This Work} \\
\multicolumn{1}{c}{Object}
 && \multicolumn{1}{c}{$H_V$$^a$}
 & \multicolumn{1}{c}{$r_n$$^b$}
 && \multicolumn{1}{c}{$H_V$$^c$}
 & \multicolumn{1}{c}{$r_n$$^d$}
 && \multicolumn{1}{l}{Reference for Previous Work}
 \\
\hline
133P$^\ddagger$ &&           15.84$\pm$0.05 & 1.9$\pm$0.3 && \multirow{2}{*}{15.88$^{+0.04}_{-0.04}$} & \multirow{2}{*}{2.03$^{+0.57}_{-0.32}$} && \citet{hsieh2009_albedos} \\
133P$^\ddagger$  &&           15.70$\pm$0.10 & 2.2$\pm$0.5 &&  &  && \citet{jewitt2014_133p} \\[-10pt]
176P$^\ddagger$ &&           15.45$\pm$0.05 & 2.0$\pm$0.2 && 15.42$^{+0.03}_{-0.03}$ & 2.29$^{+0.51}_{-0.31}$ && \citet{hsieh2009_albedos} \\
238P$^\dagger$  &&           19.40$\pm$0.07 & $\sim0.4$ && 20.53$^{+0.13}_{-0.13}$   & 0.24$^{+0.07}_{-0.04}$ && \citet{hsieh2011_238p} \\
259P$^\ddagger$ &&           20.06$\pm$0.05 & 0.30$\pm$0.02 && 19.92$^{+0.08}_{-0.08}$ & 0.32$^{+0.09}_{-0.05}$ && \citet{maclennan2012_259p} \\
288P$^{\ddagger e}$ &&           17.15$\pm$0.12 & 0.80$\pm$0.04($\times$2) && \multirow{2}{*}{17.13$^{+0.04}_{-0.04}$} & \multirow{2}{*}{\hspace{-30pt}\makecell{0.86$\pm$0.17 \\ 0.61$\pm$0.12}} && \citet{hsieh2018_238p288p} \\
288P$^{\ddagger e}$ && {17.1$-$18.0  17.9$-$18.3} & {0.64$-$0.94  0.54$-$0.67} &&  &  && \citet{agarwal2020_288p} \\
313P$^{\dagger}$ &&           17.1$\pm$0.3  & 1.00$\pm$0.15 && \multirow{2}{*}{17.77$^{+0.19}_{-0.17}$} & \multirow{2}{*}{0.85$^{+0.25}_{-0.15}$} && \citet{hsieh2015_313p} \\
313P$^{\dagger}$ &&           ---            & 0.7$\pm$0.1  &&  &  && \citet{jewitt2015_313p2} \\
324P$^{\dagger\ddagger}$ &&  18.8$\pm$0.2 & 0.55$\pm$0.05  && 18.55$^{+0.19}_{-0.17}$   & 0.59$^{+0.18}_{-0.10}$ && \citet{hsieh2014_324p} \\
358P$^\ddagger$ &&           19.9$\pm$0.2 & 0.32$\pm$0.03 && 20.15$^{+0.24}_{-0.29}$ & 0.29$^{+0.10}_{-0.06}$ && \citet{hsieh2018_358p} \\
426P$^{\dagger}$ && --- & --- && 17.41$^{+0.02}_{-0.02}$ & 1.00$^{+0.28}_{-0.15}$ && none \\
427P$^{\dagger}$ &&           ---   & 0.45$\pm$0.10 && 19.13$^{+0.23}_{-0.28}$ & 0.46$^{+0.15}_{-0.09}$ && \citet{jewitt2019_p2017s5} \\
433P$^\ddagger$ &&           16.32$\pm$0.08 & 1.6$\pm$0.2 && 16.39$^{+0.29}_{-0.38}$ & 1.61$^{+0.60}_{-0.34}$ && \citet{hsieh2021_433P} \\
P/2013 R3       &&           --- & $<$0.4 && --- & --- && \citet{jewitt2017_p2013r3} \\
P/2016 J1-A$^{\dagger}$ && --- & $\lesssim0.9$ && $>$19.95$^{+0.18}_{-0.16}$ & $<$0.31$^{+0.09}_{-0.05}$ && \citet{hui2017_p2016j1} \\
P/2016 J1-B$^{\dagger}$ && --- & $\lesssim0.4$ && $>$21.26$^{+0.23}_{-0.26}$ & $<$0.17$^{+0.06}_{-0.03}$ && \citet{hui2017_p2016j1} \\
\revision{P/2017 S9$^{\dagger}$} && \revision{---} & \revision{---} && \revision{$\geq$19.10$^{+0.35}_{-0.48}$} & \revision{$\leq$0.48$^{+0.20}_{-0.11}$} && \revision{none} \\
P/2019 A3$^{\dagger}$ && --- & --- && 19.37$^{+0.11}_{-0.10}$ & 0.41$^{+0.12}_{-0.07}$ && none \\
P/2020 O1             && 19.25$\pm$0.13 & 0.42$\pm$0.03 && --- & --- && \citet{kim2022_p2020o1} \\
P/2021 A5             && ---            & $\sim$0.15 && --- & --- && \citet{moreno2021_2019a4_2021a5} \\[2pt]
\hline
\hline
\multicolumn{9}{l}{$^a$ $V$-band absolute magnitude reported in previous work  (converted as needed from absolute magnitude} \\
\multicolumn{9}{l}{$~~~$ in originally reported filter assuming approximately solar colors).} \\
\multicolumn{9}{l}{$^b$ Nucleus radius, in km, reported in previous work.} \\
\multicolumn{9}{l}{$^c$ $V$-band absolute magnitude reported in this work (Table~\ref{table:nucleus_phase_function_params}).} \\
\multicolumn{9}{l}{$^d$ Nucleus radius, in km, reported in this work (Table~\ref{table:nucleus_phase_function_params}).} \\
\multicolumn{9}{l}{$^e$ Identified as a binary system by \citet{agarwal2017_288p}.} \\
\multicolumn{9}{l}{$^\dagger$ Phase function solution based on data reported in this work.} \\
\multicolumn{9}{l}{$^\ddagger$ Phase function solution based on data reported in previous works; see Table~\ref{table:nucleus_phase_function_params} for references} \\
\end{tabular}
\label{table:nucleus_size_comparison}
\end{table*}

\section{DISCUSSION\label{section:discussion}}

\subsection{Comparison to Previously Derived Nucleus Sizes}\label{section:discussion_prev_nucleus_sizes}

Of the \revision{15} objects for which we report nucleus sizes in this work (Table~\ref{table:nucleus_phase_function_params}), 12 have had nucleus sizes reported in previous works.  Of the revised nucleus sizes for those 12 objects we report here, five (238P, 313P, 427P, P/2016 J1-A, and P/2016 J1-B) are computed from newly reported data, six (133P, 176P, 259P, 288P, 358P, and 433P)
are re-computed from previously reported data using the Monte Carlo approach and true anomaly limits on photometric data detailed in Section~\ref{section:phase_functions} in an effort to determine more realistic uncertainties and minimize photometric contamination from faint activity, and one (324P) is computed from a combination of previously reported and newly reported data.

We list nucleus size estimates reported in the literature for MBCs, candidate MBCs, and in the case of a catastrophically disrupted MBC, the inferred precursor object, in Table~\ref{table:nucleus_size_comparison}, alongside summarized results from this work for reference.
In addition to the objects already discussed in Section~\ref{section:intro}, Table~\ref{table:nucleus_size_comparison} also lists nucleus sizes for P/2020 O1 (Lemmon-PANSTARRS), P/2021 A5 (PANSTARRS), and the aforementioned catastrophically disrupted P/2013 R3 (Catalina-PANSTARRS), all of which have been determined to have exhibited activity likely due to sublimation \citep{kim2022_p2020o1,moreno2021_2019a4_2021a5,jewitt2014_p2013r3}, and thus are considered likely MBCs for the purposes of this work.

Comparing results from this work to those of previous work listed in Table~\ref{table:nucleus_size_comparison}, we find general agreement, within uncertainties, between our computed absolute magnitudes and estimated nucleus sizes and previously reported results in almost all cases.  The main exception is the case of 238P, which as we discuss in Section~\ref{section:activity_detection_238p}, we find was actually active at the time that the photometry used to compute its previously reported phase function parameters were obtained.  In many cases, uncertainties on estimated radii also increased significantly, since we see that these uncertainties are dominated by uncertainty on the assumed albedo, and almost all previous nucleus size estimates simply used a fixed assumed albedo value without any associated uncertainties.

In the cases of 133P and 176P, \citet{hsieh2009_albedos} used the \citet{harris1998_neasurveys} Near-Earth Asteroid Thermal Model (NEATM) to simultaneously solve for effective nucleus radii and geometric R-band albedos, incorporating both optical and infrared photometry.  This of course differs from the method we use in this work for converting derived $H_V$ values to nucleus radii (Section~\ref{section:nucleus_sizes}). The method used here yields larger nucleus size estimates even using the same $H_V$ values reported in the previous work, thus explaining the larger differences in reported nucleus radii for 133P and 176P between this work and previous work than perhaps would be expected from the differences in reported $H_V$ values alone.

Meanwhile, in the case of 313P, the nucleus size estimate from \citet{jewitt2015_313p2} was based on high-spatial resolution {\it HST} image data when the object was active, and as such, the discrepancy between that size estimate and the one reported here could be due to model-dependent over-subtraction of the coma by \citet{jewitt2015_313p2}.  As such, we consider our size determination based on a full phase function fit to observations when no activity was visibly present to be the more reliable result, although we note that both values are consistent with each other within reported uncertainties.  By contrast, nucleus size estimates for 133P and 427P independently derived by \citet{jewitt2014_133p} and \citet{jewitt2019_p2017s5}, respectively, from {\it HST} data and derived here from ground-based optical data are in good agreement, and can therefore be considered high-confidence results.

\subsection{Significance of MBC Nucleus Sizes\label{section:discussion_size_implications}}

Adopting the upper limit nucleus radii determined here for P/2016 J1-A and P/2016 J1-B as their true radii for the purposes of this analysis, and assuming that they account for the majority of the mass of their original parent body, we find an approximate effective radius of the parent body of $r_n\sim0.32$~km (i.e., the radius of a spherical body with a mass equal to the total mass of the two components, which are also both assumed to be spherical).  Similarly assuming that the two components of 288P formed from the splitting of a single parent body, we find an approximate effective radius of the parent body of $r_n\sim1.0$~km.  Thus, from Table~\ref{table:nucleus_phase_function_params}, of the \revision{17} distinct MBCs or candidate MBCs for which nucleus sizes have been measured or estimated, we see that more than half (\revision{10} objects or their inferred parent bodies) have $r_n\leq0.5$~km, $\sim$80\% (\revision{14} objects or their inferred parent bodies) have $r_n\leq1.0$~km, and all have $r_n\leq2.5$~km.  Due to numerous confounding factors including small sample size, poorly constrained discovery biases, \revision{and poor constraints on the size-frequency distribution of outer main-belt asteroids at sub-km scales}, a direct comparison of the MBC size distribution to that of the background asteroid population would not be particularly meaningful.  Nonetheless, given that larger MBCs should be brighter and thus presumably easier to discover if they exist, we note what appears to be a physical preference towards small (i.e., $r_n<1$~km) MBCs.

\citet{hsieh2009_htp} listed various conditions suggesting that detectable MBCs could occupy a narrow range of sizes.  For example, if MBCs require recent impacts in order to be active (i.e., via the impact excavation of surface regolith insulating a pocket of near-surface ice from solar heating), as has been suggested \citep[e.g.,][]{hsieh2004_133p,haghighipour2016_mbcimpacts}, larger asteroids would be expected to be more likely to exhibit activity due to their larger collisional cross-sections and thus higher likelihood of impact activation.  Larger asteroids should also be able to preserve ice at greater depths and therefore for longer periods of time.

However, larger asteroids also have larger escape velocities, meaning that low-velocity dust that is ejected by many MBCs \citep[e.g., see][]{hsieh2004_133p,hsieh2009_238p} may be unable to escape the gravity of larger asteroids.  As such, even if active sublimation were taking place on those objects, ejected dust would simply fall back to the surface and never become detectable from the Earth. \citet{hsieh2009_htp} also noted that smaller objects are less effectively heated from within by $^{26}$Al, meaning that ice would be more likely to survive early radioactive heating. This last argument implicitly assumes that a particular object is primordial, however, which we now have reasons to believe may not be the case for many MBCs.

In addition to the points above, there are also other ways that size could influence which objects become observably active MBCs.  Smaller objects are destroyed by collisions on statistically shorter timescales than larger objects \citep{cheng2004_collisionalevolution,bottke2005_collisionalevolution}. As such, currently existing small asteroids should be statistically younger than larger objects, perhaps being produced by recent asteroid family forming events.  With statistically younger ages, currently existing small asteroids would be more likely than larger asteroids to contain extant near-surface ice \citep[provided that the parent bodies from which they formed contained preserved ice, perhaps deep in their interiors; e.g.,][]{prialnik2009_mbaice} and thus more likely to become exhibit present-day sublimation-driven activity.  This hypothesis is supported by a finding that MBCs are dynamically linked to known or candidate asteroid families (which are also all found to be dominated by member asteroids with low albedos and primitive taxonomic classifications) at a much higher rate than would be expected from chance given the overall family association rate found for the general main-belt asteroid population \citep{hsieh2018_activeastfamilies}.

The size-dependence of the Yarkovsky-O'Keefe-Radzievskii-Paddack (YORP) effect \citep{rubincam2000_yorp,vokrouhlicky2015_yarkovskyyorp_ast4} may also be significant, given the number of MBCs or candidate MBCs for which rapid rotation, perhaps resulting from YORP-driven spin-up, may contribute to the escape of dust particles. For these objects, the underlying primary activity driver is still believed to be due to gas drag from sublimation based on dust modeling results showing prolonged dust emission episodes, recurrent activity near perihelion, or both.  However, dust modeling results indicate low dust ejection speeds for many of these objects, suggesting that centrifugal forces from rapid rotation may facilitate the escape of particles that may not otherwise be ejected at velocities large enough to exceed each object's surface gravity \citep[e.g.,][]{jewitt2014_133p}. This possibility is supported by direct evidence of rapid rotation in several MBCs with slow estimated dust ejection velocities, including 133P \citep{hsieh2004_133p}, 427P \citep{jewitt2019_p2017s5}, 433P \citep{novakovic2022_433p}, and P/2020 O1 \citep{kim2022_p2020o1}.  There may also be other fast-rotating MBCs that have not yet been identified as such, given the difficulty of measuring rotation rates of such faint targets, where only one MBC, 176P, has actually been confirmed to date to have a slow rotation rate \citep[$P_{\rm rot}=22.23$~h;][]{hsieh2011_176p}.

Rapid rotation can also contribute to triggering sublimation-driven activity via the excavation of subsurface ice reservoirs by either the actual removal of mass into space \citep[e.g.,][]{hirabayashi2015_rotationalshedding} or simply the movement of regolith from one place to another via processes like landslides \citep[e.g.,][]{scheeres2015_spinmassloss}.  Given certain conditions \citep[e.g.,][]{marzari2011_yorpcollisions,hirabayashi2014_p2013r3}, rapid rotation can even result in the fission of an asteroid, which would clearly have the potential for exposing deeply buried ice, thus triggering activity, as might have occurred in the cases of MBCs 288P \citep{agarwal2020_288p} and P/2013 R3 \citep{jewitt2014_p2013r3}.

These considerations are important of course because smaller objects are more susceptible to spin-up via the YORP effect, with empirical studies indicating that YORP-driven spin-up becomes a significant effect for main-belt asteroids around diameters of $D\sim5$~km, which, interestingly, is the same maximum size that we happen to find for all MBC nuclei measured to date.  For instance, spin rates are found to deviate from a Maxwellian distribution for main-belt asteroids with $D\lesssim5$~km \citep{polishook2009_mbaphotometryspin}, while comparisons of numerical models to observations suggests that YORP-induced rotational disruption has a non-negligible effect on the size-frequency distribution of main-belt asteroids at diameters of $D<6$~km \citep{jacobson2014_rotationaldisruption}.  That said, while these findings suggest that YORP spin-up could be a significant contributor to MBC activity, thus favoring smaller ($D\lesssim5$~km) MBCs, further work is certainly needed to clarify the details of this contribution.  In particular, further observations to better establish the rotation period distribution of active MBCs as well as additional modeling efforts investigating the details of various YORP- and rotation-related mechanisms for triggering and maintaining activity in icy asteroids would be especially useful.

\revision{Continued efforts to determine nucleus sizes for MBCs discovered in the future either by current surveys or the Vera C.\ Rubin Observatory's upcoming Legacy Survey of Space and Time \citep[LSST;][]{ivezic2019_lsst} will also be informative for assessing the robustness of the size preference we find in this work towards small ($r_n<1$~km) MBCs, as well as possible physical reasons behind the size preference, if it is found to be real.  LSST is expected to reach a 5$\sigma$ limiting magnitude of $m=24.4$ in a single 30~s $r'$-band exposure\footnote{\url{https://smtn-002.lsst.io/}}, which is not as deep as many of the observations reported here.  As such, while LSST will provide the benefit of more frequent and regular observations, the fact that there are often only small observing windows during which small MBC nuclei are bright enough to observe but distant enough to be inactive means that future characterization efforts for at least some MBC nuclei may still require targeted observations similar to the ones reported here.  This will be especially true for efforts to obtain rotational lightcurves, which typically require data with greater and more consistent photometric precision than is normally needed for phase function determination, where these will be crucial for discerning between the YORP-driven spin-up hypothesis for MBC activity presented here and the hypothesis that the small sizes of MBCs are indicative of their formation in recent family-forming events \citep{hsieh2018_activeastfamilies}.}

\revision{In addition to obtaining direct observations of MBC nuclei from which sizes can be estimated, LSST should also help to better constrain the size-frequency distribution of the outer main asteroid belt at sub-km scales.  These results should then enable more meaningful assessments of whether the high fraction of MBC nuclei with $r_n<1$~km can be simply attributed to the abundance of outer main-belt asteroids in that size range in general, or reflects a true physical preference towards small MBCs.}

\section{SUMMARY\label{section:summary}}

In this work, we presented observations of a number of MBCs when they were expected to be inactive in order to constrain the sizes and other physical properties of their nuclei.  We report the following key results:
\begin{enumerate}
    \item{Using a Monte Carlo-style approach to phase function fitting in order to ascertain realistic uncertainties, we find best-fit $V$-band absolute magnitudes and equivalent effective nucleus radii (assuming $V$-band albedos of $p_V=0.05\pm0.02$) of
    $H_V=20.5\pm0.1$ and $r_n=(0.24\pm0.05)$~km for 238P/Read,
    $H_V=17.8\pm0.1$ and $r_n=(0.9\pm0.2)$~km for 313P/Gibbs,
    $H_V=18.6\pm0.2$ and $r_n=(0.6\pm0.1)$~km for 324P/La Sagra,
    $H_V=17.41\pm0.02$ and $r_n=(1.0\pm0.2)$~km for 426P/PANSTARRS,
    $H_V=19.1\pm0.3$ and $r_n=(0.5\pm0.1)$~km for 427P/ATLAS,
    $H_V>20.0\pm0.2$ and $r_n<(0.3\pm0.1)$~km for P/2016 J1-A (PANSTARRS),
    $H_V>21.3\pm0.3$ and $r_n<(0.17\pm0.04)$~km for P/2016 J1-B (PANSTARRS),
    \revision{$H_V\geq19.1\pm0.5$ and $r_n\leq(0.5\pm0.2)$~km for P/2017 S9 (PANSTARRS),} and
    $H_V=19.4\pm0.1$ and $r_n=(0.4\pm0.1)$~km for P/2019 A3 (PANSTARRS).
    \revision{P/2016 J1-A and P/2016 J1-B were found to be active during at least a portion of our observations, while our observations of P/2017 S9 occurred at an orbit position at which other MBCs have exhibited activity, and as such, we report their $V$-band absolute magnitudes and effective nucleus radii as lower and upper limits, respectively.}
    We note that the absolute magnitude of 427P and lower-limit absolute magnitudes of P/2016 J1-B and P/2017 S9 were derived assuming $G_r=0.18\pm0.28$ due to the limited amount of available photometric data for these objects.
    For these derivations, we emphasize the importance of including associated uncertainties for $G$ parameter and albedo values when they are assigned assumed values, as is often done in the absence of more object-specific information, in order to avoid misleadingly precise derived $H$ and $r_n$ values, which can affect the interpretation of downstream analyses like photometric activity searches.
    }
    \item{Using the same Monte Carlo approach to phase function fitting applied to new observations reported here, we also derive revised best-fit $V$-band absolute magnitudes and equivalent effective nucleus radii (including realistic uncertainties; assuming $V$-band albedos of $p_V=0.05\pm0.02$ in all cases, except for 176P, for which we use $p_V=0.06\pm0.02$) of
    $H_V=15.9\pm0.1$ and $r_n=(2.0\pm0.4)$~km for 133P/Elst-Pizarro,
    $H_V=15.42\pm0.03$ and $r_n=(2.3\pm0.4)$~km for 176P/LINEAR,
    $H_V=19.9\pm0.1$ and $r_n=(0.32\pm0.06)$~km for 259P/Garradd,
    $H_V=17.13\pm0.04$ for the combined 288P/(300163) 2006 VW$_{139}$ system and $r_n=(0.9\pm0.2)$~km and $r_n=(0.6\pm0.1)$~km for 288P's two primary components,
    $H_V=20.2\pm0.3$ and $r_n=(0.3\pm0.1)$~km for 358P/PANSTARRS, and
    $H_V=16.4\pm0.4$ and $r_n=(1.6\pm0.4)$~km for 433P/(248370) 2005 QN$_{173}$.
    }
    \item{We identify photometric evidence of activity in observations of 238P in October and November 2021. These results mark 238P's fourth consecutive active apparition during perihelion passages, where we also find a most likely activity onset date of 2021 October 18 (231 days prior to perihelion) at an approximate true anomaly of $\nu=296^{\circ}$, and a net initial mass loss rate over the period in question ($298.7^{\circ}<\nu<305.4^{\circ}$) of ${\dot M}=(0.3\pm0.1)$~kg~s$^{-1}$.  Analysis of previous observations of 238P using the new nucleus size derived in this work indicates that the object was actually active at the time of the 2010 observations used to measure its nucleus size in a previous work \citep{hsieh2011_238p}.  Within 3-$\sigma$ uncertainties, we find both 238P's estimated activity onset times and net initial mass loss rates to be comparable during perihelion approaches in 2010, 2016, and 2021.
    }
    \item{Observations of P/2016 J1-A, and P/2016 J1-B show both photometric and morphological evidence of activity in 2021 and 2022, representing the first confirmation that P/2016 J1-A and P/2016 J1-B are recurrently active, making the two fragments collectively the tenth MBC to be confirmed to be recurrently active and therefore likely to be exhibiting sublimation-driven activity.
    }
    \item{The nucleus of 313P is found to have broadband colors of $g'-r'=0.52\pm0.05$ and $r'-i'=0.22\pm0.07$, corresponding to a mean spectral slope over the $g'r'i'$ wavelength region of $S'_{gri}=(5.91\pm0.01)$~\%/100~nm.  This result is within 1-$\sigma$ of the mean visible slope of $(3.58\pm3.21)$\%/1000\AA\ found for members of the Lixiaohua \revision{asteroid} family, \revision{and in particular, is close to the mean visible slope of $(5.99\pm1.00)$\%/1000\AA\ found for T-type asteroids in the Lixiaohua family,} and so is consistent with 313P being a member of that family.
    }
    \item{We report the non-detection of P/2015 X6 (PANSTARRS) at its predicted ephemeris positions based on current orbit solutions in several observation attempts, where we conclude that the object's current nucleus size is below our detection limits (i.e., $r\lesssim0.1$~km and $r\lesssim0.3$~km for its 1-$\sigma$ and 3-$\sigma$ ephemeris uncertainty regions, respectively), and that the object may have even disintegrated following its 2015 apparition given our unsuccessful recovery attempts in 2020 even when it was expected to become active again. However, we cannot exclude the possibility that a combination of uncertainty in the astrometric measurements used to derive the object's orbit and possible non-gravitational perturbations from asymmetric mass loss since 2015 mean that the orbit solution for the nucleus is simply no longer accurate enough to recover the object.}
    \item{We find that of \revision{17} distinct MBCs or candidate MBCs for which nucleus sizes (or inferred parent body sizes in the cases of apparent split objects 288P and P/2016 J1-A/B) have been measured or estimated, more than half have $r_n<0.5$~km, $>$80\% have $r_n\lesssim1.0$~km, and all have $r_n\leq2.5$~km.  This finding points to what appears to be a strong physical preference toward small (i.e., $r_n<1$~km) MBCs, where one notable possibility is that that YORP spin-up may play a significant role in triggering and/or facilitating MBC activity.
    }
\end{enumerate}

\clearpage

\acknowledgments


\revision{We thank two anonymous reviewers for helpful comments that improved this work.}  HHH, MSPK, MMK, JP, SSS, AT, and CAT acknowledge support from the NASA Solar System Observations program (Grant 80NSSC19K0869), while the RJWs acknowledge support via a grant from the NASA Near-Earth Object Observations Program (Grant 80NSSC18K0971) to support operation of the Pan-STARRS telescopes.  The work of JP was conducted at the Jet Propulsion Laboratory, California Institute of Technology, under a contract with the National Aeronautics and Space Administration (80NM0018D0004).


We are grateful to staff at CFHT, Gemini-North, Gemini-South, LDT, Magellan, and Palomar for their assistance in obtaining observations.

This work is based on observations obtained at the Gemini Observatory, which is operated by the Association of Universities for Research in Astronomy, Inc., under a cooperative agreement with the NSF on behalf of the Gemini partnership: the National Science Foundation (United States), the National Research Council (Canada), CONICYT (Chile), Ministerio de Ciencia, Tecnolog\'{i}a e Innovaci\'{o}n Productiva (Argentina), and Minist\'{e}rio da Ci\^{e}ncia, Tecnologia e Inova\c{c}\~{a}o (Brazil).
We are particularly appreciative of Gemini Observatory's Large and Long Program (LLP) program, under which many of the observations presented here were obtained and which has been crucial for this work, as well as S.\ Margheim and S.\ Leggett for administering this program.

This work is also based on observations obtained at the Hale Telescope at Palomar Observatory as part of a continuing collaboration between the California Institute of Technology, NASA/JPL, Yale University, and the National Astronomical Observatories of China.

This work also made use of data obtained at the Lowell Discovery Telescope (LDT).  Lowell Observatory is a private, nonprofit institution dedicated to astrophysical research and public appreciation of astronomy and operates the LDT in partnership with Boston University, the University of Maryland, the University of Toledo, Northern Arizona University, and Yale University. Partial support of the LDT was provided by Discovery Communications. LMI was built by Lowell Observatory using funds from the National Science Foundation (NSF grant AST-1005313; PI: P.\ Massey).

Finally, this work made use of observations obtained with MegaPrime/MegaCam, a joint project of CFHT and CEA/DAPNIA, at the Canada-France-Hawaii Telescope (CFHT) which is operated by the National Research Council (NRC) of Canada, the Institut National des Science de l'Univers of the Centre National de la Recherche Scientifique (CNRS) of France, and the University of Hawaii. The observations at the Canada-France-Hawaii Telescope were performed with care and respect from the summit of Maunakea which is a significant cultural and historic site.

This research made use of
{\tt astropy}, a community-developed core {\tt python} package for astronomy;
{\tt ccdproc}, an {\tt astropy} package for image reduction;
{\tt sbpy}, an {\tt astropy} affiliated package for small-body planetary astronomy;
{\tt uncertainties} (version 3.0.2), a {\tt python} package for calculations with uncertainties by E.\ O.\ Lebigot (\url{http://pythonhosted.org/uncertainties/});
\revision{{\tt pyraf}, which is a product of the Space Telescope Science Institute, which is operated by AURA for NASA;}
and NASA's Astrophysics Data System Bibliographic Services.

%

\vspace{5mm}
\facilities{CFHT (MegaCam), Gemini:Gillett (GMOS-N), Gemini:South (GMOS-S), LDT (LMI), Hale (WASP), Magellan:Baade (IMACS)}


\software{{\tt astropy} \citep{astropy2018_astropy},
    {\tt astroquery} \citep{ginsburg2019_astroquery},
    {\tt ccdproc} \citep{craig2017_ccdproc},
    {\tt IRAF} \citep{tody1986_iraf,tody1993_iraf}, 
    {\tt L.A.Cosmic} \citep{vandokkum2001_lacosmic,vandokkum2012_lacosmic},
    \revision{{\tt numpy} \citep[][]{harris2020_numpy}},
    \revision{{\tt pyraf} \citep[v2.2.1;][]{stsci2012_pyraf}},
    {\tt RefCat2} \citep{tonry2018_refcat},
    {\tt sbpy} \citep{mommert2019_sbpy},
    \revision{{\tt scipy} \citep{virtanen2020_scipy}},
    {\tt uncertainties} (v3.0.2, E.\ O.\ Lebigot)
          }

\clearpage

\bibliography{main.bbl}{}
\bibliographystyle{aasjournal}

\clearpage

\appendix

\section{Composite Images of Target Objects\label{section:appendix_images}}
\setcounter{figure}{0}
\renewcommand{\thefigure}{A\arabic{figure}}

\begin{figure*}[htb!]
\centerline{\includegraphics[width=7.0in]{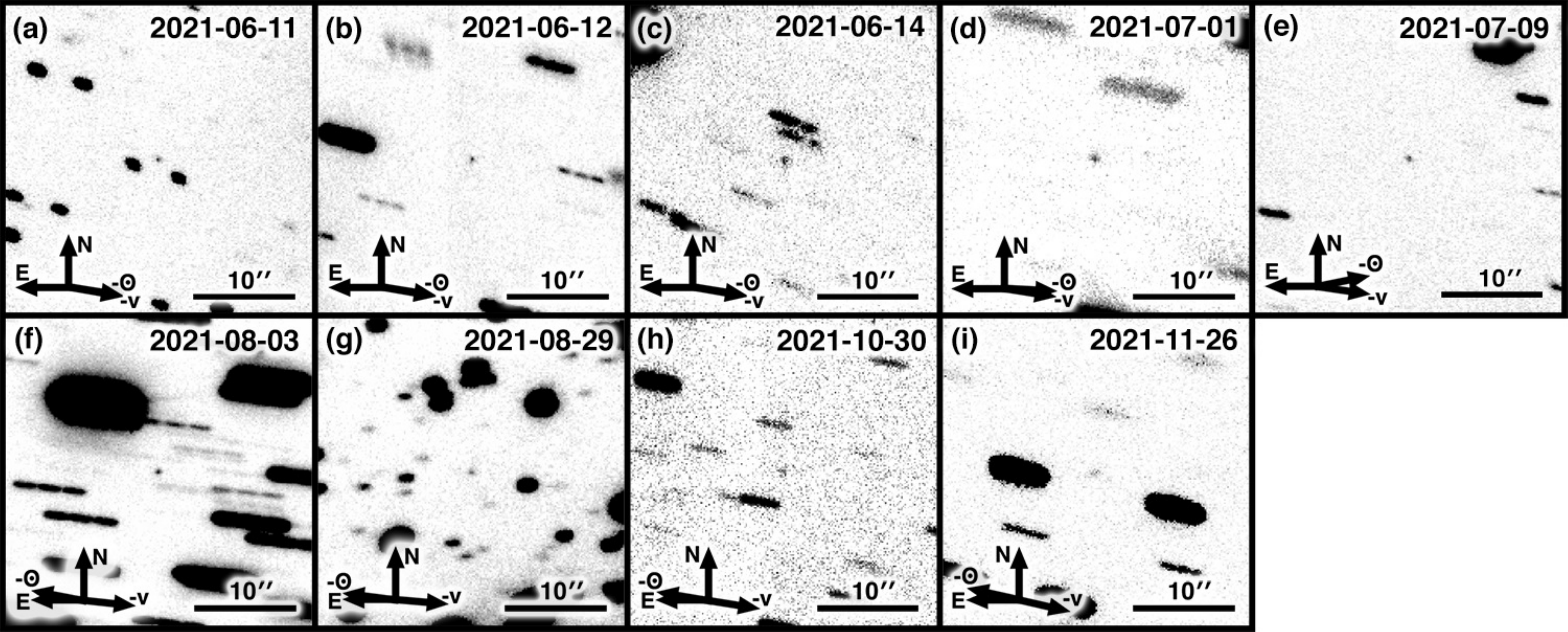}}
\caption{\small Composite images of 238P/Read constructed from data detailed in Table~\ref{table:observations}.  Scale bars indicate the size of each panel.  North (N), East (E), the antisolar direction ($-\odot$), and the negative heliocentric velocity direction ($-v$) are indicated in each panel.  The object is located at the center of each panel.\\
}
\phantomsection
\end{figure*}

\begin{figure*}[htb!]
\centerline{\includegraphics[width=7.0in]{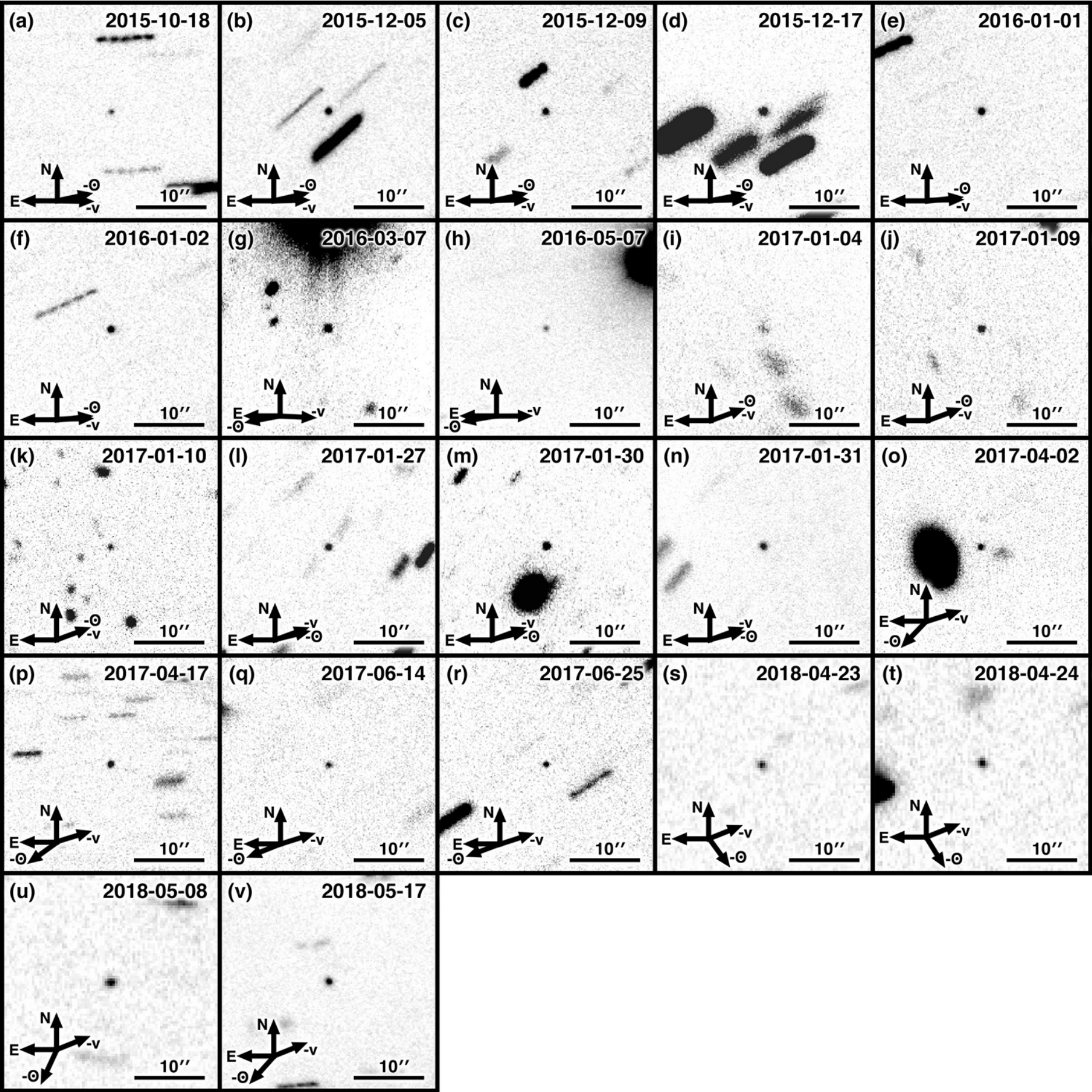}}
\caption{\small Composite images of 313P/Read constructed from data detailed in Table~\ref{table:observations}.  Scale bars indicate the size of each panel.  North (N), East (E), the antisolar direction ($-\odot$), and the negative heliocentric velocity direction ($-v$) are indicated in each panel.  The object is located at the center of each panel.\\
}
\phantomsection
\end{figure*}

\begin{figure*}[htb!]
\centerline{\includegraphics[height=1.4in]{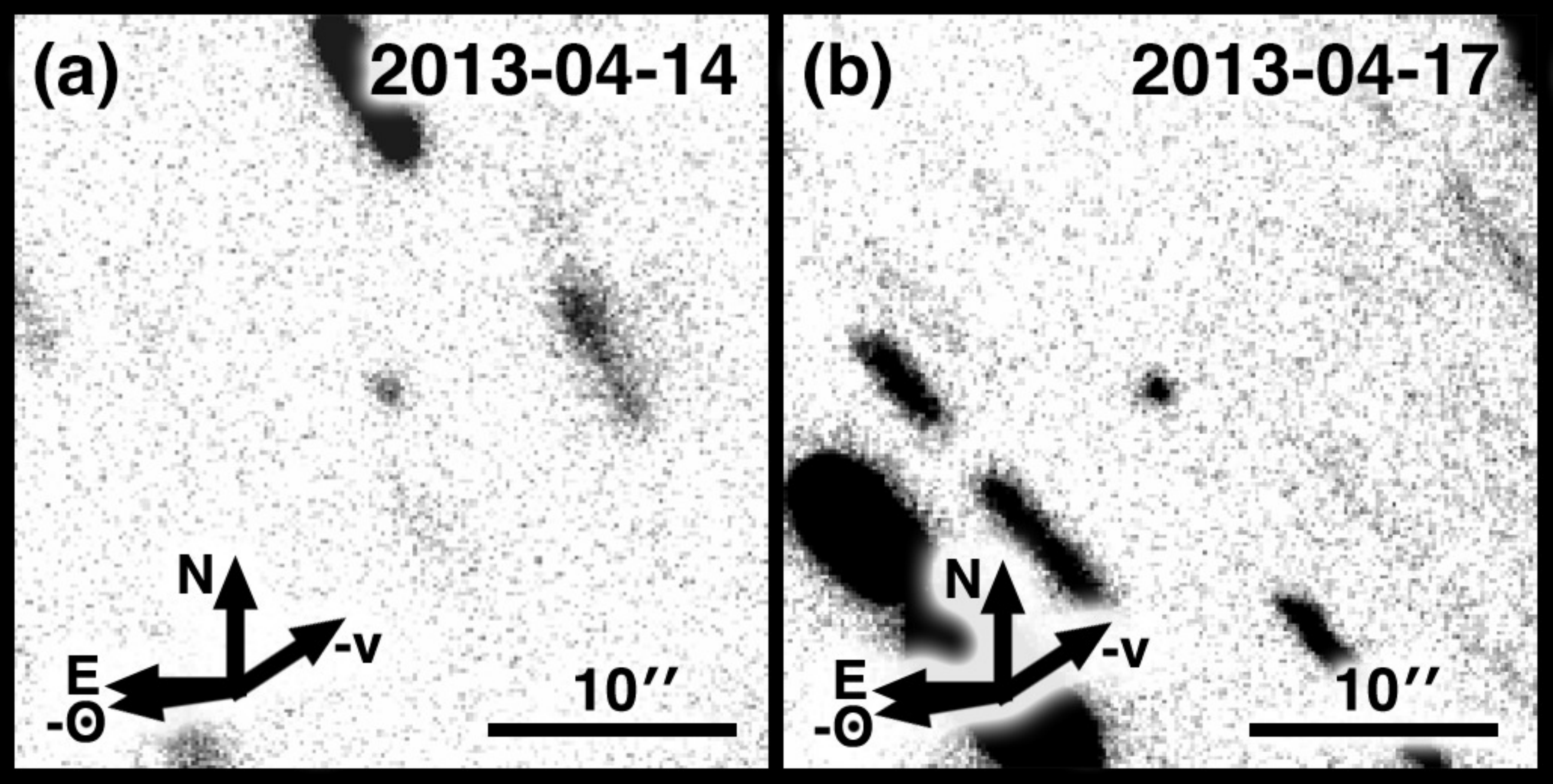}}
\caption{\small Composite images of 324P/La Sagra constructed from data detailed in Table~\ref{table:observations}.  Scale bars indicate the size of each panel.  North (N), East (E), the antisolar direction ($-\odot$), and the negative heliocentric velocity direction ($-v$) are indicated in each panel.  The object is located at the center of each panel.\\
}
\phantomsection
\end{figure*}

\begin{figure*}[htb!]
\centerline{\includegraphics[width=7.0in]{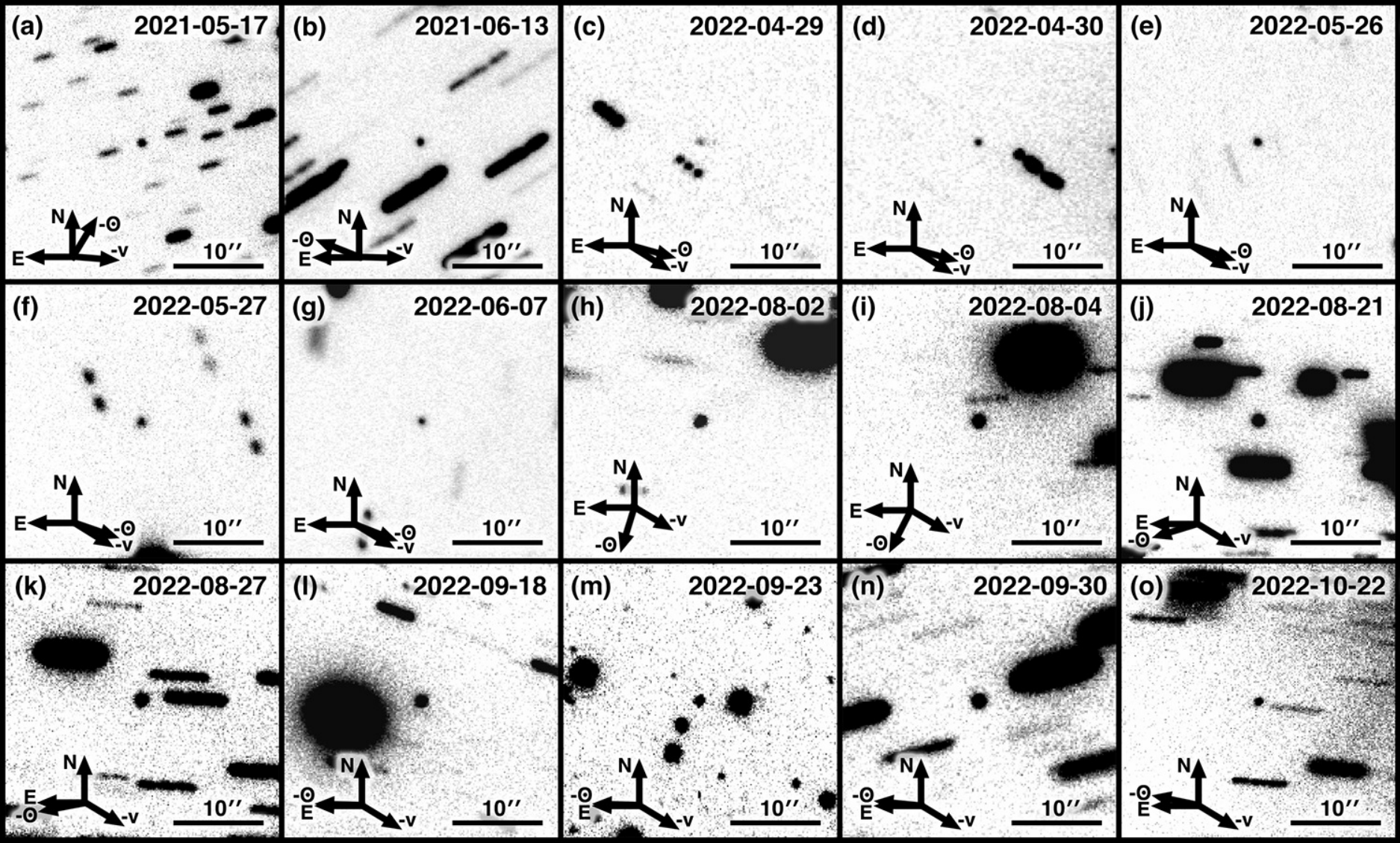}}
\caption{\small Composite images of 426P/PANSTARRS constructed from data detailed in Table~\ref{table:observations}.  Scale bars indicate the size of each panel.  North (N), East (E), the antisolar direction ($-\odot$), and the negative heliocentric velocity direction ($-v$) are indicated in each panel.  The object is located at the center of each panel.\\
}
\phantomsection
\end{figure*}

\begin{figure*}[htb!]
\centerline{\includegraphics[height=1.4in]{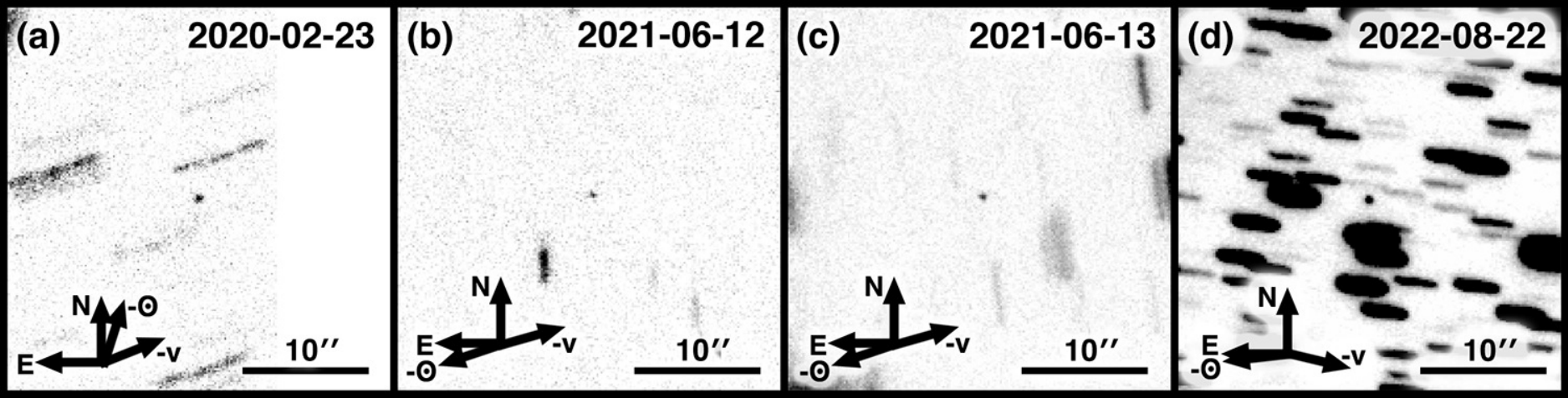}}
\caption{\small Composite images of 427P/ATLAS constructed from data detailed in Table~\ref{table:observations}.  Scale bars indicate the size of each panel.  North (N), East (E), the antisolar direction ($-\odot$), and the negative heliocentric velocity direction ($-v$) are indicated in each panel.  The object is located at the center of each panel.\\
}
\phantomsection
\end{figure*}

\begin{figure*}[htb!]
\centerline{\includegraphics[width=7.0in]{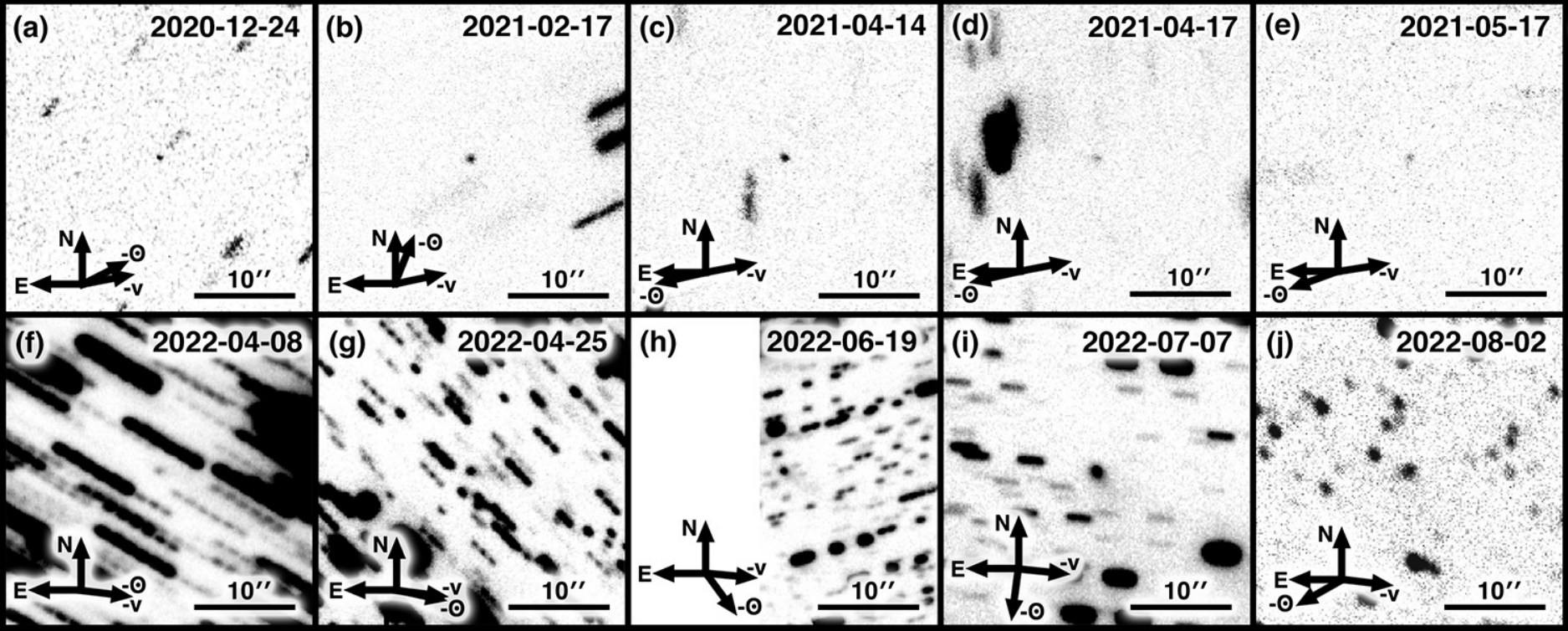}}
\caption{\small Composite images of P/2016 J1-A (PANSTARRS) constructed from data detailed in Table~\ref{table:observations}.  Scale bars indicate the size of each panel.  North (N), East (E), the antisolar direction ($-\odot$), and the negative heliocentric velocity direction ($-v$) are indicated in each panel.  The object is located at the center of each panel.\\
}
\phantomsection
\end{figure*}

\begin{figure*}[htb!]
\centerline{\includegraphics[height=1.4in]{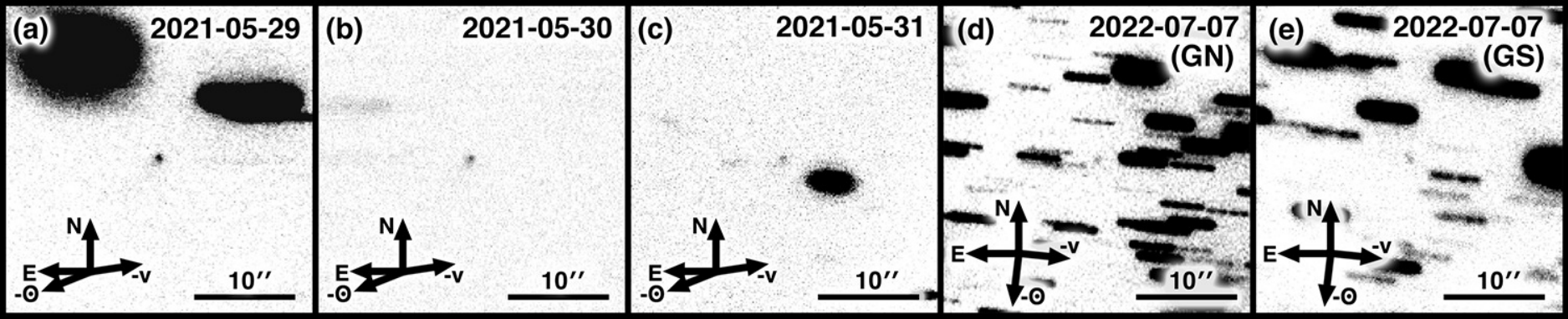}}
\caption{\small Composite images of P/2016 J1-B (PANSTARRS) constructed from data detailed in Table~\ref{table:observations}.  Scale bars indicate the size of each panel.  North (N), East (E), the antisolar direction ($-\odot$), and the negative heliocentric velocity direction ($-v$) are indicated in each panel.  The object is located at the center of each panel.\\
}
\phantomsection
\end{figure*}

\begin{figure*}[htb!]
\centerline{\includegraphics[height=1.4in]{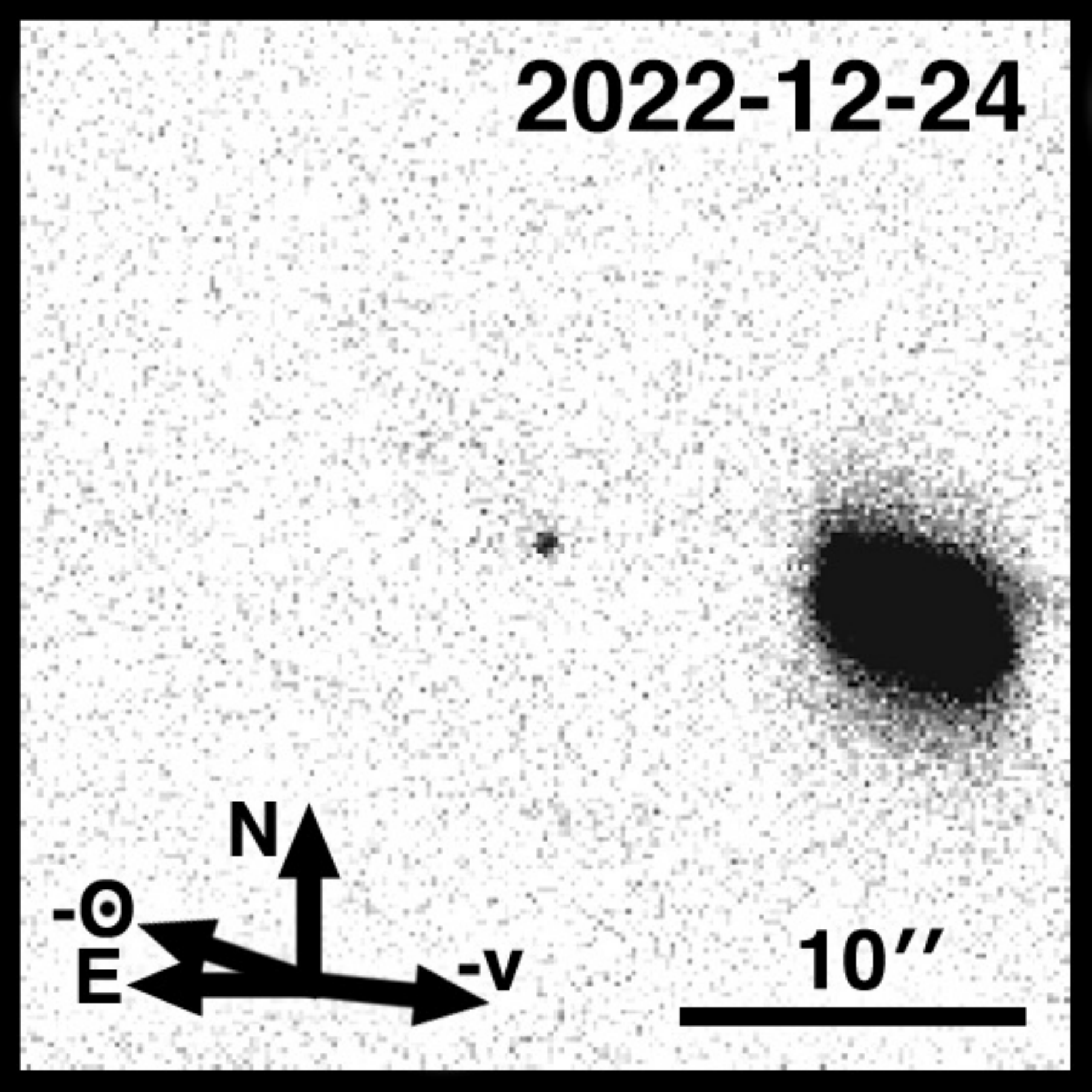}}
\caption{\small Composite images of P/2017 S9 (PANSTARRS) constructed from data detailed in Table~\ref{table:observations}.  Scale bars indicate the size of each panel.  North (N), East (E), the antisolar direction ($-\odot$), and the negative heliocentric velocity direction ($-v$) are indicated in each panel.  The object is located at the center of each panel.\\
}
\phantomsection
\end{figure*}

\begin{figure*}[htb!]
\centerline{\includegraphics[width=7.0in]{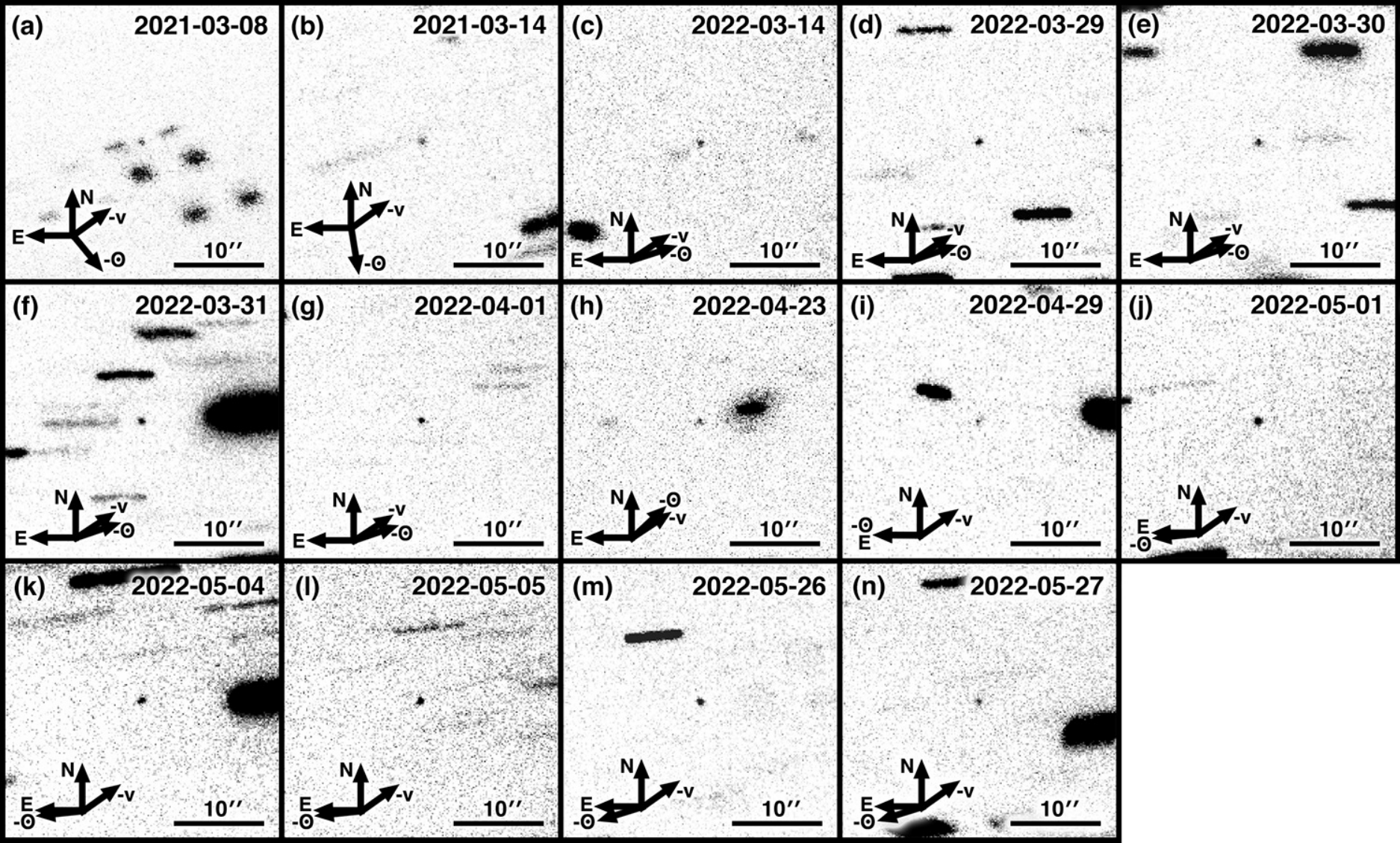}}
\caption{\small Composite images of P/2019 A3 (PANSTARRS) constructed from data detailed in Table~\ref{table:observations}.  Scale bars indicate the size of each panel.  North (N), East (E), the antisolar direction ($-\odot$), and the negative heliocentric velocity direction ($-v$) are indicated in each panel.  The object is located at the center of each panel.\\
}
\phantomsection
\end{figure*}

\clearpage

\section{Observations\label{section:appendix_obs}}
\setcounter{table}{0}
\renewcommand{\thefigure}{B\arabic{table}}

{\scriptsize{
\begin{verbatim}
================================================================================
Byte-by-byte Description of file: mbc_nucleus_observations_hsieh_psj2023.txt	
--------------------------------------------------------------------------------
   Bytes Format Units Label       Explanations					
--------------------------------------------------------------------------------
   1- 11 A11    ---   target      Target name					
  12- 23 A10    ---   utdate	     UT Date of observation (1)			
  26- 34 A9     ---   telescope   Telescope (2)					
  37- 38 I2     ---   n_exp       Number of exposures (3)			
  41- 44 I4     s     t_exp       Total exposure time (3)			
  47- 50 A4     ---   filter      Filter used for observation			
  53- 57 F5.1   deg   nu          True anomaly at time of observation		
  60- 64 F5.3   au    r_h         Heliocentric distance at time of observation	
  67- 71 F5.3   au    delta       Geocentric distance at time of observation	
  74- 77 F4.1   deg   alpha       Solar phase angle				
  80- 84 F5.2   mag   app_m       Apparent magnitude in specified filter	
  87- 90 F4.2   mag   app_m_e     1-sigma uncertainty of apparent magnitude	
  93- 97 F5.2   mag   app_m_r     Equivalent apparent r'-band magnitude for	
		                                   non-r'-band observations, assuming solar	
				                                   colors (using Holmberg et al. 2006) (3)	
 100-103 F4.2   mag   app_m_r_e   1-sigma uncertainty of equivalent apparent	
                                   r'-band magnitude (3)			
 106-110 F5.2   mag   red_m       Reduced r'-band magnitude, normalized to	
                                   r_h = 1 au and delta = 1 au (3)		
 113-116 F4.2   mag   red_m_e     1-sigma uncertainty of reduced r'-band	
                                   magnitude (3)				
 119-123 F5.2   mag   abs_m       Absolute r'-band magnitude, normalized to	
                                   r_h = 1 au, delta = 1 au, and alpha = 1 deg,	
                                   computed using best-fit phase function	
                                   parameters listed in Table 3 (3)		
 126-129 F4.2   mag   abs_m_e     1-sigma uncertainty of absolute r'-band	
                                   magnitude (3)				
 132     I1     ---   phsfn       Used to compute reported phase function	
                                   parameters (4)				
 135-136 I2     ---   reference   Reference for previously reported data (3,5)	
--------------------------------------------------------------------------------
Note (1): In YYYY-MM-DD format                                                  
								                
Note (2):							                
   Blanco =    4.0 m Victor M. Blanco Telescope;		                
   CFHT =      3.6 m Canada-France-Hawaii Telescope;		                
   du Pont =   2.5 m Irenee du Pont telescope;			                
   Gemini-N =  8.1 m Gemini North telescope;			                
   Gemini-S =  8.1 m Gemini South telescope;			                
   INT =       2.5 m Isaac Newton Telescope;			                
   KeckI =     10 m Keck I Observatory;				                
   LDT =       4.3 m Lowell Discovery Telescope;		                
   Magellan =  6.5 m Magellan Baade Telescope;			                
   NTT =       3.54 m New Technology Telescope;			                
   Palomar =   5.1 m Palomar Hale Telescope;			                
   PS1 =       1.8 m Pan-STARRS1 telescope;			                
   SkyMapper = 1.35 m SkyMapper telescope;			                
   SOAR =      4.2 m Southern Astrophysical Research telescope;	                
   Subaru =    8.2 m Subaru Telescope;				                
   UH2.2 =     University of Hawaii 2.2 m telescope;		                
   VLT =       8.2 m Very Large Telescope;			                
   WHT =       4.2 m William Herschel Telescope.		                
								                
Note (3):							                
   -1 = not applicable						                
								                
Note (4):							                
    0 = no;							                
    1 = yes.							                
								                
Note (5):                                                                       
    1 = Hsieh, H. H., et al. 2010, MNRAS, 403, 363-377;                         
    2 = Hsieh, H. H., et al. 2011, AJ, 142, 29;			                
    3 = Hsieh, H. H., et al. 2014, AJ, 147, 89;			                
    4 = Hsieh, H. H., et al. 2011, ApJ Letters, 736, L18;	                
    5 = Hsieh, H. H., et al. 2018, AJ, 156, 223;		                
    6 = MacLennan, E. M., Hsieh, H. H. 2012, ApJ Letters, 758, L3;              
    7 = Hsieh, H. H., et al. 2021, PSJ, 2, 62;			                
    8 = Hsieh, H. H., et al. 2015, ApJ Letters, 800, L16;	                
    9 = Hsieh, H. H. 2014, Icarus, 243, 16-26;			                
   10 = Hsieh, H. H., et al. 2018, AJ, 156, 39;			                
   11 = Hsieh, H. H., et al. 2021, ApJ Letters, 922, L9.                        
--------------------------------------------------------------------------------
133P         2003-09-22  KeckI      -1    -1  R     133.7  3.457  3.195  16.8  21.71  0.06  21.89  0.06  16.67  0.06  15.58  0.11  0   1
133P         2003-12-13  UH2.2      -1    -1  R     144.2  3.544  2.563   1.5  20.45  0.04  20.63  0.04  15.84  0.04  15.60  0.04  1   1
133P         2003-12-15  UH2.2      -1    -1  R     144.5  3.546  2.568   2.1  20.70  0.06  20.88  0.06  16.08  0.06  15.79  0.06  1   1
133P         2004-02-16  KeckI      -1    -1  R     153.0  3.599  3.235  15.5  21.69  0.06  21.87  0.06  16.54  0.06  15.50  0.11  1   1
133P         2004-10-10  KeckI      -1    -1  R     183.8  3.675  3.958  14.4  22.15  0.04  22.33  0.04  16.52  0.04  15.52  0.10  1   1
133P         2005-01-16  UH2.2      -1    -1  R     196.4  3.647  2.697   4.7  20.99  0.05  21.17  0.05  16.21  0.05  15.72  0.06  1   1
133P         2005-04-10  UH2.2      -1    -1  R     207.4  3.597  3.217  15.7  21.86  0.12  22.04  0.12  16.72  0.12  15.67  0.15  1   1
133P         2005-05-27  UH2.2      -1    -1  R     213.8  3.558  3.854  15.0  22.03  0.20  22.21  0.20  16.52  0.20  15.50  0.22  1   1
133P         2005-05-28  UH2.2      -1    -1  R     213.9  3.557  3.866  15.0  22.08  0.21  22.26  0.21  16.57  0.21  15.55  0.23  1   1
133P         2005-12-27  UH2.2      -1    -1  R     245.3  3.299  3.312  17.1  21.70  0.06  21.88  0.06  16.69  0.06  15.58  0.12  1   1
133P         2006-04-23  UH2.2      -1    -1  R     265.1  3.118  2.164   7.1  20.09  0.04  20.27  0.04  16.12  0.04  15.49  0.07  1   1
133P         2006-05-22  UH2.2      -1    -1  R     270.3  3.071  2.360  15.4  20.84  0.05  21.02  0.05  16.72  0.05  15.68  0.10  1   1
133P         2006-05-23  UH2.2      -1    -1  R     270.5  3.069  2.369  15.6  20.73  0.05  20.91  0.05  16.60  0.05  15.56  0.10  1   1
133P         2006-05-25  UH2.2      -1    -1  R     270.9  3.066  2.388  16.0  21.04  0.11  21.22  0.11  16.90  0.11  15.83  0.14  1   1
133P         2007-03-21  UH2.2      -1    -1  R     335.5  2.676  2.810  20.7  21.15  0.11  21.33  0.11  16.95  0.11  15.69  0.16  0   1
176P         2006-02-03  UH2.2      -1    -1  R      27.7  2.630  2.707  21.2  20.25  0.01  20.43  0.01  16.17  0.01  15.14  0.05  0   2
176P         2006-08-31  UH2.2      -1    -1  R      75.2  2.933  3.546  14.3  21.23  0.08  21.41  0.08  16.32  0.08  15.53  0.09  0   2
176P         2006-09-02  UH2.2      -1    -1  R      75.6  2.937  3.530  14.6  21.09  0.05  21.27  0.05  16.19  0.05  15.38  0.07  0   2
176P         2006-12-11  UH2.2      -1    -1  R      94.5  3.124  2.424  14.5  20.01  0.01  20.19  0.01  15.79  0.01  14.99  0.04  0   2
176P         2006-12-16  UH2.2      -1    -1  R      95.3  3.133  2.378  13.3  20.13  0.01  20.31  0.01  15.95  0.01  15.19  0.04  0   2
176P         2006-12-18  UH2.2      -1    -1  R      95.7  3.137  2.361  12.8  20.09  0.09  20.27  0.09  15.92  0.09  15.18  0.10  0   2
176P         2007-01-27  KeckI      -1    -1  R     102.6  3.211  2.227   0.8  19.50  0.01  19.68  0.01  15.41  0.01  15.28  0.01  0   2
176P         2007-02-15  UH2.2      -1    -1  R     105.7  3.246  2.326   7.5  19.87  0.01  20.05  0.01  15.66  0.01  15.13  0.03  0   2
176P         2007-02-16  UH2.2      -1    -1  R     105.9  3.248  2.334   7.8  19.93  0.01  20.11  0.01  15.71  0.01  15.16  0.03  0   2
176P         2007-03-21  UH2.2      -1    -1  R     111.2  3.307  2.718  15.4  20.71  0.01  20.89  0.01  16.12  0.01  15.29  0.04  0   2
176P         2007-03-22  UH2.2      -1    -1  R     111.3  3.309  2.732  15.5  20.82  0.01  21.00  0.01  16.22  0.01  15.38  0.04  0   2
176P         2007-05-19  UH2.2      -1    -1  R     120.2  3.407  3.637  16.1  21.57  0.05  21.75  0.05  16.28  0.05  15.42  0.07  1   2
176P         2008-06-29  NTT        -1    -1  R     173.2  3.803  3.795  15.4  21.68  0.07  21.86  0.07  16.06  0.07  15.23  0.08  1   2
176P         2008-06-30  NTT        -1    -1  R     173.3  3.804  3.810  15.3  21.70  0.05  21.88  0.05  16.07  0.05  15.24  0.07  1   2
176P         2008-07-01  NTT        -1    -1  R     173.4  3.804  3.824  15.3  21.63  0.05  21.81  0.05  16.00  0.05  15.16  0.07  1   2
176P         2009-01-23  WHT        -1    -1  R     198.1  3.765  4.012  14.1  21.47  0.10  21.65  0.10  15.76  0.10  14.96  0.11  1   2
176P         2009-05-03  INT        -1    -1  R     210.6  3.687  2.702   3.9  20.30  0.04  20.48  0.04  15.49  0.04  15.13  0.04  1   2
176P         2010-08-05  NTT        -1    -1  R     281.9  2.956  1.958   4.3  19.45  0.03  19.63  0.03  15.82  0.03  15.44  0.04  1   3
176P         2010-08-06  NTT        -1    -1  R     282.1  2.954  1.959   4.7  19.29  0.04  19.47  0.04  15.66  0.04  15.26  0.04  1   3
176P         2010-08-11  UH2.2      -1    -1  R     283.0  2.945  1.970   6.6  19.21  0.02  19.39  0.02  15.57  0.02  15.08  0.03  1   3
176P         2010-08-13  NTT        -1    -1  R     283.5  2.942  1.976   7.3  19.55  0.03  19.73  0.03  15.91  0.03  15.38  0.04  1   3
176P         2010-08-13  UH2.2      -1    -1  R     283.5  2.941  1.977   7.4  19.45  0.02  19.63  0.02  15.81  0.02  15.28  0.03  1   3
176P         2010-08-14  NTT        -1    -1  R     283.7  2.939  1.980   7.7  19.45  0.03  19.63  0.03  15.81  0.03  15.26  0.04  1   3
176P         2010-08-15  NTT        -1    -1  R     283.9  2.938  1.984   8.1  19.44  0.03  19.62  0.03  15.79  0.03  15.23  0.04  1   3
176P         2010-08-16  UH2.2      -1    -1  R     284.1  2.936  1.988   8.5  19.49  0.02  19.67  0.02  15.84  0.02  15.26  0.04  1   3
176P         2010-08-28  du Pont    -1    -1  R     286.5  2.914  2.051  12.3  19.79  0.03  19.97  0.03  16.09  0.03  15.36  0.05  1   3
176P         2010-08-29  du Pont    -1    -1  R     286.7  2.912  2.058  12.6  19.75  0.03  19.93  0.03  16.04  0.03  15.30  0.05  1   3
176P         2010-08-30  du Pont    -1    -1  R     286.9  2.911  2.065  12.9  19.54  0.02  19.72  0.02  15.83  0.02  15.08  0.04  1   3
176P         2010-08-31  du Pont    -1    -1  R     287.1  2.909  2.072  13.2  19.61  0.02  19.79  0.02  15.89  0.02  15.13  0.04  1   3
176P         2010-09-01  UH2.2      -1    -1  R     287.4  2.906  2.081  13.6  19.96  0.02  20.14  0.02  16.23  0.02  15.46  0.04  1   3
176P         2010-09-04  NTT        -1    -1  R     287.9  2.901  2.103  14.3  19.77  0.04  19.95  0.04  16.02  0.04  15.22  0.06  1   3
176P         2010-09-05  NTT        -1    -1  R     288.1  2.900  2.111  14.6  20.20  0.03  20.38  0.03  16.45  0.03  15.64  0.05  1   3
176P         2010-10-05  KeckI      -1    -1  R     294.5  2.846  2.415  19.8  20.48  0.04  20.66  0.04  16.47  0.04  15.49  0.06  0   3
176P         2010-10-23  VLT        -1    -1  R     298.3  2.816  2.624  20.7  20.81  0.16  20.99  0.16  16.65  0.16  15.64  0.17  0   3
176P         2011-06-06  Subaru     -1    -1  R     353.6  2.579  3.213  15.8  20.27  0.02  20.45  0.02  15.86  0.02  15.01  0.05  0   3
176P         2011-07-01  VLT        -1    -1  R       0.0  2.576  2.976  19.4  20.64  0.06  20.82  0.06  16.40  0.06  15.43  0.08  0   3
176P         2011-08-02  Gemini-N   -1    -1  r'      8.5  2.581  2.617  22.5  20.01  0.02  20.19  0.02  16.04  0.02  14.98  0.06  0   3
176P         2011-08-04  UH2.2      -1    -1  R       9.0  2.582  2.594  22.6  19.99  0.02  20.17  0.02  16.04  0.02  14.97  0.06  0   3
176P         2011-08-26  KeckI      -1    -1  R      14.7  2.590  2.330  22.9  19.92  0.02  20.10  0.02  16.20  0.02  15.12  0.06  0   3
176P         2011-08-28  Gemini-N   -1    -1  r'     15.3  2.591  2.306  22.9  20.07  0.02  20.25  0.02  16.37  0.02  15.29  0.06  0   3
176P         2011-08-29  Gemini-N   -1    -1  r'     15.5  2.592  2.294  22.8  20.15  0.02  20.33  0.02  16.46  0.02  15.39  0.06  0   3
176P         2011-09-25  Gemini-N   -1    -1  r'     22.4  2.608  1.987  19.9  19.94  0.02  20.12  0.02  16.55  0.02  15.56  0.05  0   3
176P         2011-10-30  UH2.2      -1    -1  R      31.3  2.639  1.711   9.6  18.82  0.02  19.00  0.02  15.73  0.02  15.10  0.04  0   3
176P         2011-12-01  PS1        -1    -1  r_P1   39.2  2.674  1.701   4.4  18.86  0.05  19.04  0.05  15.75  0.05  15.37  0.05  0   3
176P         2011-12-22  NTT        -1    -1  R      44.1  2.700  1.842  12.4  19.29  0.04  19.47  0.04  15.99  0.04  15.26  0.05  0   3
176P         2011-12-31  Gemini-N   -1    -1  r'     46.4  2.712  1.936  15.1  19.59  0.02  19.77  0.02  16.17  0.02  15.34  0.05  0   3
176P         2012-11-13  UH2.2      -1    -1  R     108.8  3.278  3.239  17.5  21.25  0.10  21.43  0.10  16.30  0.10  15.39  0.11  0   3
176P         2012-12-18  UH2.2      -1    -1  R     114.4  3.340  2.797  15.4  20.60  0.06  20.78  0.06  15.93  0.06  15.09  0.07  0   3
176P         2013-05-12  UH2.2      -1    -1  R     135.4  3.564  3.436  16.5  21.23  0.08  21.41  0.08  15.97  0.08  15.10  0.09  0   3
176P         2013-05-13  UH2.2      -1    -1  R     135.5  3.565  3.452  16.5  21.19  0.08  21.37  0.08  15.92  0.08  15.05  0.09  0   3
238P         2010-07-07  UH2.2      -1    -1  R     291.8  2.704  1.821  13.0  23.61  0.10  23.79  0.10  20.33  0.10  19.83  0.20  0   4
238P         2010-07-20  UH2.2      -1    -1  R     294.8  2.674  1.709   8.5  22.85  0.06  23.03  0.06  19.73  0.06  19.36  0.15  0   4
238P         2010-08-15  SOAR       -1    -1  R     301.1  2.616  1.608   2.6  22.34  0.05  22.52  0.05  19.40  0.05  19.22  0.09  0   4
238P         2010-09-03  UH2.2      -1    -1  R     305.9  2.576  1.643  10.7  22.0   0.4   22.18  0.40  19.05  0.40  18.61  0.43  0   4
238P         2010-09-04  NTT        -1    -1  R     306.1  2.574  1.647  11.0  22.3   0.2   22.48  0.20  19.35  0.20  18.90  0.26  0   4
238P         2010-09-05  NTT        -1    -1  R     306.4  2.572  1.651  11.4  22.3   0.2   22.48  0.20  19.34  0.20  18.89  0.26  0   4
238P         2016-07-08  CFHT       -1    -1  r'    328.5  2.439  2.095  24.4     -1    -1  22.48  0.10  18.94  0.10  18.18  0.26  0   5
238P         2016-08-06  Gemini-N   -1    -1  r'    336.9  2.405  1.742  21.6     -1    -1  21.68  0.10  18.57  0.10  17.87  0.24  0   5
238P         2016-09-05  Gemini-N   -1    -1  r'    345.8  2.381  1.467  13.1     -1    -1  20.48  0.10  17.77  0.10  17.27  0.20  0   5
238P         2016-09-06  CFHT       -1    -1  r'    346.1  2.380  1.461  12.7     -1    -1  20.48  0.10  17.78  0.10  17.29  0.20  0   5
238P         2021-06-11  Gemini-S    2   600  r'    268.3  2.987  2.099  11.3  24.85  0.10     -1    -1  20.86  0.10  20.41  0.19  1  -1
238P         2021-06-12  Gemini-S    3   900  r'    268.5  2.985  2.089  11.0  24.89  0.08     -1    -1  20.92  0.08  20.47  0.18  1  -1
238P         2021-06-14  Gemini-S    2   600  r'    268.9  2.980  2.069  10.4  24.62  0.12     -1    -1  20.67  0.12  20.24  0.20  1  -1
238P         2021-07-01  Gemini-S    2   600  r'    272.2  2.937  1.938   4.6  24.50  0.15     -1    -1  20.72  0.15  20.47  0.18  1  -1
238P         2021-07-09  Gemini-S    4   600  r'    273.8  2.917  1.902   1.6  24.32  0.07     -1    -1  20.60  0.07  20.47  0.09  1  -1
238P         2021-08-03  Gemini-S    3   900  r'    278.9  2.854  1.900   8.5  24.17  0.05     -1    -1  20.50  0.05  20.12  0.15  1  -1
238P         2021-08-29  Gemini-S    3   900  r'    284.4  2.790  2.053  16.6  24.99  0.16     -1    -1  21.20  0.16  20.62  0.25  1  -1
238P         2021-10-30  Gemini-S    1   300  r'    298.7  2.646  2.691  21.4  24.64  0.31     -1    -1  20.38  0.31  19.68  0.38  0  -1
238P         2021-11-26  Gemini-S    3   900  r'    305.4  2.588  2.960  19.0  23.98  0.15     -1    -1  19.56  0.15  18.92  0.26  0  -1
259P         2011-02-28  Gemini-N   -1    -1  r'    194.5  3.600  2.709   8.0  25.4   0.1   25.58  0.10  20.64  0.10  19.83  0.11  1   6
259P         2011-03-11  Gemini-N   -1    -1  r'    195.8  3.589  2.770  10.2  25.4   0.1   25.58  0.10  20.60  0.10  19.65  0.12  1   6
259P         2011-03-26  Gemini-N   -1    -1  r'    197.6  3.572  2.897  13.1  25.7   0.2   25.88  0.20  20.81  0.20  19.69  0.22  1   6
259P         2011-03-31  Gemini-N   -1    -1  r'    198.2  3.567  2.949  13.9  25.6   0.1   25.78  0.10  20.67  0.10  19.51  0.13  1   6
259P         2011-04-01  Gemini-N   -1    -1  r'    198.3  3.565  2.959  14.0  26.0   0.1   26.18  0.10  21.07  0.10  19.90  0.13  1   6
259P         2012-01-21  Gemini-N   -1    -1  r'    240.7  2.894  2.729  19.9  25.4   0.1   25.58  0.10  21.10  0.10  19.62  0.16  1   6
259P         2012-01-31  Gemini-N   -1    -1  r'    242.5  2.861  2.557  20.0  25.5   0.1   25.68  0.10  21.36  0.10  19.88  0.16  1   6
259P         2012-02-01  Gemini-N   -1    -1  r'    242.7  2.857  2.540  19.9  25.6   0.1   25.78  0.10  21.48  0.10  20.01  0.16  1   6
259P         2012-04-15  SOAR       -1    -1  R     258.0  2.595  1.604   4.4  23.35  0.05  23.53  0.05  20.44  0.05  19.89  0.06  1   6
259P         2012-04-15  Gemini-N   -1    -1  r'    258.0  2.595  1.604   4.4  23.2   0.1   23.38  0.10  20.29  0.10  19.74  0.11  1   6
259P         2012-05-13  Gemini-N   -1    -1  r'    264.6  2.491  1.573  12.3  23.6   0.1   23.78  0.10  20.82  0.10  19.75  0.13  1   6
259P         2012-05-15  Gemini-N   -1    -1  r'    265.1  2.483  1.578  13.1  23.5   0.1   23.68  0.10  20.72  0.10  19.60  0.13  1   6
259P         2013-08-16  Gemini-N   -1    -1  r'     80.7  2.285  2.101  26.3  24.6   0.1   24.78  0.10  21.38  0.10  19.59  0.19  0   7
288P         2012-10-18  Gemini-N   -1    -1  r'    108.0  3.118  3.226  18.0  22.42  0.05  22.60  0.05  17.59  0.05  16.77  0.10  0   5
288P         2012-11-09  Gemini-N   -1    -1  r'    111.7  3.160  2.953  18.2  22.90  0.05  23.08  0.05  18.23  0.05  17.40  0.10  0   5
288P         2012-11-10  Gemini-N   -1    -1  r'    111.9  3.162  2.940  18.2  22.73  0.05  22.91  0.05  18.07  0.05  17.24  0.10  0   5
288P         2012-11-13  Gemini-N   -1    -1  r'    112.4  3.168  2.902  18.1  22.66  0.05  22.84  0.05  18.03  0.05  17.20  0.10  0   5
288P         2012-11-14  Gemini-N   -1    -1  r'    112.6  3.170  2.890  18.0  22.85  0.05  23.03  0.05  18.22  0.05  17.40  0.10  0   5
288P         2012-11-15  Gemini-N   -1    -1  r'    112.7  3.172  2.877  18.0  22.72  0.05  22.90  0.05  18.10  0.05  17.28  0.10  0   5
288P         2012-11-20  Gemini-N   -1    -1  r'    113.6  3.181  2.815  17.6  22.80  0.05  22.98  0.05  18.22  0.05  17.41  0.10  0   5
288P         2012-12-05  Gemini-N   -1    -1  r'    116.1  3.208  2.637  15.9  22.51  0.05  22.69  0.05  18.06  0.05  17.29  0.10  0   5
288P         2012-12-14  Gemini-N   -1    -1  r'    117.5  3.225  2.541  14.2  21.77  0.05  21.95  0.05  17.39  0.05  16.68  0.09  0   5
288P         2012-12-18  UH2.2      -1    -1  R     118.2  3.232  2.503  13.4  22.21  0.05  22.39  0.05  17.85  0.05  17.17  0.09  0   5
288P         2012-12-19  UH2.2      -1    -1  R     118.3  3.234  2.494  13.1  21.85  0.05  22.03  0.05  17.50  0.05  16.83  0.09  0   5
288P         2013-01-04  CFHT       -1    -1  r'    120.9  3.262  2.374   8.7  21.66  0.05  21.84  0.05  17.40  0.05  16.88  0.08  0   5
288P         2013-01-16  CFHT       -1    -1  r'    122.8  3.282  2.325   4.7  21.48  0.05  21.66  0.05  17.25  0.05  16.89  0.06  0   5
288P         2013-01-17  CFHT       -1    -1  r'    123.0  3.284  2.323   4.4  21.64  0.05  21.82  0.05  17.41  0.05  17.07  0.06  0   5
288P         2013-01-18  CFHT       -1    -1  r'    123.1  3.286  2.321   4.0  21.55  0.05  21.73  0.05  17.32  0.05  17.00  0.06  0   5
288P         2013-05-12  UH2.2      -1    -1  R     140.0  3.457  3.551  16.5  23.2   0.1   23.38  0.10  17.94  0.10  17.16  0.13  1   5
288P         2013-05-13  UH2.2      -1    -1  R     140.2  3.458  3.566  16.4  23.0   0.1   23.18  0.10  17.73  0.10  16.95  0.13  1   5
288P         2015-04-24  CFHT       -1    -1  r'    235.0  3.307  2.438  10.3  21.61  0.05  21.79  0.05  17.26  0.05  16.68  0.08  1   5
288P         2015-05-26  CFHT       -1    -1  r'    240.0  3.252  2.239   0.5  21.25  0.05  21.43  0.05  17.12  0.05  17.03  0.05  1   5
288P         2015-05-27  CFHT       -1    -1  r'    240.1  3.250  2.237   0.7  21.13  0.05  21.31  0.05  17.01  0.05  16.89  0.05  1   5
313P         2004-09-16  Subaru      2   120  VR    108.7  3.208  3.742  14.1     -1    -1  23.48  0.20  18.09  0.20  17.20  0.23  0   8
313P         2015-10-18  CFHT        5   900  r'    101.0  3.113  3.048  18.6  23.49  0.13     -1    -1  18.60  0.13  17.55  0.19  0  -1
313P         2015-12-05  CFHT       10  1800  r'    109.1  3.225  2.506  13.6  22.94  0.03     -1    -1  18.40  0.03  17.53  0.11  0  -1
313P         2015-12-09  Magellan    3   700  r'    109.7  3.234  2.472  12.7  23.00  0.04     -1    -1  18.49  0.04  17.65  0.11  0  -1 
313P         2015-12-17  CFHT        5   900  r'    111.0  3.252  2.412  10.5  22.59  0.03     -1    -1  18.12  0.03  17.38  0.10  0  -1
313P         2016-01-01  CFHT        5   900  r'    113.4  3.286  2.342   5.7  22.38  0.03     -1    -1  17.95  0.03  17.45  0.07  0  -1
313P         2016-01-02  CFHT        5   900  r'    113.6  3.288  2.339   5.3  22.47  0.03     -1    -1  18.04  0.03  17.56  0.06  0  -1
313P         2016-03-07  Gemini-N    5   900  r'    123.3  3.426  2.792  14.2  23.23  0.03     -1    -1  18.33  0.03  17.43  0.12  0  -1
313P         2016-05-07  Gemini-N    5   900  r'    131.8  3.541  3.768  15.5  24.17  0.12     -1    -1  18.54  0.12  17.60  0.17  0  -1
313P         2017-01-04  Gemini-N    4   900  r'    161.5  3.855  3.438  14.1  23.73  0.06     -1    -1  18.12  0.06  17.23  0.13  1  -1
313P         2017-01-09  Gemini-N    4   900  r'    162.1  3.858  3.370  13.6  24.18  0.06     -1    -1  18.61  0.06  17.74  0.12  1  -1
313P         2017-01-10  Gemini-N    1   225  r'    162.2  3.859  3.357  13.5  24.13  0.15     -1    -1  18.57  0.15  17.70  0.19  1  -1
313P         2017-01-27  Gemini-N    3   675  r'    164.1  3.871  3.152  11.1  24.00  0.05     -1    -1  18.57  0.05  17.80  0.11  1  -1
313P         2017-01-30  Gemini-N    2   450  r'    164.4  3.873  3.121  10.5  23.66  0.04     -1    -1  18.25  0.04  17.51  0.10  1  -1
313P         2017-01-31  Gemini-N    4   900  r'    164.6  3.874  3.111  10.3  23.60  0.04     -1    -1  18.19  0.04  17.47  0.10  1  -1
313P         2017-04-02  Gemini-N    1   300  r'    171.4  3.904  3.010   7.5  23.35  0.04     -1    -1  18.00  0.04  17.40  0.08  1  -1
313P         2017-04-17  Gemini-N    3   900  r'    173.1  3.909  3.143  10.6  23.95  0.05     -1    -1  18.50  0.09  17.76  0.10  1  -1
313P         2017-06-14  Gemini-N    3   900  r'    179.5  3.918  3.935  14.9  24.43  0.11     -1    -1  18.49  0.11  17.57  0.16  1  -1
313P         2017-06-25  Gemini-N    3   900  r'    180.7  3.918  4.093  14.4  24.53  0.09     -1    -1  18.50  0.09  17.60  0.14  1  -1
313P         2018-04-23  LDT         2   600  r'    215.5  3.703  2.722   4.0  23.07  0.10     -1    -1  18.05  0.10  17.65  0.11  1  -1
313P         2018-04-24  LDT         2   600  r'    215.7  3.701  2.719   3.9  23.10  0.11     -1    -1  18.09  0.11  17.69  0.12  1  -1 
313P         2018-05-08  LDT         4  1200  g'    217.4  3.682  2.711   5.0     -1    -1     -1    -1     -1    -1     -1    -1  0  -1
313P         2018-05-08  LDT         2   600  r'    217.4  3.682  2.711   5.0  23.00  0.04     -1    -1  18.00  0.04  17.54  0.07  1  -1
313P         2018-05-08  LDT         2   600  i'    217.4  3.682  2.711   5.0     -1    -1     -1    -1     -1    -1     -1    -1  0  -1
313P         2018-05-17  Magellan    7  2310  r'    218.6  3.668  2.735   7.0  23.21  0.03     -1    -1  18.20  0.03  17.63  0.08  1  -1
324P         2013-03-03  Gemini-N    7  1260  r'    178.8  3.570  2.695   8.6  23.97  0.06     -1    -1  19.05  0.06  18.37  0.17  1   9
324P         2013-04-08  Gemini-N    4   720  r'    183.6  3.569  3.021  14.7  24.69  0.08     -1    -1  19.53  0.08  18.57  0.24  1   9
324P         2013-04-12  Gemini-N   10  1800  r'    184.1  3.568  3.070  15.1  24.54  0.08     -1    -1  19.34  0.08  18.37  0.25  1   9
324P         2013-04-14  Gemini-N   10  1800  r'    184.4  3.568  3.094  15.3  24.61  0.10     -1    -1  19.40  0.10  18.41  0.26  1  -1
324P         2013-04-17  Gemini-N    6  1080  r'    184.8  3.567  3.133  15.6  24.43  0.07     -1    -1  19.19  0.07  18.19  0.25  1  -1
358P         2017-07-01  Gemini-S   -1    -1  r'    284.9  2.801  2.005  15.4  24.70  0.11     -1    -1  20.95  0.11  19.53  0.27  1  10
358P         2017-07-21  Gemini-S   -1    -1  r'    289.4  2.755  1.815   9.9  24.03  0.06     -1    -1  20.54  0.06  19.49  0.17  1  10
358P         2017-09-17  Gemini-S   -1    -1  r'    302.9  2.633  1.805  15.0  24.38  0.12     -1    -1  21.00  0.12  19.60  0.26  0  10
358P         2017-09-18  Gemini-S   -1    -1  r'    303.1  2.631  1.811  15.3  24.48  0.15     -1    -1  21.09  0.15  19.67  0.28  0  10
358P         2017-09-22  Gemini-S   -1    -1  r'    304.1  2.623  1.839  16.4  24.27  0.10     -1    -1  20.85  0.10  19.36  0.28  0  10
358P         2017-10-26  Gemini-S   -1    -1  r'    312.7  2.560  2.154  22.3  25.15  0.36     -1    -1  21.44  0.36  19.58  0.52  0  10
426P         2021-05-17  Gemini-S    1   300  r'    204.1  3.638  2.666   5.1  22.78  0.04     -1    -1  17.85  0.04  17.34  0.04  1  -1
426P         2021-06-13  Gemini-N    3   900  r'    207.7  3.619  2.662   6.2  22.81  0.02     -1    -1  17.89  0.02  17.32  0.02  1  -1
426P         2022-04-29  Magellan    3   195  R     254.6  3.243  3.135  18.1     -1    -1  23.08  0.10  18.04  0.10  16.92  0.10  1  -1
426P         2022-04-30  Magellan    3   560  r'    254.7  3.241  3.120  18.1  23.33  0.07     -1    -1  18.31  0.07  17.18  0.08  1  -1
426P         2022-05-26  Gemini-S    3   900  r'    259.1  3.201  2.721  17.4  23.06  0.03     -1    -1  18.36  0.03  17.27  0.04  1  -1
426P         2022-05-27  Gemini-S    2   600  r'    259.3  3.200  2.706  17.3  23.12  0.04     -1    -1  18.43  0.04  17.34  0.05  1  -1
426P         2022-06-07  Gemini-S    3   900  r'    261.1  3.183  2.551  16.0  22.83  0.03     -1    -1  18.28  0.03  17.25  0.04  1  -1
426P         2022-08-02  Palomar     3   900  r'    271.0  3.095  2.098   4.2  21.70  0.03     -1    -1  17.64  0.03  17.19  0.03  1  -1
426P         2022-08-04  Gemini-N    3   450  r'    271.4  3.091  2.096   4.4  21.65  0.01     -1    -1  17.59  0.01  17.13  0.01  1  -1
426P         2022-08-21  Gemini-N    6   900  r'    274.5  3.065  2.125   8.4  21.98  0.01     -1    -1  17.91  0.01  17.22  0.02  1  -1
426P         2022-08-27  Gemini-S    3   900  r'    275.6  3.056  2.152  10.1  22.08  0.01     -1    -1  17.99  0.01  17.22  0.02  1  -1
426P         2022-09-18  Gemini-N   12  1800  r'    279.8  3.021  2.320  15.6  22.48  0.01     -1    -1  18.25  0.01  17.23  0.03  1  -1
426P         2022-09-23  Gemini-S    3   900  r'    280.7  3.014  2.367  16.5  22.64  0.01     -1    -1  18.37  0.01  17.32  0.03  1  -1
426P         2022-09-30  Palomar    16  4800  r'    282.1  3.003  2.441  17.6  22.82  0.02     -1    -1  18.49  0.02  17.39  0.03  1  -1
426P         2022-10-22  Gemini-S    3   900  r'    286.4  2.969  2.695  19.5  22.96  0.02     -1    -1  18.44  0.02  17.27  0.04  1  -1
427P         2020-02-23  Gemini-S    3   900  r'    171.3  4.141  3.197   4.7  24.89  0.11     -1    -1  19.28  0.11  18.88  0.11  1  -1
427P         2021-06-12  Gemini-S    3   900  r'    219.5  3.768  3.201  13.9  25.31  0.16     -1    -1  19.90  0.16  19.11  0.16  1  -1
427P         2021-06-13  Gemini-S    6  1800  r'    219.6  3.766  3.212  14.0  25.10  0.08     -1    -1  19.69  0.08  18.89  0.08  1  -1
427P         2022-08-22  Gemini-N    7  1050  r'    293.7  2.533  1.934  21.2  23.57  0.04     -1    -1  20.12  0.04  19.08  0.04  1  -1
433P         2004-07-08  CFHT       -1    -1  i'    287.5  2.740  2.028  17.7  20.72  0.03  20.84  0.03  17.12  0.03  16.19  0.03  1  11
433P         2010-06-14  PS1        -1    -1  z'    339.3  2.416  1.728  21.1  20.07  0.13  20.22  0.13  17.12  0.13  16.08  0.13  0  11
433P         2010-08-02  PS1        -1    -1  i'    353.5  2.390  1.383   3.9  19.01  0.05  19.13  0.05  16.53  0.05  16.17  0.05  0  11
433P         2010-08-05  PS1        -1    -1  r'    354.4  2.389  1.378   2.5  18.81  0.03     -1    -1  16.22  0.03  15.95  0.03  0  11
433P         2010-08-06  PS1        -1    -1  g'    354.7  2.389  1.377   2.0  19.17  0.03  18.72  0.04  16.13  0.04  15.90  0.04  0  11
433P         2010-08-31  PS1        -1    -1  i'      2.1  2.388  1.429   9.8  19.23  0.04  19.35  0.04  16.68  0.04  16.05  0.04  0  11
433P         2010-09-06  PS1        -1    -1  g'      3.8  2.388  1.463  12.2  20.08  0.05  19.63  0.05  16.91  0.05  16.18  0.05  0  11
433P         2010-10-30  PS1        -1    -1  z'     19.6  2.413  2.018  23.8  20.48  0.23  20.64  0.23  17.20  0.23  16.09  0.23  0  11
433P         2011-11-24  PS1        -1    -1  r'    109.4  3.150  2.180   4.0  20.74  0.10     -1    -1  16.56  0.10  16.19  0.10  0  11
433P         2011-11-24  PS1        -1    -1  g'    109.4  3.150  2.180   4.0  21.20  0.13  20.75  0.13  16.57  0.13  16.20  0.13  0  11
433P         2011-11-30  PS1        -1    -1  i'    110.4  3.162  2.179   1.9  20.26  0.09  20.38  0.09  16.19  0.09  15.96  0.09  0  11
433P         2011-12-01  PS1        -1    -1  g'    110.6  3.164  2.180   1.4  20.98  0.10  20.53  0.10  16.34  0.10  16.15  0.10  0  11
433P         2015-08-18  SkyMapper  -1    -1  i'    320.3  2.483  1.888  21.8  20.72  0.26  20.84  0.26  17.49  0.26  16.43  0.26  0  11
433P         2018-12-15  Blanco     -1    -1  r'    191.7  3.733  3.508  15.2  22.63  0.31     -1    -1  17.04  0.31  16.20  0.31  1  11
433P         2020-02-04  Blanco     -1    -1  r'    248.9  3.165  3.059  18.1  21.91  0.11     -1    -1  16.98  0.11  16.04  0.11  1  11
433P         2020-02-10  Blanco     -1    -1  z'    250.0  3.152  2.960  18.2  21.59  0.14  21.75  0.14  16.90  0.14  15.96  0.14  1  11
P/2016 J1-A  2020-12-24  Gemini-N    2   600  r'    260.0  3.133  2.700  17.5  25.16  0.21     -1    -1  20.52  0.21  19.60  0.33  1  -1
P/2016 J1-A  2021-02-17  Gemini-N    3   900  r'    269.9  3.009  2.059   6.3  24.23  0.07     -1    -1  20.27  0.07  19.78  0.15  1  -1
P/2016 J1-A  2021-04-14  Gemini-N    6  1800  r'    280.9  2.884  2.162  16.0  24.71  0.08     -1    -1  20.74  0.08  19.87  0.26  1  -1
P/2016 J1-A  2021-04-17  Gemini-N    3   900  r'    281.5  2.877  2.187  16.7  24.82  0.14     -1    -1  20.83  0.14  19.93  0.29  1  -1
P/2016 J1-A  2021-05-17  Gemini-N    3   900  r'    287.9  2.812  2.487  20.9  25.01  0.21     -1    -1  20.79  0.21  19.76  0.36  1  -1
P/2016 J1-A  2022-04-08  Gemini-N    7   700  r'     13.3  2.462  2.177  23.9  23.22  0.05     -1    -1  19.57  0.05  18.45  0.32  0  -1
P/2016 J1-A  2022-04-25  Gemini-N    9   900  r'     18.0  2.472  1.992  23.0  22.83  0.02     -1    -1  19.37  0.02  18.28  0.31  0  -1
P/2016 J1-A  2022-06-19  Gemini-N    4   400  r'     33.1  2.526  1.594  11.6  21.91  0.03     -1    -1  18.89  0.03  18.18  0.20  0  -1
P/2016 J1-A  2022-07-07  Gemini-S    2   600  r'     37.9  2.549  1.584   9.1  21.70  0.02     -1    -1  18.67  0.02  18.06  0.17  0  -1
P/2016 J1-A  2022-08-02  Palomar     1   300  r'     44.6  2.588  1.701  13.5  22.34  0.07     -1    -1  19.12  0.07  18.34  0.23  0  -1
P/2016 J1-B  2021-05-29  Gemini-N    4  1200  r'    290.5  2.786  2.619  21.3  23.92  0.07     -1    -1  19.60  0.07  18.60  0.37  0  -1
P/2016 J1-B  2021-05-30  Gemini-N    3   900  r'    290.8  2.783  2.630  21.3  23.94  0.07     -1    -1  19.62  0.07  18.62  0.37  0  -1
P/2016 J1-B  2021-05-31  Gemini-N    1   300  r'    291.0  2.781  2.641  21.3  24.07  0.09     -1    -1  19.74  0.09  18.74  0.37  0  -1
P/2016 J1-B  2022-07-07  Gemini-N    4   400  r'     37.9  2.549  1.584   9.1  24.70  0.14     -1    -1  21.67  0.14  21.08  0.26  1  -1
P/2016 J1-B  2022-07-07  Gemini-S    2   600  r'     37.9  2.549  1.584   9.1  24.68  0.19     -1    -1  21.65  0.19  21.06  0.29  1  -1
P/2017 S9    2022-12-24  Gemini-S    1   300  r'    338.2  2.229  2.447  23.7  23.72  0.11     -1    -1  20.04  0.11  18.96  0.40  1  -1
P/2019 A3    2021-03-08  Gemini-S    3   900  r'    172.5  3.970  3.000   3.5  24.70  0.09     -1    -1  19.32  0.09  19.11  0.13  1  -1
P/2019 A3    2021-03-14  Gemini-S    3   900  r'    173.1  3.972  3.000   3.4  24.47  0.10     -1    -1  19.09  0.10  18.89  0.14  1  -1
P/2019 A3    2022-03-14  Gemini-S    3   900  r'    213.1  3.761  3.026  11.4  25.06  0.19     -1    -1  19.78  0.19  19.34  0.26  1  -1
P/2019 A3    2022-03-29  Gemini-S    3   900  r'    214.9  3.739  2.852   8.1  24.63  0.11     -1    -1  19.49  0.11  19.14  0.18  1  -1
P/2019 A3    2022-03-30  Gemini-S    1   300  r'    215.0  3.737  2.842   7.8  24.57  0.20     -1    -1  19.44  0.20  19.10  0.25  1  -1
P/2019 A3    2022-03-31  Gemini-S    6  1800  r'    215.1  3.736  2.832   7.5  24.60  0.08     -1    -1  19.48  0.08  19.14  0.16  1  -1
P/2019 A3    2022-04-01  Gemini-S    3   900  r'    215.3  3.734  2.823   7.3  24.73  0.08     -1    -1  19.62  0.08  19.29  0.16  1  -1
P/2019 A3    2022-04-23  Gemini-S    1   300  r'    217.9  3.699  2.695   0.8  24.32  0.17     -1    -1  19.33  0.17  19.24  0.17  1  -1
P/2019 A3    2022-04-29  Magellan    1   400  R     218.7  3.689  2.684   1.3     -1    -1  24.38  0.20  19.40  0.20  19.29  0.21  1  -1
P/2019 A3    2022-05-01  Gemini-S    3   900  r'    218.9  3.686  2.683   1.9  24.42  0.06     -1    -1  19.44  0.06  19.30  0.09  1  -1
P/2019 A3    2022-05-04  Gemini-S    3   900  r'    219.3  3.681  2.684   2.8  24.41  0.13     -1    -1  19.44  0.13  19.26  0.15  1  -1
P/2019 A3    2022-05-05  Gemini-S    2   600  r'    219.4  3.679  2.685   3.1  24.52  0.20     -1    -1  19.55  0.20  19.36  0.22  1  -1
P/2019 A3    2022-05-26  Gemini-S    3   900  r'    222.1  3.642  2.768   9.2  24.90  0.09     -1    -1  19.88  0.09  19.50  0.18  1  -1
P/2019 A3    2022-05-27  Gemini-S    3   900  r'    222.3  3.639  2.780   9.7  24.57  0.20     -1    -1  19.54  0.20  19.15  0.26  1  -1
\end{verbatim}
}}



\end{CJK*}
\end{document}